\documentclass[11pt]{article}
\usepackage{eurosym}
\usepackage{amsfonts}
\usepackage{amssymb}
\usepackage{graphicx}
\usepackage{amsmath}
\usepackage{makeidx}
\usepackage{indentfirst}
\usepackage[T1]{fontenc}
\usepackage[utf8]{inputenc}

\newcounter{resultnum}[section]
\newcounter{conclusionnum}[section]
\newcounter{conditionnum}[section]
\newcounter{conjecturenum}[section]
\newcounter{examplenum}[section]
\newcounter{exercisenum}[section]
\newcounter{lemmanum}[section]
\newcounter{notationnum}[section]
\newcounter{theoremnum}[section]
\newcounter{definitionnum}[section]
\newcounter{corollarynum}[section]
\newcounter{remarknum}[section]
\newcounter{propositionnum}[section]
\newcounter{acknowledgementnum}[section]
\newcounter{algorithmnum}[section]
\newcounter{axiomnum}[section]
\newcounter{casenum}[section]
\newcounter{claimnum}[section]
\newcounter{summarynum}[section]
\newcounter{problemnum}[section]
\begin{document}

\title{Off-diagonal solutions in Einstein gravity modeling f(R) gravity\\
and dynamical dark energy vs $\Lambda $CDM cosmology }
\date{Nov 20, 2025}
\author{ \textbf{Sergiu I. Vacaru} \thanks{%
emails: sergiu.vacaru@fulbrightmail.org ; sergiu.vacaru@gmail.com } \vspace{.1 in} \\
{\small \textit{Astronomical Observatory, Taras Shevchenko National University of Kyiv, Kyiv 01601, Ukraine}} \\
{\small \textit{Department of Physics, California State University at Fresno, Fresno, CA 93740, USA}} \\
{\small \textit{\ Department of Physics, Kocaeli University, Kocaeli, 41001, T\"{u}rkiye }}
 \vspace{0.1in} 
}
\maketitle

\begin{abstract}
Modified gravity theories (MGTs) have long been studied as alternatives to
general relativity (GR) and the standard $\Lambda$CDM cosmological model.
For example, exponential $f(R)$ models often yield better fits to
observational data, suggesting that $\Lambda$CDM may be inadequate. In this
work, we argue that the gravitational and accelerating cosmology paradigm
can remain close to GR and $\Lambda$CDM if one considers broader classes of
off-diagonal cosmological solutions of the Einstein equations. These
solutions are constructed using the anholonomic frame and connection
deformation method (AFCDM), which enables the decoupling and integration of
nonlinear systems in nonholonomic dyadic variables with connection
distortions. The resulting off-diagonal Einstein manifolds and cosmological
models are characterized by nonholonomic constraints, nonlinear symmetries,
and effective cosmological constants. Such structures allow one to
approximate cosmological effects, mimic features of MGTs, and describe
gravitational polarization, local anisotropies, and dark energy and dark
matter phenomena within GR. We further show that these models can be endowed
with relativistic versions of Perelman's thermodynamic variables for
geometric flows, which we compute in general form for accelerating cosmology.

\end{abstract}

\tableofcontents

%%%%%

%%%%%

%%%

\section{Introduction}

\label{sec1}

\vskip5pt A new era in the construction and study of gravity and cosmology
theories began with the discovery of late-time cosmic acceleration \cite%
{riess98,perlmutter99}. To explain and confront the growing body of
experimental and observational data, a wide range of modified gravity
theories (MGTs) has been developed. Early works and subsequent advances on
dark energy (DE) and dark matter (DM) physics can be found in \cite%
{sotiriou10,nojiri11,capo11,clifton12,harko14,copeland06,v16plb,od24,v25},
and references therein. Over the past 25 years, physicists have been
compelled to design new cosmological frameworks either by introducing
additional sources within general relativity (GR) or by elaborating and
refining various MGTs, which provide alternative ways to accommodate
observational results. Other researchers, however, prefer to preserve the
paradigm of the standard cosmological model (the $\Lambda$CDM model),
assuming that GR remains valid but requires the inclusion of additional -
yet unknown - DE and DM components.

%%%%%

\vskip5pt In a series of recent works \cite{vv25a,vv25b,v25}, we constructed
new classes of exact and parametric generic off-diagonal solutions, which
provide effective models of DE and DM phenomena both within GR and in MGTs.
The main objective of this paper is to demonstrate how such off-diagonal
cosmological solutions in GR, involving effective cosmological constants and
polarized physical constants, can be applied to describe modern
observational data. In particular, we analyze the conditions under which
these solutions can account for baryonic acoustic oscillations (BAO) \cite%
{desi24} and the Pantheon compilation of Type Ia supernovae (SN Ia)
observations \cite{pantheon21}.

\vskip5pt These off-diagonal cosmological solutions are determined in
general form by certain classes of generating and integration functions and
(effective) generating sources \cite{vv25a,vv25b,v25}, which may, in
principle, depend on all spacetime coordinates. Each class of solutions is
characterized by nonlinear symmetries of its generating data and associated
(effective) cosmological constants. This framework enables improved
estimation of free model parameters, such as the Hubble constant, as well as
refined equations of state for DE. Consequently, it offers promising
approaches to addressing the Hubble constant tension problem \cite%
{desi24,roy24,batic24,dival21}. Observational data from SN Ia, BAO, CMB, and
related probes have already been studied extensively within MGTs,
particularly in the context of $f(R)$ gravity theories \cite%
{hu07,appl07,nojiri08,cog07}.\footnote{%
We cite here only some early works on $f(R)$ gravity (alternatively denoted
as $F(R)$). The broader bibliography on MGTs is vast, containing thousands
of papers, and a comprehensive survey is beyond the scope of this article.}
For the purposes of this paper, we focus instead on results related to
so-called exponential gravity and the $\Lambda$CDM framework \cite%
{yang10,od20}.

\vskip5pt The classical gravitational and matter field equations in GR and
MGTs form highly nonlinear systems of partial differential equations (PDEs).
For generic off-diagonal metric ansatze and Levi-Civita (LC) or other types
of (non)linear connections, such PDE systems cannot, in general, be solved
in closed analytic form, even with advanced analytic and numerical methods.
Moreover, the physical relevance of generic off-diagonal solutions (which
cannot be diagonalized through coordinate transformations in finite
spacetime regions) has often been unclear, and their applications in
cosmology and astrophysics have been considered problematic, especially
given their inherently nonlinear interactions and the presence of
nonhomogeneous and locally anisotropic dynamics. Historically, the most
important exact or parametric solutions describing black holes (BHs),
wormholes (WHs), and cosmological models were obtained using diagonalizable
metric ansatz with high symmetries (spherical, cylindrical, etc.), typically
depending on a single spatial or temporal coordinate and involving rotation
or Killing symmetries. In GR, the reduction of Einstein's equations to
systems of nonlinear ordinary differential equations (ODEs), along with the
corresponding physically significant solutions, is reviewed in the classical
references \cite{hawking73,misner73,wald82,kramer03}.

\vskip5pt In \cite{vv25a,vv25b}, we explained in detail why it is essential
to study generic off-diagonal configurations and relativistic G.
Perelman-type \cite{perelman1} thermodynamics within GR (see also \cite%
{gheorghiuap16} for applications to geometric flows and $R^{2}$-gravity). On
a four-dimensional Lorentzian spacetime manifold, a generic off-diagonal
metric is characterized by six independent coefficients, each depending on
all spacetime coordinates (four of the ten coefficients of a symmetric
metric can always be eliminated, consistent with the Bianchi identities). By
contrast, prescribing a diagonal metric ansatz with at most four independent
coefficients, while imposing smoothness and symmetry conditions, reduces the
(modified) Einstein equations to systems of second-order ODEs. This
restriction, however, precludes the construction of more general classes of
solutions, including off-diagonal configurations with additional degrees of
freedom governed by nonlinear PDEs.

\vskip5pt The anholonomic frame and connection deformation method (AFCDM),
developed in our works beginning in 1988, provides a systematic geometric
and analytic approach for constructing generic off-diagonal solutions in GR
and MGTs. Reviews of its applications and subsequent developments can be
found in \cite{sv11,bsssvv25}, while more recent results for Einstein
gravity and nonmetric Einstein--Dirac systems are presented in \cite%
{vv25a,vv25b,v25}. The AFCDM enables the decoupling and integration of
certain nonlinear PDE systems in general form, without reducing them to
ODEs. Importantly, such off-diagonal solutions in 4D gravity involve two
additional degrees of freedom, even in Einstein's theory, which allow us to
model new observational data and explore nonlinear off-diagonal
gravitational and (effective) matter field interactions. This framework
opens the way for constructing new models of nonlinear classical and quantum
theories, locally anisotropic thermodynamics, and for investigating
nonlinear and parametric effects in DE and DM physics.

\vskip5pt The main \textbf{Hypothesis} of this work is that \textit{%
accelerating cosmological models and the various DE and DM effects usually
attributed to MGTs (for instance, to exponential $f(R)$ gravity) can instead
be modeled by \textbf{target} generic off-diagonal solutions in GR with
effective cosmological constants}. Such target metrics are constructed as
nonholonomic/off-diagonal deformations and connection distortions of certain 
\textbf{primary} MGT configurations, e.g., of exponential type.

The resulting off-diagonal cosmological models mimic $\Lambda$CDM cosmology
asymptotically (at early times), when exponential terms become negligible,
though residual parametric deformations may persist. However, the subsequent
nonholonomic and off-diagonal cosmological evolution can differ essentially
from that predicted by $\Lambda$CDM or other MGTs.

In our approach, GR and the standard cosmological paradigm are not
fundamentally altered. Instead, the generating and integration data for such
solutions can be chosen to reproduce observational data with high accuracy.
For appropriate nonholonomic constraints and small parametric deformations,
the off-diagonal cosmological metrics can effectively reproduce predictions
of exponential or other classes of MGTs. Moreover, such solutions may encode
data from nonmetric gravity theories or from nonassociative star-product
deformations of string gravity studied in \cite{v25,bsssvv25}. We argue that
the effects of these general MGTs can be equivalently modeled on 4-d
Einstein manifolds with effective geometric-flow driven cosmological
constants $\Lambda (\tau )$ and corresponding $\tau $-families of
off-diagonal cosmological metrics in GR.\footnote{%
Here $\tau $ denotes a positive, temperature-like parameter.}

In this framework, possible DE and DM configurations arise from the
off-diagonal terms of the metric and from nonlinear, locally anisotropic
polarizations of physical constants. The Bekenstein-Hawking paradigm is not
applicable to describe the thermodynamic properties of such accelerating,
locally anisotropic cosmological configurations.

\vskip5pt The paper is organized as follows: Section \ref{sec2} provides
geometric preliminaries on how cosmological solutions in MGTs can be
equivalently modeled via off-diagonal GR configurations with effective
cosmological constants. We also compute generalized Perelman-type
thermodynamic variables, which are essential for selecting "more optimal"
cosmological solutions in GR and MGTs. Section \ref{sec3} briefly discusses
how SN Ia, BAO, and other observational data can be described by generic
off-diagonal solutions. The resulting DE and cosmological models, together
with their equations of state (EoS), are analyzed, discussed, and confronted
with observations. Finally, conclusions are presented in Section \ref{sec4}.

%%%%%%

\section{Generic off-diagonal cosmological solutions in GR and geometric
flows}

\label{sec2} The formulation of GR in nonholonomic dyadic variables with
distortions of connections allows the application of the AFCDM for
generating off-diagonal physically important solutions. Such cosmological or
quasi-stationary solutions are characterized by relativistic generalizations
of G. Perelman thermodynamics \cite{vv25a,vv25b}; see also \cite%
{v25,bsssvv25}, for more general constructions concerning MGTs with
nonmetricity or nonsymmetric metrics, nonassociative and noncommutative
nonholonomic, of Finsler-like variables, etc. We provide the necessary
geometric preliminaries in the first subsection. Then (in the next
subsections, by applying the AFCDM), we construct new classes of
off-diagonal cosmological metrics in GR with effective cosmological
constants. Certain conditions on the nonholonomic vacuum structure and
effective generating sources for such solutions, which encode the main
features of \textit{the exponential f(R) gravity}, are formulated. %%%%%%

\subsection{Nonholonomic 2+2 spacetime splitting and distortion of
connections}

Let us consider a 4-d Lorentz spacetime nonholonomic manifold $\mathbf{V}$
of signature $(+++-)$ enabled with a (formal) nonlinear connection,
N-connection, structure defined as a nonholonomic fibered $2+2$ distribution 
$\mathbf{N}:\ T\mathbf{V}=h\mathbf{V}\oplus v\mathbf{V.}$ Such a Whitney sum 
$\oplus $ defines a conventional 2-d horizontal (h) and vertical (v)
non-integrable (equivalently, nonholonomic or anholonomic) splitting with
local coefficients 
\begin{equation}
\mathbf{N}(u)=N_{i}^{a}(x,y)dx^{i}\otimes \partial /\partial y^{a},
\label{ncon}
\end{equation}
when local coordinates $x=\{x^{i}\}$ and $y=\{y^{a}\}$ are labeled by
abstract or coordinate indices running values $i,j,...=1,2$ and $%
a,b,...=3,4, $ for $y^{4}=ct$ is a time-like coordinate (we can always
consider that the velocity of light is $c=1$). We consider that $\mathbf{V}$
is a pseudo-Riemannian manifold of necessary smooth class in any point $%
u=\{u^{\alpha }\}=\{x^{i},y^{a}\},$ for $\alpha ,\beta ,...=1,2,3,4.$ The
coefficients $N_{i}^{a}$ allow us to introduce locally some N-adapted frames
and, respectively, coframes: 
\begin{eqnarray}
\mathbf{e}_{\nu } &=&(\mathbf{e}_{i},e_{a})=(\mathbf{e}_{i}=\partial
/\partial x^{i}-\ N_{i}^{a}(u)\partial /\partial y^{a},\ e_{a}=\partial
_{a}=\partial /\partial y^{a}),\mbox{ and  }  \label{nader} \\
\mathbf{e}^{\mu } &=&(e^{i},\mathbf{e}^{a})=(e^{i}=dx^{i},\ \mathbf{e}%
^{a}=dy^{a}+\ N_{i}^{a}(u)dx^{i}).  \label{nadif}
\end{eqnarray}%
For instance, a N-elongated basis (\ref{nader}) satisfies certain
nonholonomic relations $[\mathbf{e}_{\alpha },\mathbf{e}_{\beta }]=\mathbf{e}%
_{\alpha }\mathbf{e}_{\beta }-\mathbf{e}_{\beta }\mathbf{e}_{\alpha
}=W_{\alpha \beta }^{\gamma }\mathbf{e}_{\gamma }.$ The (antisymmetric)
nontrivial anholonomy coefficients are computed $W_{ia}^{b}=\partial
_{a}N_{i}^{b},W_{ji}^{a}=\Omega _{ij}^{a}=\mathbf{e}_{j}\left(
N_{i}^{a}\right) -\mathbf{e}_{i}(N_{j}^{a}),$ where $\Omega _{ij}^{a}$
define the coefficients of an N-connection curvature $\Omega $.\footnote{%
A N-adapted base $\mathbf{e}_{\alpha }\simeq \partial _{\alpha }=\partial
/\partial u^{\alpha }$ is holonomic if and only if all anholonomy $W_{\alpha
\beta }^{\gamma }$ coefficients vanish. If so, the usual partial derivatives 
$\partial _{\alpha }$ can be considered using certain coordinate transforms.
We shall typically use boldface labels of geometric objects (like $\mathbf{A}%
=\{\mathbf{A}_{\ \beta }^{\alpha }\}$) to emphasize that such geometric/
physical objects are adapted to an N-connection structure, and called, in
brief, d-objects, or d-vectors, d-tensors.}

Any metric structure $\mathbf{g}$ on $\mathbf{V}$ can be written
equivalently as a d--metric or, respectively, in a coordinate base, 
\begin{equation}
\ \mathbf{g}=(hg,vg)=\ g_{ij}(x,y)\ e^{i}\otimes e^{j}+\ g_{ab}(x,y)\ 
\mathbf{e}^{a}\otimes \mathbf{e}^{b}=\underline{g}_{\alpha \beta
}(u)du^{\alpha }\otimes du^{\beta },  \label{dm}
\end{equation}%
for $hg=\{\ g_{ij}\}$ and $\ vg=\{g_{ab}\}.$ We compute the off-diagonal
coefficients if we introduce the coefficients of (\ref{nadif}) into (\ref{dm}%
) with a corresponding regrouping for a coordinate dual basis: 
\begin{equation}
\underline{g}_{\alpha \beta }=\left[ 
\begin{array}{cc}
g_{ij}+N_{i}^{a}N_{j}^{b}g_{ab} & N_{j}^{e}g_{ae} \\ 
N_{i}^{e}g_{be} & g_{ab}%
\end{array}%
\right] .  \label{ansatz}
\end{equation}
Such a (d-) metric $\mathbf{g}=\{\underline{g}_{\alpha \beta }\}$ is generic
off--diagonal if the anholonomy coefficients $W_{\alpha \beta }^{\gamma }$
are not all zero. %%%%%%

Let us summarize some definitions and results of \cite{vv25a,vv25b} which
are important for this work: A \textbf{\ d--connection} $\mathbf{D}=(hD,vD) $
is a linear (equivalently, affine) connection preserving under parallelism
the N--connection splitting (\ref{ncon}). Using a $\mathbf{D,}$ we define a
covariant N--adapted derivative $\mathbf{D}_{\mathbf{X}}\mathbf{Y.}$ Such
constructions can be performed for a d--vector field $\mathbf{Y}=hY+vY$ in
the direction of a d--vector $\mathbf{X}=hX+vC.$ For N--adapted frames (\ref%
{nader}) and (\ref{nadif}), any covariant d-derivative $\mathbf{D}_{\mathbf{X%
}}\mathbf{Y}$ can be computed as in GR \cite{misner73} and, in a more
general sense as in metric-affine gravity and various MGTs \cite%
{hehl95,v25,bsssvv25}. The N-adapted coefficients involve respective h- and
v-indices, 
\begin{equation}
\mathbf{D}=\{\mathbf{\Gamma }_{\ \alpha \beta }^{\gamma }=(L_{jk}^{i},\acute{%
L}_{bk}^{a};\acute{C}_{jc}^{i},C_{bc}^{a})\},\mbox{ where }hD=(L_{jk}^{i},%
\acute{L}_{bk}^{a})\mbox{ and }vD=(\acute{C}_{jc}^{i},C_{bc}^{a}).
\label{hvdcon}
\end{equation}
%%%%%%

Any d--connection $\mathbf{D}$ is characterized by three fundamental
geometric d-objects, 
\begin{eqnarray}
\mathcal{T}(\mathbf{X,Y}) &:=&\mathbf{D}_{\mathbf{X}}\mathbf{Y}-\mathbf{D}_{%
\mathbf{Y}}\mathbf{X}-[\mathbf{X,Y}],\mbox{ torsion d-tensor,  d-torsion};
\label{fundgeom} \\
\mathcal{R}(\mathbf{X,Y})&:=&\mathbf{D}_{\mathbf{X}}\mathbf{D}_{\mathbf{Y}}-%
\mathbf{D}_{\mathbf{Y}}\mathbf{D}_{\mathbf{X}}-\mathbf{D}_{\mathbf{[X,Y]}},%
\mbox{ curvature d-tensor, d-curvature};  \notag \\
\mathcal{Q}(\mathbf{X}) &:=&\mathbf{D}_{\mathbf{X}}\mathbf{g,}%
\mbox{nonmetricity d-fields, d-nonmetricity}.  \notag
\end{eqnarray}%
We note that a LC connection $\nabla $ is not a d-connection because it does
not preserve an h- and v-decomposition under parallel transports.
Nevertheless, if we consider a zero distortion d-tensor, $\mathbf{Z,}$ for $%
\mathbf{D=}\nabla +\mathbf{Z,}$ i.e. $\mathbf{D\rightarrow }\nabla ,$ we can
compute similar distortions and geometric objects like $\ ^{\nabla }\mathcal{%
T}(\mathbf{X,Y}):=\nabla _{\mathbf{X}}\mathbf{Y}-\nabla _{\mathbf{Y}}\mathbf{%
X}-[\mathbf{X,Y}]=0\,$ and $\ ^{\nabla }\mathcal{Q}(\mathbf{X}):=\ ^{\nabla }%
\mathbf{D}_{\mathbf{X}}\mathbf{g}=0$, but $\ ^{\nabla }\mathcal{R}\neq 0$ is
just that for the pseudo-Riemannian geometry. %%%%%%%%

For any d-metric structure $\mathbf{g}$ (\ref{dm}), we can define two
important linear connection structures and a respective canonical distortion
relation: 
\begin{eqnarray}
(\mathbf{g,N}) &\rightarrow &\left\{ 
\begin{array}{cc}
\mathbf{\nabla :} & \mathbf{\nabla g}=0;\ _{\nabla }\mathcal{T}=0,\ 
\mbox{\
the LC--connection }; \\ 
\widehat{\mathbf{D}}: & \widehat{\mathbf{Q}}=0;\ h\widehat{\mathcal{T}}=0,v%
\widehat{\mathcal{T}}=0,\ hv\widehat{\mathcal{T}}\neq 0,%
\mbox{ the canonical
d-connection}.%
\end{array}%
\right.  \label{twocon} \\
&\rightarrow &\widehat{\mathbf{D}}[\mathbf{g}]=\nabla \lbrack \mathbf{g}]+%
\widehat{\mathcal{Z}}[\mathbf{g}],  \label{canondist}
\end{eqnarray}%
where $\widehat{\mathcal{Z}}[\mathbf{g}]=\{\widehat{\mathbf{Z}}_{\
\alpha\beta }^{\gamma }[\mathbf{g,N}]\}$ is the canonical distortion
d-tensor. In \cite{vv25a,vv25b}, we proved in detail that GR can be defined
equivalently using both types of geometric data $[\mathbf{g},\nabla ]$ and
(or) $[\mathbf{g},\mathbf{N},\widehat{\mathbf{D}}].$ The priority of hat
variables is that they a allow to decouple and integrate the Einstein
equations with nontrivial N-connection structure $\mathbf{N}$ "absorbing" in
a sense the off-diagonal terms in $\mathbf{g}=\{\underline{g}_{\alpha \beta
}\}$ from (\ref{dm}). It should be noted that the distortions (\ref%
{canondist}) involve a canonical d-torsion structure, $\widehat{\mathcal{T}}%
=\{\widehat{\mathbf{T}}_{\ \alpha \beta }^{\gamma }\}$, as we stated in (\ref%
{twocon}). We do not need additional sources (spin-like as in
Einstein-Cartan gravity, or an H-field as in string gravity) for the
canonical d-torsion $\widehat{\mathcal{T}}=\{\widehat{\mathbf{T}}_{\ \alpha
\beta }^{\gamma }\}$. We can include the distortions of the Ricci tensor as
certain effective matter sources in the Einstein equations for $[\mathbf{g}%
,\nabla ]$. An alternative variant for extracting LC configurations is to
impose additional constraints on generating and integration functions for
respective solutions (see next subsection), which result in zero distortion
d-tensors, 
\begin{equation}
\widehat{\mathbf{Z}}=0,\mbox{ which is equivalent to }\ \widehat{\mathbf{D}}%
_{\mid \widehat{\mathcal{T}}=0}=\nabla .  \label{lccond}
\end{equation}
%%%%

The Einstein equations in GR can be written equivalently in hat variables,
which is more convenient for general decoupling and integration in generic
off-diagonal form, 
\begin{eqnarray}
\widehat{\mathbf{R}}_{\ \ \beta }^{\alpha } &=&\widehat{\mathbf{\Upsilon }}%
_{\ \ \beta }^{\alpha },  \label{cdeq1} \\
\widehat{\mathbf{T}}_{\ \alpha \beta }^{\gamma } &=&0,%
\mbox{ if we extract
LC configuations with }\nabla .  \label{lccond1}
\end{eqnarray}%
All coefficients are defined in N-adapted frames (\ref{nader}) and (\ref%
{nadif}). The equations (\ref{lccond1}) are equivalent to (\ref{lccond}),
when the induced nonholonomic d-torsion $\widehat{\mathcal{T}}=\{\widehat{%
\mathbf{T}}_{\ \alpha \beta }^{\gamma }[\mathbf{g,N,}\widehat{\mathbf{D}}]\}$
is defined as in (\ref{fundgeom}). This system of nonlinear PDEs can be
derived in an abstract geometric form as in \cite{misner73} but using $%
\widehat{\mathbf{D}}$ and respective N-adapted frame transforms and
distortions of geometric d-objects. %%%%%%

We emphasize that the nonholonomic canonical gravitational equations (\ref%
{cdeq1}) can be proven in N-adapted variational form. We can introduce
conventional gravitational and matter fields Lagrange densities, $\ ^{g}L(%
\widehat{\mathbf{R}}is)$ ($\widehat{\mathbf{R}}is$ is the Ricci scalar for $%
\widehat{\mathbf{D}},$ similarly as in GR with $\ ^{g}L(R)$). We can
postulate a $\ ^{m}L(\varphi ^{A},\mathbf{g}_{\beta \gamma }),$ when the
stress-energy d-tensor of matter fields $\varphi ^{A}$ (labelled by a
general index $A$) is defined and computed as in GR but with respective
dyadic decompositions, 
\begin{equation}
\mathbf{T}_{\alpha \beta }=-\frac{2}{\sqrt{|\mathbf{g}_{\mu \nu }|}}\frac{%
\delta (\ ^{m}L\sqrt{|\mathbf{g}_{\mu \nu }|})}{\delta \mathbf{g}^{\alpha
\beta }}.  \label{emdt}
\end{equation}%
Defining $T:=\mathbf{g}^{\alpha \beta }\mathbf{T}_{\alpha \beta }$ and
certain effective sources determined by distortions of Ricci d-tensors, we
can consider $\widehat{\mathbf{Y}}[\mathbf{g,}\widehat{\mathbf{D}}]\simeq \{%
\mathbf{T}_{\alpha\beta }-\frac{1}{2}\mathbf{g}_{\alpha \beta }T\}.$ In
various physical theories like \cite{v25}, we can postulate more general $\
^{m}L,$ for instance, depending on some covariant/spinor derivatives. For
our purposes, we consider (effective) sources $\widehat{\mathbf{Y}}[\mathbf{%
g,}\widehat{\mathbf{D}}]=\{\Upsilon _{~\delta }^{\beta }(x,y)\}$
parameterized as : 
\begin{equation}
\widehat{\Upsilon }_{~\delta }^{\beta }=diag[\Upsilon _{\alpha }:\Upsilon
_{~1}^{1}=\Upsilon _{~2}^{2}=~^{h}\Upsilon (x^{k});\Upsilon
_{~3}^{3}=\Upsilon _{~4}^{4}=~^{v}\Upsilon (x^{k},y^{a})].  \label{esourc}
\end{equation}%
For some general classes of energy-momentum tensors, we can define
respective frame/coordinate transforms if such conditions are not satisfied
for a $\Upsilon _{\beta \delta }$). Such assumptions stated that we generate
off-diagonal solutions for certain classes of nonholonomic transforms and
constraints when the effective sources are determined by \textbf{two
generating sources} $\ ^{h}\Upsilon (x^{k})$ and $\ ^{v}\Upsilon
(x^{k},y^{a})$. It imposes certain nonholonomic constraints on the dynamics
of $\mathbf{T}_{\alpha \beta }$ (\ref{emdt}), with possible (effective)
cosmological constant $\Lambda $ and a conventional splitting of constants
into h- and v-components. Such constraints may involve distortion d-tensors $%
\widehat{\mathbf{Z}}[\mathbf{g}]$ and other values included in $\widehat{%
\mathbf{Y}}.$ %%%%%

We emphasize that a parametrization (\ref{esourc}) allows us to decouple and
integrate in general forms the geometric flows and gravitational and matter
field equations. Such constructions are possible if we consider that $%
\widehat{\mathbf{Y}}[\mathbf{g,}\widehat{\mathbf{D}},\kappa ]$ contains a
small parameter $\kappa $, or if the gravitational and matter field dynamics
is subjected to certain convenient classes of constraints, trapping
hypersurface conditions, ellipsoid symmetries etc. \cite%
{v16plb,vv25a,vv25b,v25,bsssvv25}. In such cases, the solutions can be
constructed exactly or recurrently using power decompositions on a small
constant (it can be a deformation one, or an additional physical constant) $%
\kappa ^{0},\kappa ^{1},\kappa ^{2},...$ We say that the corresponding
classes of solutions are exact or parametric; for simplicity, we can study
only linear dependencies on $\kappa ^{0}$ and $\kappa ^{1}.$ %%%%%%

Finally, we note that the conservation laws for (\ref{cdeq1}) can be written
in a form with $\widehat{\mathbf{D}}^{\beta }\widehat{\mathbf{\Upsilon }}_{\
\ \beta }^{\alpha }\neq 0,$ which is different from the Einstein and
energy-momentum tensors written in standard form in GR. Non-zero covariant
divergences are typical for nonholonomic systems, and if the constraints (%
\ref{lccond1}) are not imposed. This is similar to the nonholonomic
mechanics; the conservation laws are not standard ones. Using distortions
relations, we can rewrite (\ref{cdeq1}) in terms of $\nabla ,$ when $\nabla
^{\beta }E_{\ \ \beta }^{\alpha }=\nabla ^{\beta }T_{\ \ \beta }^{\alpha
}=0. $ We conclude that there are no conceptual problems with the
formulation of GR and the definition of conservation laws for matter fields
using two different linear connections (\ref{twocon}), which are defined by
the same metric structure $\mathbf{g}$. We can use $\widehat{\mathbf{D}}$ to
find off-diagonal solutions and then to constrain the integral varieties to
extract LC-configurations. %%%%%

\subsection{Generating off-diagonal cosmological solutions using the AFCDM}

The application of the AFCDM for generating off-diagonal cosmological
solutions is explained in detail in \cite{vv25a}, particularly in formulas
(74) and (77) and in Table 3 of Appendix B.3 of that work. The goal of this
subsection is to re-formulate certain results in a form suitable for
constructing cosmological solutions of the Einstein equations (\ref{cdeq1})
written in canonical dyadic variables, taking into account that the
conditions (\ref{lccond1}) can always be imposed additionally when it is
necessary to extract LC configurations. We follow the same conventions and
notations as in \cite{vv25a} for constructing d-metric target off-diagonal
Einstein cosmological spacetimes, where the primary metrics in the present
work are chosen for the exponential $f(R)$ model introduced in \cite{od24}.
It is worth noting that in many other papers on MGTs, the notation $F(R)$ is
used instead.

%%%%%%
%%%%%%

\subsubsection{General decoupling of the Einstein equations for canonical
ansatz in N-adapted frames}

For constructing locally anisotropic cosmological solutions, we can employ
an off-diagonal (in coordinate frames) canonical ansatz for the d-metric: 
\begin{eqnarray}
\underline{\mathbf{g}} &=&g_{i}(x^{k})dx^{i}\otimes dx^{i}+\underline{h}%
_{3}(x^{k},t)\underline{\mathbf{e}}^{3}\otimes \underline{\mathbf{e}}^{3}+%
\underline{h}_{4}(x^{k},t)\underline{\mathbf{e}}^{4}\otimes \underline{%
\mathbf{e}}^{4},  \notag \\
&&\underline{\mathbf{e}}^{3}=dy^{3}+\underline{n}_{i}(x^{k},t)dx^{i},\ 
\underline{\mathbf{e}}^{4}=dy^{4}+\underline{w}_{i}(x^{k},t)dx^{i}=dt+%
\underline{w}_{i}(x^{k},t)dx^{i}.  \label{dmc}
\end{eqnarray}%
In N-adapted frames (\ref{nadif}) and for corresponding local coordinates,
this metric exhibits an explicit Killing symmetry along the space-like
direction $\partial _{3}$ and highlights a generic dependence on the
time-like coordinate $y^{4}=t$. In our works, we use underlined symbols
(such as $\underline{h}_{a},\ \underline{w}_{i},\ \underline{n}_{i}$) to
emphasize quantities that depend explicitly on the time-like coordinate,
characterizing locally anisotropic configurations. Such underlining can be
omitted when treating more general or, for instance, quasi-stationary
configurations. The corresponding N-connection coefficients are
parameterized as $\underline{N}_{i}^{3}=\underline{n}_{i}(x^{k},t)$ and $%
\underline{N}_{i}^{4}=\underline{w}_{i}(x^{k},t)$ and and the d-metric
coefficients take the general form $\underline{\mathbf{g}}_{\alpha \beta
}=[g_{ij}(x^{\kappa }),\underline{g}_{ab}(x^{\kappa },t)],$ where all
functions are assumed to belong to the necessary smooth class.

In this subsection, we outline certain general decoupling and integration
properties using ansatz of type (\ref{dmc}). To generate quasi-stationary
d-metrics, one can modify the N-adapted coefficients, for instance, by
performing the substitutions $\underline{h}_{4}(x^{k},t)\rightarrow
h_{3}(x^{k},y^{3}), \underline{h}_{3}(x^{k},t)\rightarrow h_{4}(x^{k},y^{3})$
and $\underline{n}_{i}(x^{k},t)\rightarrow w_{i}(x^{k},y^{3}), \underline{w}%
_{i}(x^{k},t)\rightarrow n_{i}(x^{k},y^{3}).$ Such N-adapted dual space
``time symmetries can be introduced only for generic off-diagonal
configurations admitting respective Killing symmetries along $\partial_{4}$
or $\partial_{3}$. Even under these restrictions, one can still investigate
the main geometric and physical properties of generic off-diagonal
cosmological metrics.

It should be noted that the AFCDM can be extended to more general classes of
d-metrics, as discussed in \cite{vv25a,bsssvv25}. However, such
generalizations typically lead to more cumbersome expressions and require
more sophisticated geometric techniques.

Furthermore, by imposing additional nonholonomic constraints and
deformations, one can generate new classes of exact solutions to systems of
nonlinear PDEs, which can be interpreted either within GR or as deformations
to various types of MGTs (through the introduction of alternative effective
sources). The coefficients of a d-metric $\underline{\mathbf{g}}_{\alpha
\beta }(x^{k},t)$ depend generically on three of the four spacetime
coordinates. Therefore, such an ansatz provides almost direct solutions of
the field equations (\ref{cdeq1}), without reducing the problem to solving
simplified systems of nonlinear ODEs.

%%%%%%%

A tedious computation of the nontrivial coefficients of the canonical Ricci
d-tensor $\widehat{\mathbf{R}}_{\ \ \beta }^{\alpha }[\underline{\mathbf{g}}%
]=$ $\widehat{\underline{\mathbf{R}}}_{\ \ \beta }^{\alpha }$ for the
off-diagonal cosmological ansatz $\underline{\mathbf{g}}$ (\ref{dmc}) is
similar to that presented in \cite{vv25a,bsssvv25}. For such locally
anisotropic cosmological configurations, the nonholonomic Einstein equations
(\ref{cdeq1}) with effective sources of type (\ref{esourc}), 
\begin{equation}
\widehat{\Upsilon }_{~\delta }^{\beta }\rightarrow \widehat{\underline{%
\Upsilon }}_{~\delta }^{\beta }=diag[\underline{\Upsilon }_{\alpha
}:\Upsilon _{~1}^{1}=\Upsilon _{~2}^{2}=~^{h}\Upsilon (x^{k});\underline{%
\Upsilon }_{~3}^{3}=\underline{\Upsilon }_{~4}^{4}=~^{v}\underline{\Upsilon }%
(x^{k},t)],  \label{esourcc}
\end{equation}
can be written in the form: 
\begin{eqnarray}
\widehat{R}_{1}^{1} &=&\widehat{R}_{2}^{2}=\frac{1}{2g_{1}g_{2}}[\frac{%
g_{1}^{\bullet }g_{2}^{\bullet }}{2g_{1}}+\frac{(g_{2}^{\bullet })^{2}}{%
2g_{2}}-g_{2}^{\bullet \bullet }+\frac{g_{1}^{\prime }g_{2}^{\prime }}{2g_{2}%
}+\frac{\left( g_{1}^{\prime }\right) ^{2}}{2g_{1}}-g_{1}^{\prime \prime
}]=-\ ^{h}\Upsilon ,  \notag \\
\underline{\widehat{R}}_{3}^{3} &=&\underline{\widehat{R}}_{4}^{4}=\frac{1}{2%
\underline{h}_{3}\underline{h}_{4}}[\frac{\left( \underline{h}_{3}^{\diamond
}\right) ^{2}}{2\underline{h}_{3}}+\frac{\underline{h}_{3}^{\diamond }%
\underline{h}_{4}^{\diamond }}{2\underline{h}_{4}}-\underline{h}%
_{3}^{\diamond \diamond }]=-\ ^{v}\underline{\Upsilon },  \label{riccist2} \\
\underline{\widehat{R}}_{3k} &=&\frac{\underline{h}_{3}}{2\underline{h}_{4}}%
\underline{n}_{k}^{\diamond \diamond }+\left( \frac{3}{2}\underline{h}%
_{3}^{\diamond }-\frac{\underline{h}_{3}}{\underline{h}_{4}}\underline{h}%
_{4}^{\diamond }\right) \frac{\ \underline{n}_{k}^{\diamond }}{2\underline{h}%
_{4}}=0;  \notag \\
\underline{\widehat{R}}_{4k} &=&\frac{\ \underline{w}_{k}}{2\underline{h}_{3}%
}[\underline{h}_{3}^{\diamond \diamond }-\frac{\left( \underline{h}%
_{3}^{\diamond }\right) ^{2}}{2\underline{h}_{3}}-\frac{(\underline{h}%
_{3}^{\diamond })(\underline{h}_{4}^{\diamond })}{2\underline{h}_{4}}]+\frac{%
\underline{h}_{3}^{\diamond }}{4\underline{h}_{3}}(\frac{\partial _{k}%
\underline{h}_{3}}{\underline{h}_{3}}+\frac{\partial _{k}\underline{h}_{4}}{%
\underline{h}_{4}})-\frac{\partial _{k}(\underline{h}_{4}^{\diamond })}{2%
\underline{h}_{4}}=0  \notag
\end{eqnarray}%
In these formulas, for example, $\underline{h}_{3}^{\diamond }=\partial _{4}%
\underline{h}_{3}=\partial _{t}\underline{h}_{3},$ when $\underline{h}%
_{3}^{\ast }=\partial _{3}\underline{h}_{3}=0$ (such assumptions are
necessary if we construct non-degenerate solutions, which can be always
satisfied by choosing corresponding N-adapted frames and systems of
coordinates). We note that $\partial _{3}h_{3}$ can be not zero for
quasi-stationary configurations, or if we consider other systems of frames/
coordinates.

The equations (\ref{riccist2}) can be written in a more compact symbolic
form if we express $g_{i}=e^{\psi (x^{k})}$ and introduce the coefficients 
\begin{equation}
\underline{\alpha }_{i}=\underline{h}_{3}^{\diamond }\partial _{i}(%
\underline{\varpi }),\underline{\beta }=\underline{h}_{3}^{\diamond }(%
\underline{\varpi })^{\diamond }\mbox{  and }\underline{\gamma }=(\ln \frac{|%
\underline{h}_{3}|^{3/2}}{|\underline{h}_{4}|})^{\diamond },  \label{coeff}
\end{equation}%
for$\underline{\varpi }=\ln |\underline{h}_{3}^{\diamond }/\sqrt{|\underline{%
h}_{3}\underline{h}_{4}}|,$ where $\underline{\Psi }=\exp (\underline{\varpi 
})$ can be considered in next subsection as a \textbf{generating function.}
This way, we represent the nonlinear system (\ref{riccist2}) in the form: 
\begin{eqnarray}
\psi ^{\bullet \bullet }+\psi ^{\prime \prime } &=&2\ ^{h}\Upsilon ,
\label{eq1} \\
(\underline{\varpi })^{\diamond }\underline{h}_{3}^{\diamond } &=&2%
\underline{h}_{3}\underline{h}_{4}\ ^{v}\underline{\Upsilon },  \label{e2a}
\\
\ \underline{n}_{k}^{\diamond \diamond }+\underline{\gamma }\underline{n}%
_{k}^{\diamond } &=&0,  \label{e2b} \\
\underline{\beta }\underline{w}_{j}-\underline{\alpha }_{j} &=&0.
\label{e2c}
\end{eqnarray}%
Any solution of this system of nonlinear PDEs is a solution of (\ref{cdeq1})
parameterized as locally anisotropic cosmological ansatz (\ref{dmc}) for
canonically parameterized effective sources (\ref{esourcc}). The equations (%
\ref{eq1}) and (\ref{e2a}) involve respectively two \textbf{generating
sources} $\ ^{h}\Upsilon (x^{k})$ and $\ ^{v}\underline{\Upsilon }(x^{k},t).$
It should be noted here that instead of a cosmological type ansatz (\ref{dmc}%
) we can consider quasi-stationary ones with explicit dependence on $y^{3}$
and when the generic off-diagonal solutions of (\ref{cdeq1}) do not depend
on $y^{4}=t.$ The above procedure can be used for such systems (in abstract
geometric form omitting underlying of symbols, changing $\diamond
\rightarrow \ast $). Respective nonlinear systems of PDEs for
quasi-stationary configurations are studied in details, for instance, in 
\cite{vv25a} (see formulas (30) - (35) in that partner work). These reflects
certain nonlinear symmetries and duality properties of such nonholonomic
Einstein systems and their generic off-diagonal solutions which will be
considered in next subsection.

%%%%%

Let us explain the general decoupling property of the above systems of
equations for generic off-diagonal cosmologic configurations: The equation (%
\ref{eq1}) is a standard 2-d Poisson equation with source $2\ ^{h}\Upsilon $
(it is the same as for quasi-stationary configurations). It can be a 2-d
wave equation if we consider h-metrics with signature, for instance, $(+,-)$
but we shall not analyze such models in this work. Prescribing any data $(%
\underline{h}_{4},\ ^{v}\underline{\Upsilon }),$ we can search a coefficient 
$\underline{h}_{3}$ as a solution of a second order on $\partial _{t}$
nonlinear PDE (\ref{e2a}). Inversely, we can prescribe a couple $(\underline{%
h}_{3},\ ^{v}\underline{\Upsilon })$ when a coefficient $\underline{h}_{4}$
is a solution of a first-order nonlinear PDE. At the end of this subsection,
we show how using a generating function $\underline{\Psi }(x^{k},t)$, such
equations can be integrated in explicit form. So, having defined in some
general forms $\underline{h}_{3}(x^{k},t)$ and $\underline{h}_{4}(x^{k},t),$
we can compute respective coefficients $\underline{\alpha }_{i}$ and $%
\underline{\beta }$ for (\ref{e2c}). Such linear equations for $\underline{w}%
_{j}(x^{k},t)$ can be solved in general form. This means that such equations
and respective unknown functions are decoupled from the rest of the system
of nonlinear equations. At the forth step, we can solve (\ref{e2b}) and find 
$\underline{n}_{k}(x^{k},t).$ We have to perform two general integrations on
the time-like coordinate $t$ for any $\underline{\gamma }(x^{k},t)$
determined by $\underline{h}_{3}(x^{k},t)$ and $\underline{h}_{4}(x^{k},t)$
as we explained above. So, solving step-by-step four equations (\ref{eq1}) -
(\ref{e2c}), we can generate off-diagonal cosmological solutions of
(modified) Einstein equations written in canonical nonholonomic dyadic
variables and using respective distortions of connections. This can be done
in explicit form by using the general decoupling property for the
off-diagonal metric ansatz (\ref{dmc}) and respective generating sources (%
\ref{esourcc}). %%%%%
%%%%

\subsubsection{Relativistic W-entropy for geometric flows of nonholonomic
Einstein systems}

In this generalized framework, the introduction of a $\tau $-family of
d-metrics $\widehat{\underline{\mathbf{g}}}(\tau )=\widehat{\underline{%
\mathbf{g}}}(\tau ,r,\theta ,t)$ represents a natural extension of the
canonical constructions used in NESs. Here the parameter $\tau $ plays the
role of a flow (or a "temperature-like") evolution parameter, analogous to
the one introduced by G. Perelman in the theory of Ricci flows, but
generalized to relativistic and nonholonomic setting \cite%
{perelman1,gheorghiuap16,vv25a,vv25b,v25,bsssvv25}. This approach allows us
to interpret the evolution of geometric and physical quantities -- metric
coefficients, nonlinear connection (N-connection) structures, effective
sources, etc. -- as thermodynamic processes driven by geometric flows. Such $%
\tau$-dependent nonholonomic configurations are used to model statistical
ensembles of quasi-stationary cosmological geometries, even when no global
horizons or holographic boundaries are present.

In canonical dyadic variables adapted to the nonholonomic structure of the
manifold $\mathcal{W}$, one can define a relativistic generalization of
Perelman's W-functional, denoted by the hat version 
\begin{equation}
\widehat{\mathcal{W}}(\tau )=\int_{t_{1}}^{t_{2}}\int_{\Xi }\left( 4\pi \tau
\right) ^{-2}e^{-\widehat{\zeta }(\tau )}\sqrt{|\mathbf{g}(\tau )|}\delta
^{4}u[\tau (f(\widehat{\mathbf{R}}sc(\tau ))+|\widehat{\mathbf{D}}(\tau )%
\widehat{\zeta }(\tau )|^{2}+\widehat{\zeta }(\tau )-4],  \label{wf1}
\end{equation}%
where the function $\widehat{\zeta }(\tau )=\widehat{\zeta }(\tau
,x^{i},y^{a})$ denotes a normalization function used to define the
integration measure in the corresponding geometric or physical functional.
However, in concrete geometric or physical models, it can be prescribed
explicitly to ensure the desired normalization conditions or to encode
specific thermodynamic or field-theoretic properties of the system. In
expression (\ref{wf1}), we consider $t$-families of not intersecting 3-d
hypersurfaces $\Xi $, which determine closed 4-d regions $U\subset \mathbf{V}
$. All geometric quantities involved are assumed to be of sufficient smooth
class so that the corresponding functionals $\widehat{\mathcal{W}}(\tau )$
are well-defined. The symbol $\delta ^{4}u$, used instead of $d^{4}u$,
states that the integration is performed with respect to the N-elongated
differentials defined in (\ref{nadif}). The positive, temperature--like
evolution parameter $\tau \subset \lbrack \tau _{0}\leq \tau _{1}]$
parameterizes the family of geometric configurations considered in the flow. 
%%%%%%

The corresponding relativistic geometric flow equations can be derived in
canonical dyadic variables either by using abstract geometric methods, as
developed in \cite{misner73,bsssvv25,gheorghiuap16}, or by performing an
N-adapted variational calculus with respect to the functional $\widehat{%
\mathcal{W}}(\tau )$.

The nonholonomic structure can be prescribed in specific forms when the
relativistic geometric evolution is governed by nonlinear systems of PDEs of
the type $\widehat{\mathbf{R}}_{\ \ \beta }^{\alpha }(\tau )=\widehat{%
\mathbf{\Upsilon }}_{\ \ \beta }^{\alpha }(\tau ),$ which describe $\tau $%
-families of nonholonomic Einstein equations (\ref{cdeq1}). Although we do
not elaborate on such details here, it is worth emphasizing that for
self-similar configurations corresponding to a fixed parameter $\tau _{0}$,
one obtains nonholonomic Ricci solitons, which are equivalent to
off-diagonal Einstein spaces with certain effective cosmological constants.
Using the AFCDM, we can then construct various classes of off-diagonal
cosmological solutions to the respective $\tau $-families of nonlinear PDE
systems (\ref{eq1}) -- (\ref{e2c}), as will be demonstrated in the next
subsection.

Finally, we note that locally anisotropic cosmological d-metrics (\ref{dmc})
subjected to above conditions (\ref{eq1}) - (\ref{e2c}) can be represented
equivalently in local coordinate form using generic off-diagonal ansatz $%
\mathbf{\hat{g}}=\underline{\widehat{g}}_{\alpha \beta }(u)du^{\alpha
}\otimes du^{\beta }$ (\ref{ansatz}), when%
\begin{equation*}
\widehat{\underline{g}}_{\alpha \beta }=\left[ 
\begin{array}{cccc}
g_{1}+(\underline{N}_{1}^{3})^{2}\underline{h}_{3}+(\underline{N}%
_{1}^{4})^{2}\underline{h}_{4} & \underline{N}_{1}^{3}\underline{N}_{2}^{3}%
\underline{h}_{3}+\underline{N}_{1}^{4}\underline{N}_{2}^{4}\underline{h}_{4}
& \underline{N}_{1}^{3}\underline{h}_{3} & \underline{N}_{1}^{4}\underline{h}%
_{4} \\ 
\underline{N}_{2}^{3}\underline{N}_{1}^{3}\underline{h}_{3}+\underline{N}%
_{2}^{4}\underline{N}_{1}^{4}\underline{h}_{4} & g_{2}+(\underline{N}%
_{2}^{3})^{2}\underline{h}_{3}+(\underline{N}_{2}^{4})^{2}\underline{h}_{4}
& \underline{N}_{2}^{3}\underline{h}_{3} & \underline{N}_{2}^{4}\underline{h}%
_{4} \\ 
\underline{N}_{1}^{3}\underline{h}_{3} & \underline{N}_{2}^{3}\underline{h}%
_{3} & \underline{h}_{3} & 0 \\ 
\underline{N}_{1}^{4}\underline{h}_{4} & \underline{N}_{2}^{4}\underline{h}%
_{4} & 0 & \underline{h}_{4}%
\end{array}%
\right]
\end{equation*}%
\begin{equation}
=\left[ 
\begin{array}{cccc}
e^{\psi }+(\underline{n}_{1})^{2}\underline{h}_{3}+(\underline{w}_{1})^{2}%
\underline{h}_{4} & \underline{n}_{1}\underline{n}_{2}\underline{h}_{3}+%
\underline{w}_{1}\underline{w}_{2}\underline{h}_{4} & \underline{n}_{1}%
\underline{h}_{3} & \underline{w}_{1}\underline{h}_{4} \\ 
\underline{n}_{1}\underline{n}_{2}\underline{h}_{3}+n_{1}n_{2}\underline{h}%
_{4} & e^{\psi }+(\underline{n}_{2})^{2}\underline{h}_{3}+(\underline{w}%
_{2})^{2}\underline{h}_{4} & \underline{n}_{2}\underline{h}_{3} & \underline{%
w}_{2}\underline{h}_{4} \\ 
\underline{n}_{1}\underline{h}_{3} & \underline{n}_{2}\underline{h}_{3} & 
\underline{h}_{3} & 0 \\ 
\underline{w}_{1}\underline{h}_{4} & \underline{w}_{2}\underline{h}_{4} & 0
& \underline{h}_{4}%
\end{array}%
\right] .  \label{qeltorsoffd}
\end{equation}%
Constructing exact or parametric solutions for such an ansatz is not
possible if we work directly with the LC connection $\underline{\nabla }%
(x^{k},t)$. The AFCDM prescribes using the canonical d-connection $\widehat{%
\mathbf{D}}(x^{k},t)$ for decoupling and generating solutions. Then, certain
LC configurations can be extracted by imposing additional nonholonomic
constraints (\ref{lccond1}), when $\widehat{\underline{\mathbf{T}}}_{\
\alpha \beta }^{\gamma }(x^{k},t)=0$. In explicit form, we explain this
procedure in next subsection. The formulas for quasi-stationary analogs of
off-diagonal metrics (\ref{qeltorsoffd}) (when the geometric objects and
coefficients are not underlined, and with generic dependence on $%
(x^{k},y^{3}),$ but not on $t$) are provided in \cite{vv25a,bsssvv25}. %%%%

\subsection{Generating off-diagonal cosmological solutions}

%%%%%%%

To generate off-diagonal cosmological solutions of relativistic geometric
flow equations, we use $\tau $-families of ansatz (\ref{dmc}) which can be
parameterized in such a canonical d-form 
\begin{eqnarray}
\widehat{\underline{\mathbf{g}}}(\tau ) &=&(g_{i}(\tau ),g_{b}(\tau ),%
\underline{N}_{i}^{3}(\tau )=\underline{n}_{i}(\tau ),\underline{N}_{i}^{4}=%
\underline{w}_{i}(\tau ))  \label{lacosm1} \\
&=&g_{i}(\tau ,r,\theta )dx^{i}\otimes dx^{i}+\underline{g}_{3}(\tau
,r,\theta ,t)\underline{\mathbf{e}}^{3}\otimes \underline{\mathbf{e}}^{3}+%
\underline{g}_{4}(\tau ,r,\theta ,t)\underline{\mathbf{e}}^{4}\otimes 
\underline{\mathbf{e}}^{4},  \notag \\
&&\mbox{ for }\underline{\mathbf{e}}^{3}(\tau )=d\phi +\underline{n}%
_{i}(\tau ,r,\theta ,t)dx^{i},\ \underline{\mathbf{e}}^{4}(\tau )=dt+%
\underline{w}_{i}(\tau ,r,\theta ,t)dx^{i}.  \notag
\end{eqnarray}%
Such a d-metric Killing symmetry is on the angular coordinate $\varphi ,$
when $\partial _{\varphi }$ transforms into zero the N-adapted coefficients
of such a d-metric. This simplifies substantially the application of the
AFCDM for generating off-diagonal solutions. In principle, we can construct
more general classes of solutions including also the $\varphi $-dependence
(and various types of functionals depending on all spacetime coordinates),
but such formulas are much cumbersome and need more sophisticated geometric
methods and involve additional problems on physical interpretation, etc.,
see discussion in \cite{vv25a,vv25b,bsssvv25}. %%%%%%%

For a fixed $\tau =\tau _{0}$ and self-similar configurations, d-metrics of
type (\ref{lacosm1}) define nonholonomic cosmological Ricci solitons which
include as particular cases some (modified) Einstein equations with
effective cosmological constants. Nontrivial matter fields $\mathbf{T}%
_{\alpha \beta }$ (\ref{emdt}) can be added for respective nonholonomic
distortions and nonlinear transforms, which allow us to consider the system
of nonlinear PDEs (\ref{cdeq1}) as an example of nonholonomic Ricci soliton,
when the cosmological solutions can be extended on (and characterized
additionally) a temperature-like parameter. This is necessary for
constructing and applying the G. Perelman thermodynamics (see the end of
this section) for such generic off-diagonal solutions because they can't be
studied in the framework of the Bekenstein-Hawking paradigm \cite{bek2,haw2}%
. Hereafter, we shall omit to write in explicit form the dependence on $\tau 
$ if that does not result in ambiguities. %%%%%

\subsubsection{Equivalent forms of $\protect\tau $-families of off-diagonal
cosmological solutions}

We can integrate in general form $\tau $-families of nonlinear PDEs (\ref%
{eq1}) - (\ref{e2c}) by using off-diagonal cosmological ansatz (\ref{lacosm1}%
) with N-adapted coefficients:%
\begin{eqnarray}
g_{1}(\tau ) &=&g_{2}(\tau )=e^{\psi (\tau )},  \label{dmncoef} \\
\underline{g}_{3}(\tau ) &=&g_{3}^{[0]}(\tau )-\int dt\frac{[\underline{\Psi 
}^{2}(\tau )]^{\diamond }}{4~^{v}\underline{\Upsilon }(\tau )},\underline{g}%
_{4}(\tau )=\frac{[\underline{\Psi }^{\diamond }(\tau )]^{2}}{4(\ ~^{v}%
\underline{\Upsilon }(\tau ))^{2}\{g_{3}^{[0]}(\tau )-\int dt[\underline{%
\Psi }^{2}(\tau )]^{\diamond }/4\ ~^{v}\underline{\Upsilon }(\tau )\}}; 
\notag \\
\ \underline{n}_{k}(\tau ) &=&\ _{1}n_{k}(\tau )+\ _{2}n_{k}(\tau )\int dt%
\frac{[(\underline{\Psi }(\tau ))^{2}]^{\diamond }}{4(\ ~^{v}\underline{%
\Upsilon }(\tau ))^{2}|g_{3}^{[0]}(\tau )-\int dt[\underline{\Psi }^{2}(\tau
)]^{\diamond }/4\ ~^{v}\underline{\Upsilon }(\tau )|^{5/2}},\ \underline{w}%
_{k}(\tau )=\frac{\partial _{i}\underline{\Psi }(\tau )}{\underline{\Psi }%
^{\diamond }(\tau )}.  \notag
\end{eqnarray}%
In these formulas, we consider such\ $\tau $-families of integration and
generating data:%
\begin{eqnarray}
&&%
\begin{array}{ccccc}
\mbox{integration} &  & g_{3}^{[0]}(\tau )=g_{3}^{[0]}(\tau ,x^{i}) &  & %
\mbox{chosen to describe  observational data}; \\ 
\mbox{functions:} &  & 
\begin{array}{c}
\ _{1}n_{k}(\tau )=\ _{1}n_{k}(\tau ,x^{i}) \\ 
\ _{2}n_{k}(\tau )=\ _{2}n_{k}(\tau ,x^{i})%
\end{array}
&  & 
\begin{array}{c}
\mbox{chosen to describe  observational data}, \\ 
\mbox{can be zero for LC-configurations};%
\end{array}
\\ 
&  &  &  &  \\ 
\begin{array}{c}
\mbox{horizontal generating} \\ 
\mbox{functions and sources:}%
\end{array}
&  & 
\begin{array}{c}
\psi (\tau )=\psi (\tau ,x^{i}), \\ 
\ ^{h}\Upsilon (\tau )=\ ^{h}\Upsilon (\tau ,x^{i}),%
\end{array}
&  & 
\begin{array}{c}
\mbox{solutions of 2-d Poisson eqs.} \\ 
\psi ^{\bullet \bullet }(\tau )+\psi ^{\prime \prime }(\tau )=2\
^{h}\Upsilon (\tau );%
\end{array}
\\ 
&  &  &  &  \\ 
\begin{array}{c}
\mbox{vertical generating} \\ 
\mbox{functions}:%
\end{array}
&  & \underline{\Psi }(\tau )=\underline{\Psi }(\tau ,x^{i},t), &  & %
\mbox{chosen to describe observational data}; \\ 
&  &  &  &  \\ 
\begin{array}{c}
\mbox{vertical generating} \\ 
\mbox{sources}:%
\end{array}
&  & ~^{v}\underline{\Upsilon }(\tau )=~^{v}\underline{\Upsilon }(\tau
,x^{i},t), &  & \mbox{chosen to describe observational data}.%
\end{array}
\label{intgendata} \\
&&  \notag
\end{eqnarray}

The d-metrics (\ref{dmncoef}) possed certain space and time duality
properties which allows us to transform cosmological configurations into
quasi-stationary ones, and inverse. For instance, we model $\tau $-evolution
of quasi-stationary NESs if the v-partial derivatives are changed in the
form: $\ast \rightarrow \diamond $, i.e. $\partial _{3}\rightarrow \partial
_{4},$ for 
\begin{equation}
~^{v}\underline{\Upsilon }(\tau )=~^{v}\underline{\Upsilon }(\tau
,x^{i},t)\rightarrow ~^{v}\Upsilon (\tau )=~^{v}\Upsilon (\tau ,x^{i},y^{3})%
\mbox{ and }\underline{\Psi }(\tau )=\underline{\Psi }(\tau
,x^{i},t)\rightarrow \Psi (\tau )=\Psi (\tau ,x^{i},y^{3}).
\label{dualprinc1}
\end{equation}%
Various examples of such physically important off-diagonal quasi-stationary
solutions for locally anisotropic BHs, BTs, WHs, and other type
configurations are studied in \cite{vv25a,vv25b,bsssvv25}.

We can generate off-diagonal cosmological solutions for NESs (\ref{cdeq1})
if we fix a $\tau =\tau _{0}$ in (\ref{dmncoef}). Even in such cases, the
integration and generating data (\ref{intgendata}) define locally
anisotropic cosmological configurations modelling nonlinear gravitational
and (effective) matter field interactions in a nontrivial gravitational
vacuum background of GR, or in MGTs. Such cosmological scenarios are
possible because the AFCDM allows us to find directly solutions of systems
of nonlinear PDEs not transforming them by additional assumptions into
systems of nonlinear ODEs. The solutions of ODEs determined by integration
constants and they offer more limited possibilities in explaining, for
instance, recent cosmological observational data. %%%%%

\subsubsection{Nonlinear symmetries and polarization functions for
cosmological geometric flows}

The $\tau $-families of off-diagonal locally anisotropic cosmological
solutions (\ref{dmncoef}) possess such nonlinear symmetries : 
\begin{eqnarray}
\frac{\lbrack \underline{\Psi }^{2}(\tau )]^{\diamond }}{\ ^{v}\underline{%
\Upsilon }(\tau )} &=&\frac{[\underline{\Phi }^{2}(\tau )]^{\diamond }}{%
\underline{\Lambda }(\tau )},\mbox{ which can be
integrated as  }  \label{nonlinsym1} \\
\underline{\Phi }^{2}(\tau ) &=&\ \underline{\Lambda }(\tau )\int dt(\ ^{v}%
\underline{\Upsilon }(\tau ))^{-1}[\underline{\Psi }^{2}(\tau )]^{\diamond }%
\mbox{ and/or
}\underline{\Psi }^{2}(\tau )=(\underline{\Lambda }(\tau ))^{-1}\int dt(\
^{v}\underline{\Upsilon }(\tau ))[\underline{\Phi }^{2}(\tau )]^{\diamond }.
\notag
\end{eqnarray}%
Such formulas allows us to transform partially the (effective) matter
sources $~^{v}\underline{\Upsilon }(\tau )$ into certain $\tau $-running
(effective) cosmological constants $\ \underline{\Lambda }(\tau ),$ but
re-defining the generation functions $\underline{\Psi }^{2}(\tau)\rightarrow 
\underline{\Phi }^{2}(\tau ).$ Such nonlinear symmetries of off-diagonal
cosmological solutions can be defined for nonholonomic Ricci solitons for $%
\tau _{0}$ and in GR if the LC conditions (\ref{lccond1}) are imposed
additionally (more details and formulas are presented below). Similar
nonlinear symmetries can be derived for quasi-stationary configurations (\ref%
{dualprinc1}), for respective $(\Psi ^{2}(\tau ),\ \ ^{v}\underline{\Upsilon 
}(\tau ))\rightarrow (\Phi ^{2}(\tau ),\Lambda (\tau ))$ using $\ast $ and
integration on $y^{3}.$

In \cite{vv25a,vv25b,bsssvv25}, rigorous proofs state that nonlinear
symmetries (\ref{nonlinsym1}) can be formulated using different type
parameterizations of generating data for $v$-coefficients of d-metrics: 
\begin{eqnarray}
(\underline{\Psi }(\tau ),\ ^{v}\underline{\Upsilon }(\tau ))
&\leftrightarrow &(\underline{\widehat{\mathbf{g}}}(\tau ),\ ^{v}\underline{%
\Upsilon }(\tau ))\leftrightarrow (\underline{\eta }_{\alpha }(\tau )\ 
\underline{\mathring{g}}_{\alpha }\sim (\underline{\zeta }_{\alpha }(\tau
)(1+\kappa \underline{\chi }_{\alpha }(\tau ))\underline{\mathring{g}}%
_{\alpha },\ ^{v}\underline{\Upsilon }(\tau ))\leftrightarrow
\label{nonlintrsmalp} \\
(\underline{\Phi }(\tau ),\underline{\ \Lambda }(\tau )) &\leftrightarrow &(%
\underline{\mathbf{g}}(\tau ),\ \underline{\Lambda }(\tau ))\leftrightarrow (%
\underline{\eta }_{\alpha }(\tau )\ \underline{\mathring{g}}_{\alpha }\sim (%
\underline{\zeta }_{\alpha }(\tau )(1+\kappa \underline{\chi }_{\alpha
}(\tau ))\underline{\mathring{g}}_{\alpha },\ \underline{\Lambda }(\tau )). 
\notag
\end{eqnarray}%
In these formulas $\underline{\ \Lambda }(\tau )$ is an effective $\tau $%
-running cosmological and $\kappa $ is a small parameter $0\leq \kappa <1,$
which can be used for describing small nonholonomic and off-diagonal
deformations, $\underline{\zeta }_{\alpha }(\tau )=\underline{\zeta }%
_{\alpha }(\tau ,x^{k},t)$ and $\underline{\chi }_{\alpha }(\tau )=%
\underline{\chi }_{\alpha }(\tau ,x^{k},t)$ are respective
re-parametrization and small polarization functions. For instance, using 3-d
space spherical coordinates, we can express the off-diagonal solutions in
different forms using different generating functions and generating sources $%
(\underline{\Psi }(\tau ,r,\theta ,t),^{v}\underline{\Upsilon }%
(\tau,r,\theta ,t)),$ or $\left( \underline{\Phi }(\tau ,r,\theta ,t),\ ^{v}%
\underline{\Upsilon }(\tau ,r,\theta ,t), \underline{\Lambda }(\tau )\right) 
$. A generating function $\underline{g}_{3}(\tau ,r,\theta ,t)=\eta
_{3}(\tau ,r,\theta ,t)\underline{\mathring{g}}_{\alpha }$ can be defined
for additional assumptions on parameterizations as 
\begin{equation*}
\underline{\eta }_{3}(\tau )\simeq \underline{\eta }(\tau ,r,\theta
,t)\simeq \underline{\zeta }_{3}(\tau ,r,\theta ,t)(1+\kappa \underline{\chi 
}_{3}(\tau ,r,\theta ,t))
\end{equation*}%
using a small parameter $\kappa $ for describing small parametric
nonholonomic and/or off-diagonal deformations. %%%%%%%

$\tau $-families of h-components of cosmological d-metrics (\ref{dmncoef})
(and, in general, of (\ref{lacosm1}) \ and (\ref{dm})) can be parameterized
as 
\begin{eqnarray}
\psi (\tau ) &\simeq &\psi (\tau ,x^{k}(r,\theta ))\simeq \psi _{0}(\tau
,x^{k}(r,\theta ))(1+\kappa \ _{\psi }\chi (\tau ,x^{k}(r,\theta ))),%
\mbox{
for }\   \label{htransf} \\
\ \eta _{2}(\tau ) &\simeq &\eta _{2}(\tau ,x^{k}(r,\theta ))\simeq \zeta
_{2}(\tau ,x^{k}(r,\theta ))(1+\kappa \chi _{2}(\tau ,x^{k}(r,\theta ))),%
\mbox{ we can
consider }\ \eta _{2}=\ \eta _{1}.  \notag
\end{eqnarray}%
In these formulas, $\psi (\tau )$ and $\eta _{2}(\tau )=\ \eta _{1}(\tau )$
can be such way chosen to be related to the solutions of $\tau $-families of
2-d Poisson equations, $\partial _{11}^{2}\psi (\tau )+\partial
_{22}^{2}\psi (\tau )=2\ ^{h}\Upsilon (\tau ,x^{k}),$ to define solutions of
the h-components of (\ref{cdeq1}) with sources of type (\ref{esourc}), which
via 2-d frame transforms can be related to $\tau $-running cosmological
constants, $\ ^{h}\Upsilon \simeq \Lambda $ (in GR, we can fix $\Lambda =%
\underline{\Lambda }(\tau _{0})$). %%%%%%

To generate $\tau $-families of cosmological solutions of (\ref{cdeq1}) the
data defining nonlinear symmetries (\ref{nonlintrsmalp}) must be solutions
of such differential or integral equations:%
\begin{eqnarray}
\partial _{t}[\underline{\Psi }^{2}] &=&-\int dt\ ^{v}\underline{\Upsilon }%
\partial _{t}\underline{g}_{3}\simeq -\int dt\ ^{v}\underline{\Upsilon }%
\partial _{t}(\underline{\eta }_{3}\ \underline{\mathring{g}}_{3})\simeq
-\int dt\ ^{v}\underline{\Upsilon }\partial _{t}[\underline{\zeta }%
_{3}(1+\kappa \ \underline{\chi }_{3})\ \underline{\mathring{g}}_{3}], 
\notag \\
\underline{\Phi }^{2} &=&-4\ \underline{\Lambda }\underline{g}_{3}\simeq -4\ 
\underline{\Lambda }\underline{\eta }_{3}\underline{\mathring{g}}_{3}\simeq
-4\ \underline{\Lambda }\ \underline{\zeta }_{3}(1+\kappa \underline{\chi }%
_{3})\ \underline{\mathring{g}}_{3},  \label{nonlinsymcosm}
\end{eqnarray}%
where, for simplicity, $\tau $-dependencies are omitted. In the next
formulas, we shall not write "$(\tau ),$ or $\tau ,..."$ if that will not
result in ambiguities (and supposing that we can always consider $\tau $%
-families of NESs and respective solutions). For constructing relativistic
thermodynamic models characterizing respective cosmological configurations,
the $\tau $-dependence is typically important to be written in explicit form.

In terms of $\eta $-polarization functions stated in spherical coordinates,
the off-diagonal solutions of type (\ref{dmncoef}) can be written as 
\begin{eqnarray}
d\widehat{\underline{s}}^{2}(\tau ) &=&\ ^{\eta }\widehat{\underline{g}}%
_{\alpha \beta }(\tau ,r,\theta ,t;\underline{\mathring{g}}_{\alpha };\psi ,%
\underline{\eta }_{3};\ \underline{\Lambda }(\tau ),\ ^{v}\underline{%
\Upsilon }(\tau ))du^{\alpha }du^{\beta }=e^{\psi
}[(dx^{1})^{2}+(dx^{2})^{2}]  \label{lacosm2} \\
&&+(\underline{\eta }\underline{\mathring{g}}_{3})\{d\varphi +[\ _{1}n_{k}+\
_{2}n_{k}\int dt\frac{[\partial _{t}(\underline{\eta }\underline{\mathring{g}%
}_{3})]^{2}}{|\int dt\ ^{v}\underline{\Upsilon }\partial _{t}(\underline{%
\eta }\underline{\mathring{g}}_{3})|\ (\underline{\eta }\underline{\mathring{%
g}}_{3})^{5/2}}]dx^{k}\}^{2}  \notag \\
&&-\frac{[\partial _{t}(\underline{\eta }\ \underline{\mathring{g}}_{3})]^{2}%
}{|\int dt\ ^{v}\underline{\Upsilon }\partial _{t}(\underline{\eta }%
\underline{\mathring{g}}_{3})|\ \eta \mathring{g}_{3}}\{dt+\frac{\partial
_{i}[\int dt\ ^{v}\underline{\Upsilon }\ \partial _{t}(\underline{\eta }%
\underline{\mathring{g}}_{3})]}{\ ^{v}\underline{\Upsilon }\partial _{t}(%
\underline{\eta }\underline{\mathring{g}}_{3})}dx^{i}\}^{2}.  \notag
\end{eqnarray}%
Such locally anisotropic cosmological d-metrics are determined by two
generating functions 
\begin{equation}
\psi (\tau )\simeq \psi (\tau ,x^{k})\mbox{ and }\underline{\eta }\ (\tau
)\simeq \underline{\eta }_{3}(\tau ,x^{k},t),  \label{etapolgen}
\end{equation}%
as in (\ref{htransf}) and (\ref{nonlinsymcosm}), where $x^{k}=x^{k}(r,%
\theta) $. The d-metrics (\ref{lacosm2}) are also determined by $\tau $%
-families of integration and generating data as in (\ref{intgendata}). Such
values can be redefined in N-adapted frames as $\tau $-families of $(\psi ,%
\underline{\eta };\ \underline{\Lambda }, \ ^{v}\underline{\Upsilon }%
,_{1}n_{k},\ _{2}n_{k})$ have to be chosen in explicit form to describe
certain observational data in modern cosmology and DE and DM physics. We can
consider some primary cosmological data defined by a diagonal $\underline{%
\mathring{g}}_{\alpha }$ (for instance, defined by a FRLW metric and a $%
\Lambda $CDM model) and model further nonholonomic geometric flow and
off-diagonal deformations into certain target locally anisotropic
cosmological d-metrics $\ ^{\eta }\widehat{\underline{g}}_{\alpha \beta
}(\tau ,x^{k},t)$ (\ref{lacosm2}). Here we note that we have to impose more
special classes of such generating data to satisfy the LC-conditions (\ref%
{lccond1}) as we describe in \cite{vv25a,vv25b,v25,bsssvv25} for GR and
various types of MGTs.\footnote{%
The labels for the metric tensors, or d-tensors, are stated following such
principles: the left up or low $\eta $ states that the solutions are
generated by $\eta $-polarization functions from a prime d-metric $%
\underline{\mathbf{\mathring{g}}}$ by using N-adapted hat variables. We
shall write, for instance, that $\ ^{\eta }\widehat{\mathbf{R}}is$ is
determined by some $\eta $-deformation data $(\ ^{\eta }\widehat{\mathbf{g}}%
, \ ^{\eta }\widehat{\mathbf{D}});$ we can underline such values as $\
^{\eta }\underline{\widehat{\mathbf{R}}}is$ and $(\ ^{\eta }\widehat{%
\underline{\mathbf{g}}},\ ^{\eta }\widehat{\underline{\mathbf{D}}})$ to
emphasize that such cosmological geometric d-objects involve a generic $t$%
-dependence. For small $\kappa $-parametric dependence as in the Appendix,
see formulas (\ref{paramsoliton}), corresponding labels are of type $(\
^{\chi }\widehat{\mathbf{g}},\ ^{\chi }\widehat{\mathbf{D}})$ or $(\ ^{\chi }%
\widehat{\underline{\mathbf{g}}},\ ^{\chi }\widehat{\underline{\mathbf{D}}}%
). $ Such geometric d-objects involve corresponding primary data: $(\widehat{%
\mathbf{\mathring{g}}},\widehat{\mathbf{\mathring{D}}})$ or $(\widehat{%
\underline{\mathbf{\mathring{g}}}},\widehat{\underline{\mathbf{\mathring{D}}}%
}).$ In GR and many MGTs, one considers geometric data when $R$ is the Ricci
scalar of $(g,\nabla );$ we can write $(\underline{g},\underline{\nabla }),$
or $\mathring{R}$ for $(\mathring{g},\mathring{\nabla}),$ etc.} In
principle, we can consider that the off-diagonal cosmological solutions with
nonholonomic canonical deformations and distortions of the Einstein
equations (\ref{cdeq1}) are described by d-metrics of type (\ref{lacosm2}),
when nonholonomic induced torsion d-fields can always be transformed into
zero by corresponding subclasses of generating data. %%%%%%

\subsubsection{Effective $\protect\tau $--running cosmological constants and
the principle of space and time duality}

For $\underline{\Phi }^{2}(\tau )=-4\ \underline{\Lambda }(\tau )\underline{g%
}_{3}(\tau ),$ we can transform (\ref{lacosm2}) in form determined by
generating data $(\underline{\Phi }(\tau ),\underline{\Lambda }(\tau ))$
without $\eta $-polarizations, which may be useful, for instance, for
computing the G. Perelman thermodynamic variables in the next (sub)
sections. Even in such cases, the contributions of the generating sources $%
(\ ^{h}\underline{\Upsilon }(\tau ),\ ^{v}\underline{\Upsilon }(\tau ))$
can't be eliminated from all coefficients of a d-metric. We note that $%
\partial _{t}\underline{g}_{3}(\tau )= -\partial _{t}[\underline{\Psi }%
^{2}(\tau )]/4\ ^{v}\underline{\Upsilon }(\tau )\ $ is a partial derivative
on time of the first formula in (\ref{nonlintrsmalp}). This allows us to
introduce a new $\tau $-family of generating functions $\underline{\Psi }%
(\tau )$ and express (\ref{lacosm2}) without effective cosmological
constants $\underline{\Lambda }(\tau ).$ So, the AFCDM allows us to
construct off-diagonal solutions and transforms certain matter field sources
into effective cosmological constants, when the corresponding degrees of
freedom are absorbed by off-diagonal terms and effective polarizations of
physical constants. %%%%%%%%

We can extend in abstract and N-adapted index forms the GR and MGTs to
relativistic geometric flows and generate respective solutions by
introducing a formal dependence on $\tau $ (in respective formulas from the
previous sections) and considering new effective sources 
\begin{equation}
\widehat{\underline{\mathbf{J}}}(\tau )=\widehat{\underline{\mathbf{\Upsilon 
}}}(\tau )-\frac{1}{2}\partial _{\tau }\underline{\mathbf{g}}(\tau )=[\ ^{h}%
\widehat{\underline{\mathbf{J}}}(\tau ),\ ^{v}\widehat{\underline{\mathbf{J}}%
}(\tau )]=\widehat{\underline{\mathbf{J}}}_{\alpha }(\tau )=[\ J_{i}(\tau
)=\ \Upsilon _{i}(\tau )-\frac{1}{2}\partial _{\tau }g_{i}(\tau ),\ 
\underline{J}_{a}(\tau )=\underline{\Upsilon }_{a}(\tau )-\frac{1}{2}%
\partial _{\tau }\underline{g}_{a}(\tau )],  \label{effrfs}
\end{equation}%
where $\underline{\mathbf{g}}(\tau )=[g_{i}(\tau ),\underline{g}_{a}(\tau ),%
\underline{N}_{i}^{a}(\tau )]$ as in (\ref{dm}) and $\ \underline{\Upsilon }%
_{\alpha }(\tau )$ are $\tau $-families of type (\ref{esourc}). We also can
consider $\tau $-families of formulas (\ref{htransf}), (\ref{nonlinsymcosm})
and (\ref{etapolgen}) and $\tau $-running $\underline{\Lambda }(\tau ),$
with functional dependencise $\underline{\mathbf{g}}_{\alpha }[\underline{%
\Phi }(\tau )]\simeq \underline{\ \mathbf{g}}_{\alpha }[\underline{\eta }%
_{3}(\tau )].$ Using $\widehat{\underline{\mathbf{J}}}(\tau )$ (\ref{effrfs}%
) in formulas for $\tau $-families of d-metrics of type (\ref{lacosm2}), we
generate off-diagonal solutions for relativistic flow equations written in
the form (see details in \cite{vv25a}): 
\begin{equation}
\widehat{\underline{\mathbf{R}}}_{\ \ \beta }^{\alpha }[\underline{\Phi }%
(\tau ),\widehat{\underline{\mathbf{J}}}(\tau )]=\underline{\Lambda }(\tau )%
\mathbf{\delta }_{\ \ \beta }^{\alpha }.  \label{cdeq1b}
\end{equation}%
These formulas can be derived in variational N-adapted form by using the
W-entropy functional (\ref{wf1}). Prescribing certain effective $\widehat{%
\underline{\mathbf{J}}}(\tau )$, we impose certain nonholonomic constraints
on $\partial _{\tau }\underline{\mathbf{g}}(\tau ),$ i.e. on the
nonholonomic geometric evolution as follows from (\ref{effrfs}). It is not
possible to solve such constraints in general forms. Still, we can always
derived certain parametric formulas for decompositions on a small parameter
and in vicinity of a $\tau _{0}.$ For a fixed $\tau =\tau _{0},$ the
cosmological soltions system of nonlinear PDEs (\ref{cdeq1b}) transforms
into corresponding ones for the nonholonomic Einstein equations (\ref{cdeq1}%
). Here, we note that the canonical Ricci d-scalar is computed in abstract
form as $\widehat{\mathbf{R}}sc(\tau )=4\underline{\Lambda }(\tau ).$ 
%%%%%%

In this work, we study off-diagonal cosmological solutions in GR and
respective $\tau $-families of canonical nonholonomic Einstein equations. In 
\cite{vv25a,vv25b} (for nonassociative and noncommutative MGTs, see \cite%
{bsssvv25}), we stated in abstract symbolic form, the \textbf{principle of
space and time duality } of generic off-diagonal configurations with one
Killing symmetry on a space-like $\partial _{3}$ or time-like $\partial
_{t}. $ This can be used for mutual transforms of cosmological
configurations into quasi-stationary ones. This principle can be formulated
in terms of such $\tau $-families of nonholnomic N-adapted transforms: 
\begin{eqnarray*}
y^{3} &\longleftrightarrow &y^{4}=t,g_{3}(\tau
,x^{k},y^{3})\longleftrightarrow \underline{g}_{4}(\tau ,x^{k},t),g_{4}(\tau
,x^{k},y^{3})\longleftrightarrow \underline{g}_{3}(\tau ,x^{k},t), \\
N_{i}^{3}(\tau ) &=&w_{i}(\tau ,x^{k},y^{3})\longleftrightarrow
N_{i}^{4}(\tau )=\underline{n}_{i}(\tau ,x^{k},t),N_{i}^{4}(\tau
)=n_{i}(\tau ,x^{k},y^{3})\longleftrightarrow N_{i}^{3}(\tau )=\underline{w}%
_{i}(\tau ,x^{k},t).
\end{eqnarray*}%
Corresponding duality conditions have to be considered for prime d-metrics
and respective generating functions, generating sources and gravitational
polarization functions, and the integration functions. In explicit forms,
the duality transforms can be stated using formulas (\ref{dualprinc1}) for
generating sources (\ref{esourc}) and effective cosmological constants (they
can involve or not $\tau $-parametric dependencies): 
\begin{equation*}
\begin{array}{ccc}
\begin{array}{c}
(\Psi ,~^{v}\Upsilon )\leftrightarrow (\mathbf{g},\ ~^{v}\Upsilon
)\leftrightarrow \\ 
(\eta _{\alpha }\ \mathring{g}_{\alpha }\sim (\zeta _{\alpha }(1+\kappa \chi
_{\alpha })\mathring{g}_{\alpha },~^{v}\Upsilon )\leftrightarrow%
\end{array}
& \Longleftrightarrow & 
\begin{array}{c}
(\underline{\Psi },\ ~^{v}\underline{\Upsilon })\leftrightarrow (\underline{%
\mathbf{g}},\ ~^{v}\underline{\Upsilon })\leftrightarrow \\ 
(\underline{\eta }_{\alpha }\ \underline{\mathring{g}}_{\alpha }\sim (%
\underline{\zeta }_{\alpha }(1+\kappa \underline{\chi }_{\alpha })\underline{%
\mathring{g}}_{\alpha },\ ~^{v}\underline{\Upsilon })\leftrightarrow%
\end{array}
\\ 
\begin{array}{c}
(\Phi ,\ \Lambda )\leftrightarrow (\mathbf{g},\ \Lambda )\leftrightarrow \\ 
(\eta _{\alpha }\ \mathring{g}_{\alpha }\sim (\zeta _{\alpha }(1+\kappa \chi
_{\alpha })\mathring{g}_{\alpha },\ \Lambda ),%
\end{array}
& \Longleftrightarrow & 
\begin{array}{c}
(\underline{\Phi },\ \underline{\Lambda })\leftrightarrow (\underline{%
\mathbf{g}},\ \underline{\Lambda })\leftrightarrow \\ 
(\underline{\eta }_{\alpha }\ \underline{\mathring{g}}_{\alpha }\sim (%
\underline{\zeta }_{\alpha }(1+\kappa \underline{\chi }_{\alpha })\underline{%
\mathring{g}}_{\alpha },\ \underline{\Lambda }).%
\end{array}%
\end{array}%
\end{equation*}%
In this work, the geometric constructions are performed for cosmological
(i.e. underlined) configurations. %%%%%

\subsubsection{Different forms and generating data for off-diagonal
cosmological solutions}

For simplicity, we omit writing the explicit dependence on a temperature
like parameter $\tau .$

\paragraph{Off-diagonal cosmological solutions with generating sources 
\newline
}

Putting together the N-adapted coefficients (\ref{dmncoef}) and redefining
the generating data for effective sources (\ref{effrfs}), the $\tau $%
-families of off-diagonal cosmological soutions of \ (\ref{cdeq1b}) but
parametrized as d-metrics adapted to solutions of (\ref{cdeq1}) we obtain
such quadratic line elements: 
\begin{eqnarray}
d\underline{s}^{2}(\tau ) &=&e^{\psi }[(dx^{1})^{2}+(dx^{2})^{2}]
\label{cosmsc} \\
&&+\{g_{3}^{[0]}-\int dt\frac{[\underline{\Psi }^{2}]^{\diamond }}{4~^{v}%
\underline{J}}\}\{dy^{3}+[\ _{1}n_{k}+\ _{2}n_{k}\int dt\frac{[(\underline{%
\Psi })^{2}]^{\diamond }}{4(\ ~^{v}\underline{J})^{2}|g_{3}^{[0]}-\int dt[%
\underline{\Psi }^{2}]^{\diamond }/4\ ~^{v}\underline{J}|^{5/2}}]dx^{k}\} 
\notag \\
&&+\frac{[\underline{\Psi }^{\diamond }]^{2}}{4(\ ~^{v}\underline{J}%
)^{2}\{g_{3}^{[0]}-\int dt[\underline{\Psi }^{2}]^{\diamond }/4\ ~^{v}%
\underline{J}\}}(dt+\frac{\partial _{i}\underline{\Psi }}{\underline{\Psi }%
^{\diamond }}dx^{i})^{2}.  \notag
\end{eqnarray}%
In nonexplicit forms, the generating functionals and generating \ sources
can be written in the forms $(\underline{\Psi }^{2},~^{v}\underline{J}%
)\simeq (\underline{\tilde{\Psi}}^{2},~^{v}\underline{\Upsilon }),$ etc.

\paragraph{Off-diagonal cosmological solutions with effective cosmological
constants \newline
}

The quadratic elements for cosmological solutions (\ref{dmncoef}), i.e. (\ref%
{cosmsc}), can be written in an equivalent form using generating data $(\
^{v}\underline{J},\underline{\Phi },\underline{\Lambda })$ stated by
formulas (\ref{nonlinsym1}): 
\begin{eqnarray}
d\widehat{s}^{2}(\tau ) &=&\widehat{g}_{\alpha \beta }(x^{k},t,\ ^{v}%
\underline{J},\underline{\Phi },\underline{\Lambda })du^{\alpha }du^{\beta
}=e^{\psi }[(dx^{1})^{2}+(dx^{2})^{2}]  \label{offdiagcosmcsh} \\
&&\{g_{3}^{[0]}-\frac{\underline{\Phi }^{2}}{4\underline{\Lambda }}%
\}\{dy^{3}+[\ _{1}n_{k}+\ _{2}n_{k}\int dt\frac{\underline{\Phi }^{2}[%
\underline{\Phi }^{\diamond }]^{2}}{|\underline{\Lambda }\int dt\ \ ^{v}%
\underline{J}[\underline{\Phi }^{2}]^{\diamond }\ |h_{4}^{[0]}-\underline{%
\Phi }^{2}/4\underline{\Lambda }|^{5/2}}]\}  \notag \\
&&-\frac{\underline{\Phi }^{2}[\underline{\Phi }^{\diamond }]^{2}}{|%
\underline{\Lambda }\int dt\ ^{v}\underline{J}[\underline{\Phi }%
^{2}]^{\diamond }|[g_{4}^{[0]}-\underline{\Phi }^{2}/4\underline{\Lambda }]}%
\{dt+\frac{\partial _{i}\ \int dt\ ^{v}\underline{J}\ [\underline{\Phi }%
^{2}]^{\diamond }}{\ ^{v}\underline{J}\ [(\ \underline{\Phi }%
)^{2}]^{\diamond }}dx^{i}\}^{2}.  \notag
\end{eqnarray}%
In these formulas, the integration data are similar to (\ref{intgendata})
when $\tau $-running effective cosmological constants $\underline{\Lambda }%
(\tau )$ are introduced additionally to transform $(\underline{\Psi }(\tau
),\ ^{v}\underline{J}(\tau ))$ $\rightarrow (\underline{\Phi }(\tau ),\ 
\underline{\Lambda }(\tau )).$ The coefficients of d-metrics $\underline{%
\mathbf{\hat{g}}}[\underline{\Phi },\ ^{v}\underline{J},\underline{\Lambda }%
] $ (\ref{offdiagcosmcsh}) keep certain memory about $\ ^{v}\underline{J}\,\ 
$stated in $\underline{\mathbf{\hat{g}}}[\Psi ,\ \ ^{v}\Upsilon ]$ (\ref%
{cosmsc}). Nevertheless, the possibility of introducing effective $%
\underline{\Lambda }(\tau )$ simplifies the method of computing G. Perelman
thermodynamic variables, which will be used the end of section \ref{sec3}. 
%%%%%

\paragraph{Using a d-metric coefficient as a generating function for
off-diagonal cosmological solutions \newline
}

Taking the partial derivative on $y^{3}$ of respective formula from (\ref%
{nonlinsymcosm}) allows us to write $\underline{h}_{3}^{\diamond }=-[%
\underline{\Psi }^{2}]^{\diamond }/4\ ^{v}\underline{J}.$ If we prescribe $%
\underline{h}_{3}(\tau ,x^{i},t)$ and $\ ^{v}\underline{J}(\tau ,x^{i},t)$,
we can compute (up to an integration function) a generating function $\ 
\underline{\Psi }$ which satisfies the equation $[\underline{\Psi }%
^{2}]^{\diamond }=\int dt\ ^{v}\underline{J}\underline{h}_{3}^{\diamond }.$
This generation function is as in (\ref{cosmsc}). But in equivalent form, we
can consider as generating data directly a couple $(\underline{h}_{3},\ ^{v}%
\underline{J})$ and work with quadratic elements 
\begin{eqnarray}
d\widehat{s}^{2}(\tau ) &=&\underline{\widehat{g}}_{\alpha \beta }(\tau
,x^{k},t;h_{4},\ \ ^{v}\underline{J})du^{\alpha }du^{\beta }=e^{\psi
}[(dx^{1})^{2}+(dx^{2})^{2}]+\underline{h}_{3}\{dy^{3}+
\label{offdsolgenfgcosmc} \\
&&\lbrack \ _{1}n_{k}+\ _{2}n_{k}\int dt\frac{(\underline{h}_{3}^{\diamond
})^{2}}{|\int dt[\ ^{v}\underline{J}\ \underline{h}_{3}]^{\diamond }|\ (%
\underline{h}_{3})^{5/2}}]dx^{k}\}-\frac{(\underline{h}_{3}^{\diamond })^{2}%
}{|\int dt[\ \ ^{v}\underline{J}\underline{h}_{3}]^{\ast }|\ \underline{h}%
_{3}}\{dt+\frac{\partial _{i}[\int dt(\ ^{v}\underline{J})\ \underline{h}%
_{3}^{\diamond }]}{\ ^{v}\underline{J}\ \underline{h}_{3}^{\diamond }}%
dx^{i}\}^{2}.  \notag
\end{eqnarray}

The nonlinear symmetries (\ref{nonlinsym1}) and (\ref{nonlinsymcosm}) allow
ua to perform similar computations related to (\ref{offdiagcosmcsh}).
Expressing $\underline{\Phi }^{2}(\tau )=-4\ \underline{\Lambda }(\tau )%
\underline{h}_{3}(\tau ),$ we can eliminate $\underline{\Phi }$ from the
nonlinear element and generate a solution of type (\ref{offdsolgenfgcosmc})
which are determined by the generating data $(\underline{h}_{3};\underline{%
\Lambda },\ ^{v}\underline{J}).$ %%%%%%

\subsubsection{Constraints on generating functions and sources for
extracting LC cosmologies}

We can extract LC configurations defined by generic off-diagonal
cosmological metrics in explicit form if we impose additionally zero
conditions for the canonical d-torsion (\ref{lccond}). By straightforward
computations (such details are typically contained for dual quasi-stationary
configurations in \cite{bsssvv25,vv25a,vv25b}) we can check that the LC
conditions (\ref{lccond1}) are satisfied, if the coefficients of the
N--adapted frames and the $v$--components of d--metrics are subjected
additionally to the conditions: 
\begin{eqnarray}
\underline{n}_{k}(\tau ,x^{i}) &=&0,\partial _{i}\underline{n}_{j}(\tau
,x^{k})=\partial _{j}\underline{n}_{i}(\tau ,x^{k})\mbox{ and }\underline{n}%
_{i}^{\diamond }(\tau ,x^{k})=0;  \notag \\
\ \underline{w}_{i}^{\diamond }(\tau ,x^{i},t) &=&\underline{\mathbf{e}}%
_{i}(\tau )\ln \sqrt{|\ \underline{h}_{4}(\tau ,x^{i},t)|},\underline{%
\mathbf{e}}_{i}(\tau )\ln \sqrt{|\ \underline{h}_{3}(\tau ,x^{i},t)|}%
=0,\partial _{i}\underline{w}_{j}(\tau )=\partial _{j}\underline{w}_{i}(\tau
).  \label{zerot1}
\end{eqnarray}%
The solutions for such $w$- and $n$-functions depend on the class of vacuum
or non--vacuum cosmological metrics which we are generating using $%
\underline{h}_{3}(\tau ,x^{i},t)$ or, considering nonlinear symmetries
involving also $\ ^{v}\underline{J}(\tau ,x^{i},t)$, working with generating
data $\underline{\Psi }$ or $\underline{\Phi }^{2}(\tau )$. To solve this
problem, we can follow, for example, such two steps:

If we prescribe a generating function $\underline{\Psi }=\underline{\check{%
\Psi}}(\tau ,x^{i},t)$ for which $[\partial _{i}(\underline{\check{\Psi}}%
)]^{\diamond }=\partial _{i}(\underline{\check{\Psi}})^{\diamond },$ we can
solve the equations for $\underline{w}_{j}(\tau )$ from (\ref{zerot1}). This
is possible in explicit form if $\ ^{v}\underline{\Upsilon }=const,$ or $\
^{v}\underline{\Upsilon }(\tau )$, when the effective source is expressed as
a functional $\ ^{v}\underline{\Upsilon }(\tau ,x^{i},t)=\ \ ^{v}\underline{%
\Upsilon }[\underline{\check{\Psi}}(\tau )].$

Then, we can solve the conditions $\partial _{i}\underline{w}_{j}(\tau
)=\partial _{j}\underline{w}_{i}(\tau )$ if we chose a generating function $%
\ \underline{\check{A}}(\tau)=\underline{\check{A}}(\tau ,x^{k},t)$ and
define 
\begin{equation*}
\underline{w}_{i}(\tau ,x^{i},t)=\underline{\check{w}}_{i}(\tau
,x^{i},t)=\partial _{i}\ \underline{\check{\Psi}}/(\underline{\check{\Psi}}%
)^{\diamond }=\partial _{i}\underline{\check{A}}(\tau ,x^{k},t).
\end{equation*}%
The equations for $n$-functions in (\ref{zerot1}) are solved by any $\check{n%
}_{i}(\tau ,x^{k})=\partial _{i}[\ ^{2}n(\tau ,x^{k})].$ Any set of
functions $(\underline{\check{A}}(\tau ,x^{k},t),\ ^{2}n(\tau ,x^{k}))$
allows us to generate in explicit form $\tau $-families of off-diagonal
cosmological solutions in GR. In a more general framework, one can interpret
(\ref{zerot1}) as nonholonomic constraints defining the structures of
certain N-connection coefficients. These coefficients, in turn, determine
integral varieties for the corresponding cosmological NESs, possibly in
non-explicit form. %%%%%%

Putting together the above coefficients, we can write respective $\tau $%
-families of quadratic elements for cosmological off-diagonal solutions with
zero canonical d-torsion in such a form: 
\begin{eqnarray}
d\check{s}^{2}(\tau ) &=&\check{g}_{\alpha \beta }(\tau ,x^{k},t)du^{\alpha
}du^{\beta }  \label{offdcosmlc} \\
&=&e^{\psi }[(dx^{1})^{2}+(dx^{2})^{2}]+\{g_{3}^{[0]}-\int dt\frac{[%
\underline{\check{\Psi}}^{2}]^{\diamond }}{4(\ ^{v}\underline{\Upsilon }[%
\underline{\check{\Psi}}])}\}\{dy^{3}+\partial _{i}[\ ^{2}n]dx^{i}\}^{2} 
\notag \\
&&+\frac{[\underline{\check{\Psi}}^{\diamond }]^{2}}{4(\ ^{v}\underline{%
\Upsilon }[\underline{\check{\Psi}}])^{2}\{g_{4}^{[0]}-\int dt[\underline{%
\check{\Psi}}^{2}]^{\diamond }/4\ ^{v}\underline{\Upsilon }[\underline{%
\check{\Psi}}]\}}\{dt+[\partial _{i}(\underline{\check{A}})]dx^{i}\}^{2}. 
\notag
\end{eqnarray}%
Similar constraints on generation functions as in (\ref{zerot1}), with
re-defined nonlinear symmetries allow us to extract LC configurations for
all classes of quasi-stationary solutions in GR \cite{vv25a,vv25b}. %%%%%

To extract LC configurations from generic off-diagonal and inhomogeneous
cosmological solutions in GR, one starts by prescribing suitable generating
and integration data of type (\ref{cosmsc}), (\ref{offdiagcosmcsh}), or (\ref%
{offdsolgenfgcosmc}). These data define d-metrics similar to (\ref%
{offdcosmlc}), which can then be constrained nonholonomically to satisfy
zero torsion conditions, yielding LC-compatible configurations. The
resulting d-metrics encode effective sources that can mimic modified gravity
contributions while remaining within the framework of GR. The physical
interpretation of such solutions depends critically on the choice of local
and global symmetries, nonlinear structures, and possible nontrivial
polarizations, which may necessitate revisiting conventional cosmological
principles and thermodynamic formulations.

%%%%%%

Equation (\ref{offdcosmlc}) can be reformulated in nonholonomic variables to
obtain expressions such as (\ref{lacosm2}), using prime data $\underline{%
\mathring{g}}_{\alpha}$ from standard GR or from well-defined MGT
cosmologies. The resulting off-diagonal metrics preserve their core physical
meaning but acquire nontrivial polarizations and $\tau$-dependent (running)
constants that can be constrained by observations. In the next section, we
discuss methods for testing these solutions and identifying when they remain
compatible with GR. %%%%%%

\section{Testing off-diagonal cosmology and geometric thermodynamic variables%
}

\label{sec3} The nonholonomic Einstein equations (\ref{cdeq1}), and
generalizations to $\tau $-families of systems of nonlinear PDEs
nonholonomic Einstein equations (\ref{cdeq1b}) were derived using abstract
geometric methods as in \cite{misner73} for canonical geometric data $(%
\widehat{\mathbf{g}},\widehat{\mathbf{D}}).$ Such formulas can also be
derived in N-adapted variational forms using distortions of connections \cite%
{v25,bsssvv25}. For geometric flows, the variational calculus is performed
using certain modified $F$- and $W$-functionals introduced by G. Perelman 
\cite{perelman1}, see \cite{vv25a,vv25b}. Using off-diagonal cosmological
solutions, we can model in "almost" equivalent forms different MGTs, for
instance, defined by with such (effective) actions:%
\begin{eqnarray}
\ ^{\eta }S &=&\int \ ^{\eta }\delta ^{4}u\sqrt{\mid \ ^{\eta }\mathbf{g}%
_{\alpha \beta }\mid }\left[ \frac{\ ^{\eta }\widehat{\mathbf{R}}is}{16\pi G}%
+\ ^{m}L(\varphi ^{A},\ ^{\eta }\mathbf{g}_{\beta \gamma })\right] \simeq
\label{act1} \\
\ ^{\chi }S &=&\int \ ^{\chi }\delta ^{4}u\sqrt{\mid \ ^{\ \chi }\mathbf{g}%
_{\alpha \beta }\mid }\left[ \frac{\ ^{\chi }\widehat{\mathbf{R}}is}{16\pi G}%
+\ ^{m}L(\varphi ^{A},\ ^{\chi }\mathbf{g}_{\beta \gamma })\right] \simeq
\label{act2} \\
\mathring{S} &=&\int \mathring{\delta}^{4}u\sqrt{\mid \mathbf{\mathring{g}}%
_{\alpha \beta }\mid }\left[ \frac{f(\mathring{R})}{16\pi G}+\ ^{m}L(\varphi
^{A},\ \mathbf{\mathring{g}}_{\beta \gamma })+\ ^{e}L(\varphi ^{A},\ ^{\eta }%
\mathbf{g}_{\beta \gamma },\ ^{\eta }\widehat{\mathcal{Z}})\right] .
\label{act3}
\end{eqnarray}%
In these formulas, $G$ is the Newtonian gravitational constant, and $\ ^{m}L$
is the matter Lagrangian generating the energy-momentum tensor $\mathbf{T}%
_{\alpha \beta }$ (\ref{emdt}). The term $\ ^{\eta }\delta ^{4}u$ is defined
using the N-elongated differentials (\ref{nadif}). The effective Lagrange
density (\ref{act1}) is computed from the primary Lagrangians $\ ^{m}L$ and $%
\ ^{\eta }\widehat{\mathbf{R}}is$, , via canonical distortions (\ref%
{canondist}). The so-called $f(\mathring{R})$ gravity has been extensively
studied, as reviewed in \cite{sotiriou10, nojiri11, capo11, clifton12,
harko14, copeland06, v25, bsssvv25}. We emphasize that, in general, the
three actions (\ref{act1}), (\ref{act2}), and (\ref{act3}) define three
distinct classes of modified gravity theories, each with its own system of
modified Einstein equations. Nevertheless, one can impose suitable
nonholonomic geometric conditions on the generating data and their
distortions so that certain classes of off-diagonal solutions become
equivalent (or nearly equivalent, up to small parametric deformations),
allowing them to describe in similar form a variety of physical processes
and observational data. In this way, various nonlinear cosmological systems
and off-diagonal configurations can be modeled as exact or parametric
solutions in GR with effective sources and nonlinear symmetries determined
by distorted connections adapted to N-connection structures. %%%%%%

The goal of this section is to demonstrate that recent observational data in
modern cosmology can be equivalently described by modified gravity theories
of type (\ref{act1}) and (\ref{act3}) when expressed using appropriate
classes of off-diagonal cosmological solutions of type (\ref{lacosm2}). We
also note that theories involving the functional $\ ^{\chi }S$ (\ref{act2})
can be constructed as $\kappa $-parametric decompositions of $\ ^{\eta }S$ (%
\ref{act1}) (see the corresponding solutions in the Appendix). Via the
associated nonlinear symmetries, such models can be mapped to certain
off-diagonal cosmological configurations within GR. Importantly, these
physical scenarios lie beyond the standard Bekenstein-Hawking framework \cite%
{bek2,haw2}. This motivates the introduction, and computation at the end of
this section, of the corresponding cosmological versions of G. Perelman's
type thermodynamic variables \cite%
{perelman1,gheorghiuap16,vv25a,vv25b,v25,bsssvv25}. %%%%%%%

\subsection{Effective modelling of f(R) cosmology by off-diagonal solutions
in GR}

%%%%%%

The off-diagonal cosmological solutions of type (\ref{lacosm2}) are
constructed in general form for the corresponding system of nonlinear PDEs (%
\ref{cdeq1}), determined by generating and integration data $(\psi ,%
\underline{\eta }_{3};\ \underline{\Lambda },\ ^{v}\underline{\Upsilon }%
,_{1}n_{k},\ _{2}n_{k})$ and by certain prime d-metrics $\underline{\mathbf{%
\mathring{g}}}[\breve{a}]$ (\ref{pflrw}). Under additional assumptions, such
solutions can encode or reproduce recent observational data for accelerating
cosmology, similarly to the known diagonal configurations in MGTs \cite{od24}%
. Using the AFCDM, one can generate cosmological solutions for $\tau $%
-families of NESs, from which LC-configurations are extracted by imposing
the nonholonomic constraints (\ref{lccond1}). Because these nonlinear and
physically important PDE systems are solved directly in an off-diagonal
form, the resulting models possess additional functional degrees of freedom
supplied by the generating and integration functions. This framework is
therefore substantially more general than diagonal ansatz constructions,
where (modified) Einstein equations reduce to nonlinear ODE systems with
solutions determined only by integration constants. %%%%%

We can model off-diagonal geometric and cosmological evolution (\ref{lacosm2}%
) of a prime Friedman-Lama\^{\i}tre-Robrtson-Walker (FLRW) metric 
\begin{equation*}
d\breve{s}^{2}=\underline{\breve{g}}_{\alpha ^{\prime }\beta ^{\prime
}}du^{\alpha ^{\prime }}du^{\beta ^{\prime }}=\breve{a}%
^{2}(t)((dx^{1})^{2}+(dx^{2})^{2}+(dy^{3})^{2})-dt^{2},
\end{equation*}%
with a scaling parameter, $\underline{\breve{g}}_{\alpha ^{\prime }\beta
^{\prime }}=diag[\breve{a}^{2},\breve{a}^{2},\breve{a}^{2},-1]$ and
coordinates $u^{\alpha ^{\prime }}=(x^{1},x^{2},x^{3},t).$ To apply the
AFCDM without introducing coordinate singularities, we first perform a
frame/coordinate transformation $\underline{\mathring{g}}_{\alpha \beta
}(u^{\gamma })=e_{\ \alpha }^{\alpha ^{\prime }}e_{\ \beta }^{\beta ^{\prime
}}\underline{\breve{g}}_{\alpha ^{\prime }\beta ^{\prime }}(u^{\gamma
^{\prime }}),$ for $u^{\gamma }=(r,\theta ,\varphi ,t),$ expressed in a
conventional N-adapted form (\ref{dm}) (but for prime data),%
\begin{equation}
\underline{\mathbf{\mathring{g}}}[\breve{a}]=(\mathring{g}_{i}(r,\theta ),%
\underline{\mathring{g}}_{a}(r,\theta ,t),\underline{\mathring{N}}%
_{i}^{a}(r,\theta ,t)),  \label{pflrw}
\end{equation}%
which is a functional on the scaling cosmological function $\breve{a},$
which depends on the type of Friedman equations we postulate in a
cosmological theory with diagonalizable metrics. In GR, both $\underline{%
\mathbf{\mathring{g}}}[\breve{a}]$ and $\underline{\breve{g}}_{\alpha
^{\prime }\beta ^{\prime }}$ satisfy Einstein equations with, for example,
perfect fluid type source of type (\ref{emdt}), $\mathbf{\breve{T}}_{\alpha
^{\prime }\beta ^{\prime }}=diag(\breve{p},\breve{p},\breve{p},-\breve{\rho}%
),$ where $\breve{p}$ and $\breve{\rho}$ are the pressure and energy
density. Small parametric off-diagonal geometric flow and cosmological
evolution are encoded by formula (\ref{paramsoliton}), leading to d-metrics
of type (\ref{lacosm2}) which describe off-diagonal geometric and
cosmological evolution. These models can be analyzed using respective
relativistic geometric flow thermodynamics, as discussed in the next
section. %%%%%

We can introduce a set of $\eta $-polarization functions $(\underline{\eta }%
_{\alpha }(u^{\alpha }),\underline{\eta }_{i}^{a}(u^{\alpha }))$, which
define nonholonomic deformations of a prime cosmological d-metric $%
\underline{\mathbf{\mathring{g}}}=(\underline{\mathring{g}}_{\alpha
}(u^{\alpha }),\underline{\mathring{N}}_{i}^{a}(u^{\alpha }))$ into a target
cosmological d-metric $\underline{\mathbf{g}}=(\underline{g}_{\alpha
}(r,\theta ,t),\underline{N}_{i}^{a}(r,\theta ,t)),$ by means of relations $(%
\underline{g}_{\alpha }=\underline{\eta }_{\alpha }\underline{\mathring{g}}%
_{\alpha }, \underline{N}_{i}^{a}=\underline{\eta }_{i}^{a}\underline{%
\mathring{N}}_{i}^{a}).$ We parameterize local spherical coordinates as $%
u^{\alpha}=(x^{i},y^{a})=(r,\theta ,\varphi ,t)$ and use underline symbols
to emphasize dependence on the time-like variable $u^{4}=y^{4}=t.$ The
resulting d-metric $\widehat{\underline{\mathbf{g}}}=\underline{\mathbf{g}}%
(r,\theta ,t)$, see (\ref{dm}), defines an exact (or parametric) solution of
the nonholonomic Einstein equations (\ref{cdeq1}) for a prescribed effective
matter source encoded via the generating functions (\ref{esourc}). Thus, a
family of cosmological configurations is characterized by the canonical data 
$(\widehat{\underline{\mathbf{g}}},\underline{\widehat{\mathbf{D}}}, \ ^{h}%
\widehat{\mathbf{\Upsilon }},\ ^{v}\underline{\widehat{\mathbf{\Upsilon }}}, 
\underline{\Lambda })$ where the d-objects are determined by the chosen
generating and integration functions, effective/generating sources, and an
effective cosmological constant $\underline{\Lambda }.$ %%%%%%

Let us explain how we can model a $f(\mathring{R})$ cosmology by using
off-diagonal solutions (\ref{lacosm2}) of nonholonomic Einstein equations.
For this, we must consider prime metrics defined necessary types of
cosmological scale functions $\breve{a}(t)$ in $\underline{\mathbf{\mathring{%
g}}}[\breve{a}],$ for $\mathring{a}(t)$ in $\underline{\mathbf{\mathring{g}}}%
[\mathring{a}].$ We define such primary data \ (for $\underline{\mathring{%
\nabla}}$ and any trivial N-connection $\underline{\mathring{N}}_{i}^{a}$
and a nontrivial cosmological constant $\mathring{\Lambda}$): 
\begin{eqnarray*}
\mathring{f}_{R}&:= &\frac{df(\mathring{R})}{d\mathring{R}}%
\mbox{ for a
prime modification of the Einstein vacaruum when }\widehat{\mathbf{\mathring{%
R}}}is=\mathring{R}=\breve{R}=4\mathring{\Lambda}; \\
\mathring{H}&:= &\frac{\mathring{a}^{\diamond }}{\mathring{a}}%
\mbox{ is the
conventional prime Hubble constant}.
\end{eqnarray*}%
In these formulas, following our conventions from \cite%
{vv25a,vv25b,v25,bsssvv25}, $\mathring{a}^{\diamond}:=\partial \mathring{a}%
/\partial t,$ where $\mathring{\Lambda}$ is a cosmological constant used in
GR, with possible extensions to certain MGTs. In general, the physical
meaning of a $\mathring{\Lambda}$ can be different from $\underline{\Lambda }
$ in (\ref{nonlintrsmalp}), or a $\tau $-running $\underline{\Lambda }(\tau) 
$ (\ref{cdeq1b}). We can prescribe a prime value $\mathring{\Lambda}=%
\underline{\Lambda }$, or model a $\tau $-evolution determined by geometric
flow equations (\ref{cdeq1b}) when $\mathring{\Lambda}=\underline{\Lambda }
(\tau ),$ for $\tau =\tau _{0}.$ The continuity equations, $\underline{%
\mathring{\nabla}}^{\alpha ^{\prime }}\mathbf{\mathring{T}}_{\alpha ^{\prime
}\beta ^{\prime }}=0$ for a prime energy-momentum tensor $\mathbf{\mathring{T%
}}_{\alpha ^{\prime }\beta ^{\prime }}=diag(\mathring{p},\mathring{p}, 
\mathring{p}, -\mathring{\rho})$ and the LC-connection $\underline{\mathring{%
\nabla}}$ defined by $\underline{\mathbf{\mathring{g}}}[\mathring{a}]$ are
obtained in the usual form:%
\begin{equation}
\mathring{\rho}^{\diamond }=-3\mathring{H}(\mathring{\rho}+\mathring{p}).
\label{primconserv}
\end{equation}%
%
%
%
%
%%%%%%

The prime FLRW equations can be written in standard form for the $f(%
\mathring{R})\,$-modified gravity (see \cite{od24} and references therein)%
\begin{equation}
\frac{d\mathring{H}}{d\log \mathring{a}}=\frac{\mathring{R}}{6\mathring{H}}-2%
\mathring{H},\mbox{ for }\mathring{R}=6\mathring{H}^{\diamond }+12\mathring{H%
}^{2},\ \frac{d\mathring{R}}{d\log \mathring{a}}=(3\mathring{f}_{RR})^{-1}%
\mathring{H}^{-2}(8\pi G\mathring{\rho}+\mathring{f}_{R}(\frac{\mathring{R}}{%
2}-\mathring{H}^{2})-\frac{\mathring{f}}{2}).  \label{scalingf}
\end{equation}%
Such formulas can be used to define cosmological models for certain
exponential models, for instance, given by 
\begin{equation}
f(\mathring{R})=\mathring{R}-2\mathring{\Lambda}(1-e^{-\beta \mathcal{%
\mathring{R}}^{\alpha }}),\mbox{ for a normalized Ricci scalar }\mathcal{%
\mathring{R}=}\frac{\mathring{R}}{2\mathring{\Lambda}},  \label{mgrmod}
\end{equation}%
where $\beta $ and $\alpha $ are dimensionless constants to be determined by
experimental data in some diagonal limits of prime metrics. %%%%%

Generic off-diagonal cosmological $\eta \,$-deformations to d-metrics (\ref%
{lacosm2}) transform the scale factor and local bases (and all above
formulas, with conventional $\mathring{H}\rightarrow H,\mathring{R}
\rightarrow \widehat{\mathbf{R}}is, \mathring{\Lambda}\rightarrow \underline{%
\Lambda }(\tau ),$ etc.), 
\begin{equation}
\mathring{a}\rightarrow a(x^{i},t)=\ ^{\eta }a(x^{i},t)\mathring{a}%
\mbox{
and }\partial _{\alpha }=(\partial _{i},\partial _{3},\partial
_{t})=(\partial _{i},\underline{\mathbf{\mathring{e}}}_{3},\underline{%
\mathbf{\mathring{e}}}_{4})\rightarrow \underline{e}_{\alpha }.
\label{etascalfact}
\end{equation}%
The N-adapted frame and d-metric coefficients for such transforms are
computed using formulas 
\begin{eqnarray}
d\widehat{\underline{s}}^{2} &=&\ ^{\eta }\widehat{\underline{g}}_{\alpha
\beta }du^{\alpha }du^{\beta }=\ ^{\eta }a^{2}(x^{i},t)\mathring{a}^{2}[(%
\underline{e}^{1})^{2}+(\underline{e}^{2})^{2}+(\underline{e}^{3})^{2}]-(%
\underline{e}^{4})^{2},\mbox{ for any}  \label{lacosm3} \\
&&\ ^{\eta }a(x^{i},t)\mathring{a}\underline{e}^{i} = e^{\psi }dx^{i},%
\mbox{
for a corresponding change of local bases/coordinates with  }i=1,2;  \notag
\\
&& \ ^{\eta }a(x^{i},t)\mathring{a}\underline{e}^{3} = (\underline{\eta }%
\underline{\mathring{g}}_{3})^{1/2}\{d\varphi +[\ _{1}n_{k}+\ _{2}n_{k}\int
dt\frac{[\partial _{t}(\underline{\eta }\underline{\mathring{g}}_{3})]^{2}}{%
|\int dt\ ^{v}\underline{\Upsilon }\partial _{t}(\underline{\eta }\underline{%
\mathring{g}}_{3})|\ (\underline{\eta }\underline{\mathring{g}}_{3})^{5/2}}%
]dx^{k}\},  \notag \\
&& \underline{e}^{4} =\frac{[\partial _{t}(\underline{\eta }\ \underline{%
\mathring{g}}_{3})]}{(|\int dt\ ^{v}\underline{\Upsilon }\partial _{t}(%
\underline{\eta }\underline{\mathring{g}}_{3})|\ \eta \mathring{g}_{3})^{1/2}%
}\{dt+\frac{\partial _{i}[\int dt\ ^{v}\underline{\Upsilon }\ \partial _{t}(%
\underline{\eta }\underline{\mathring{g}}_{3})]}{\ ^{v}\underline{\Upsilon }%
\partial _{t}(\underline{\eta }\underline{\mathring{g}}_{3})}dx^{i}\}. 
\notag
\end{eqnarray}%
Both representations (\ref{lacosm2}) and (\ref{lacosm3}) describe equivalent
classes of off-diagonal cosmological metrics: The first one is more
convenient for generating new classes of solutions, but the second one can
be used for computing and comparing physical properties of certain
cosmological parameters. For instance, we can analyze how a $a(x^{i},t)$
involves $\eta $-polarizations of prime $\mathring{a}$, and such physical
effects are described in a nontrivial gravitational vacuum determined by a
correspondingly constructed nonholonomic (dual) frame $\underline{e}^{\alpha
}.$ %%%%%%%

Such formulas may be extended for $\tau $-families of cosmological solutions
with $a(\tau ,x^{i},t)=\ ^{\eta }a(\tau,x^{i},t)\mathring{a}$ (\ref%
{etascalfact}) and related cosmological parameters. Considering respective
N-adapted frame structures, we can generalize (\ref{scalingf}) to 
\begin{eqnarray}
\frac{dH}{d\log a} &=&\frac{dH}{d\log a}=\frac{\widehat{R}sc}{6H}-2H=\frac{%
\mathring{R}+\widehat{Z}sc}{6\ ^{\eta }H\mathring{H}}-2\ ^{\eta }H\mathring{H%
},  \notag \\
\mbox{ for } H &:=&\frac{a^{\diamond }}{a}=\frac{(\ ^{\eta }a\mathring{a}%
)^{\diamond }}{\ ^{\eta }a\mathring{a}}=\ ^{\eta }H\mathring{H},%
\mbox{ where
}\ ^{\eta }H=1+\frac{\ ^{\eta }a^{\diamond }}{\ ^{\eta }a\mathring{H}}.
\label{offdcosmp}
\end{eqnarray}%
Some locally anisotropic $a(\tau ,x^{i},t)$ and $H(\tau ,x^{i},t)$ satisfy
more sophisticate conservation laws than (\ref{primconserv}) because of
nonholonomic variables. The distortion relation $\widehat{R}sc=\mathring{R}+ 
\widehat{Z}sc$ can be computed using for formula (\ref{canondist}). Here, we
note that the formula (\ref{mgrmod}) for MGT and respective prime
cosmological metrics do not have "simple" analogs with $f(\mathring{R}%
)\rightarrow f(\widehat{R}sc)$ even such transforms can be computed by using
(\ref{nonlintrsmalp}) with $\kappa $-parametric decompositions. In this
work, we use classes of prime and target solutions with $f(\mathring{R})$,
which can be $\eta $-deformed in off-diagonal solutions of NESs and their
geometric flows. %%%%%%

\subsection{Testing the off-diagonal GR and MGTs models with observational
data}

Certain models of f(R) gravity considered in the literature \cite{od24} (see
also references therein) were analyzed for $\alpha =1$. For $\alpha >0$, the
above prime models reduce to the usual $\Lambda$CDM scenario in the limits $%
\beta \rightarrow +\infty $ and/or $\mathring{R}\rightarrow +\infty $ (the
latter limit corresponding to early-time cosmology). In this way, both GR
and MGT cosmological models can be examined at early times, since the
earliest observational data arise from the Cosmic Microwave Background
(CMB), at redshift $z\simeq 1100.$ By contrast, the SN Ia, CC, BAO, and
related observational data \cite{desi24,roy24,batic24,dival21} are located
at much lower redshifts, with $z\leq 2.4.$ The authors of \cite{od24,od24a}
estimate that for $\mathcal{\mathring{R}}<10^{13}$, the prime cosmological
epoch corresponds to $z< 10^{4},$ during which any inflationary
contributions are negligible. This follows from the fact that at the end of
inflation the normalized prime scalar satisfies $\mathcal{\mathring{R}}_{0}%
\mathcal{\sim } 10^{85}$; that is, the quantity defined in (\ref{mgrmod}) is
many orders of magnitude larger during the inflationary phase. %%%%%%

\subsubsection{Cosmological parameters for $\Lambda $CDM, f(R) and
off-diagonal metrics}

The matter domination epoch $z<10^{4}$ begins with a pressureless matter
containing baryons and DM with prime densities and respective cosmological
evolution: 
\begin{equation}
\ _{m}\mathring{\rho}=\ _{b}\mathring{\rho}+\ _{DM}\mathring{\rho}%
\mbox{
and, using }(\ref{primconserv}),\mathring{\rho}=\ _{m}^{0}\mathring{\rho}\ 
\mathring{a}^{-3}+\ _{r}^{0}\mathring{\rho}\ \mathring{a}^{-4},  \label{dmd}
\end{equation}%
where $\ _{m}^{0}\mathring{\rho}$ and $\ _{r}^{0}\mathring{\rho}$ are prime
energy densities of matter and radiation stated at present time $t_{0}$ for $%
\mathring{a}(t_{0})=1.$ We can incorporate the Planck data and reduce the
number of free prime parameters as in \cite{od24,od24a} by specifying $\ _{r}%
\mathring{X}=\ _{r}^{0}\mathring{\rho}/\ _{m}^{0}\mathring{\rho}=2,9656\cdot
10^{-4}.$ The prime MGT with cosmological data (\ref{scalingf}) and (\ref%
{mgrmod}) in the limit $\mathring{R}\rightarrow +\infty $ (more precisely,
for $\beta $ $\mathcal{\mathring{R}}^{\alpha }\gg 1$) transforms into the $%
\Lambda $CDM model. Therefore, without loss of generality, we may assume
that for prime configurations the Hubble parameter and the Ricci scalar
asymptotically approach their $\Lambda$CDM values in the form:%
\begin{equation}
\frac{\mathring{H}^{2}}{(\ _{0}^{\ast }H)^{2}}=\frac{\ _{m}^{\ast }\mathring{%
\Omega}}{\mathring{a}^{3}}(1+\frac{_{r}\mathring{X}}{\mathring{a}})+\
_{\Lambda }^{\ast }\mathring{\Omega}\mbox{ and }\frac{\mathring{R}}{2%
\mathring{\Lambda}}=2+\frac{\ _{m}^{\ast }\mathring{\Omega}}{2\ _{\Lambda
}^{\ast }\mathring{\Omega}\mathring{a}^{3}}.  \label{primecosmpar}
\end{equation}%
In these formulas, $\ {0}^{\ast }H$ denotes the Hubble constant, while ${m}%
^{\ast }\mathring{\Omega}$ and $_{\Lambda }^{\ast }\mathring{\Omega}$ are,
respectively, the ratios of the matter density and the cosmological constant
parameters (see details in \cite{od24a}) for a $\Lambda$CDM model. In our
case, such a $\Lambda$CDM background is used to mimic both the $f(R)$
configurations and the (prime) off-diagonal cosmological solutions. %%%%%

For target off-diagonal cosmological configurations (\ref{lacosm3}) and (\ref%
{offdcosmp}), the cosmological parameters (\ref{primecosmpar}) can be
consistently expressed in N-adapted frames and generalized in the form: 
\begin{equation}
\frac{H^{2}}{(\ _{0}^{\ast }H)^{2}}=\frac{\ _{m}^{\ast }\Omega }{a^{3}}(1+%
\frac{_{r}X}{a})+\ _{\Lambda }^{\ast }\Omega \mbox{ and }\frac{\widehat{R}sc%
}{2\underline{\Lambda }}=2+\frac{\ _{m}^{\ast }\Omega }{2\ _{\Lambda }^{\ast
}\Omega a^{3}}.  \label{targcosmpar}
\end{equation}%
In general, the values of the cosmological $\Lambda$CDM parameters differ
for the prime exponential $f(R)$ gravity and the corresponding off-diagonal
modeling in GR. We assume that these solutions and theories match the values
in (\ref{primecosmpar}) and (\ref{targcosmpar}) at redshifts $10^{3} \leq z
\leq 10^{5}$, where various modifications, geometric flows, and off-diagonal
cosmological evolution scenarios can lead to deviations from the standard $%
\Lambda$CDM evolution. This behavior can be described by formulas of the
form: 
\begin{eqnarray}
_{m}\mathring{\Omega}(\ \mathring{H})^{2} &=&\ _{m}^{\ast }\mathring{\Omega}%
(\ _{0}^{\ast }H)^{2}=\frac{8\pi G}{3}\ _{m}^{0}\mathring{\rho}(t_{0})%
\mbox{
and }\ _{\Lambda }\mathring{\Omega}(\ \mathring{H})^{2}=\ _{m}^{\ast }%
\mathring{\Omega}(\ _{0}^{\ast }H)^{2}=\frac{\mathring{\Lambda}}{3},%
\mbox{
in f(R) gravity };  \label{3modcomp} \\
_{m}\Omega (\tau )(\ H(\tau ))^{2} &=&\ _{m}^{\ast }\mathring{\Omega}(\
_{0}^{\ast }H)^{2}=\frac{8\pi G}{3}\ _{m}^{0}\rho (\tau ,t_{0})\mbox{ and }
\ _{\Lambda }\Omega (\tau )(\ H(\tau ))^{2}=\ _{m}^{\ast }\mathring{\Omega}%
(\ _{0}^{\ast }H)^{2}=\frac{\underline{\Lambda }(\tau )}{3},%
\mbox{
off-diagonal }.  \notag
\end{eqnarray}
%%%%%%%

Introducing normalized Hubble parameters, $\ ^{n}H:=H/\ _{0}^{\ast }H\,$ and 
$\ ^{n}\mathring{H}:=\mathring{H}/\ _{0}^{\ast }H$, and using (\ref{3modcomp}%
), the cosmological equations (\ref{scalingf}) can be equivalently expressed
in the form: 
\begin{equation}
\frac{d\ ^{n}\mathring{H}}{d\log \mathring{a}}=\ _{m}^{\ast }\mathring{\Omega%
}\frac{\mathcal{\mathring{R}}}{\ ^{n}\mathring{H}}-2\ ^{n}\mathring{H},\frac{%
d\ \mathcal{\mathring{R}}}{d\log \mathring{a}}=\frac{\frac{\ _{m}^{\ast }%
\mathring{\Omega}}{\mathring{a}^{3}}(1+\frac{_{r}\mathring{X}}{\mathring{a}}%
)+\ _{\Lambda }^{\ast }\mathring{\Omega}[1-(1+\alpha \beta \mathcal{%
\mathring{R}}^{\alpha })e^{-\beta \mathcal{\mathring{R}}^{\alpha }}]}{\alpha
\beta (\alpha \beta \mathcal{\mathring{R}}^{\alpha }+1-\alpha )\mathcal{%
\mathring{R}}^{\alpha -2}e^{-\beta \mathcal{\mathring{R}}^{\alpha }}(\ ^{n}%
\mathring{H}^{2}-1+\alpha \beta \mathcal{\mathring{R}}^{\alpha -1}e^{-\beta 
\mathcal{\mathring{R}}^{\alpha }}).}  \label{scalingfn}
\end{equation}%
This system of equations can be solved numerically to obtain approximate
solutions. At an initial point, the factor satisfies $\epsilon \sim
e^{-\beta \mathcal{\mathring{R}}^{\alpha }}\sim 10^{-9},$ see details in 
\cite{od24a,od24}. Using the formulas (\ref{primecosmpar}), the equations (%
\ref{scalingfn}) allow us to compute an initial value%
\begin{equation*}
(\log \mathring{a})_{ini}=-\frac{1}{3}\log \{\frac{\ _{\Lambda }^{\ast }%
\mathring{\Omega}}{\ _{m}^{\ast }\mathring{\Omega}}\left[ \left( -\frac{\log
\epsilon }{\beta }\right) ^{1/\alpha }-2\right] \}.
\end{equation*}%
This implies that certain parametric solutions $\ ^{n}\mathring{H}(\mathring{%
a})$ and $\mathcal{\mathring{R}}(\mathring{a})$ can be obtained numerically
for this particular f(R) gravity. These solutions allow one to define the
Hubble parameter $\mathring{H}(\mathring{a})$ or, equivalently, $\mathring{H}%
(z),$ for $z=\mathring{a}^{-1}-1$ and $\mathring{a}(t_{0})=1,$ via solutions
of (\ref{3modcomp}). In this way, the cosmological evolution predicted by
such MGTs can be compared with experimental data. %%%%%%%%

The f(R) cosmological configurations can be embedded into cosmological
off-diagonal solutions within GR. This embedding can be evaluated explicitly
for small $\kappa $-deformations and $\chi $-polarizations as discussed in
the Appendix, see formulas (\ref{paramsoliton}). Specifically, we can
express this via efficient dependencies as $\mathring{H}(\mathring{a}%
)\rightarrow H(a)\simeq \mathring{H}(\mathring{a})+\kappa ^{\lbrack
1]}H(a(\tau ,r,\theta )),$ which allows for effective, small, locally
anisotropic polarizations of the cosmological constant.

More generally, using (\ref{offdcosmp}) and the corresponding nonholonomic
configurations, one can model general off-diagonal cosmological $\eta $%
-deformations in GR, which can differ significantly from the prime $\Lambda $%
CDM structure. The physical properties of such configurations cannot
generally be understood in closed form, even when local N-adapted
parameterizations (\ref{lacosm3}) are introduced. Nevertheless, these
cosmological models can be characterized thermodynamically by employing a
relativistic generalization \cite{vv25a,vv25b} of Perelman's approach \cite%
{perelman1}, as discussed at the end of this section. %%%%%%%

\subsubsection{N-adapted cosmological observational data and off-diagonal
solutions}

We discuss fitting an off-diagonal cosmological solution (\ref{lacosm3}) to
observational data. Unlike diagonal f(R) models, the effective scale factor $%
a(\tau ,x^{i},t)$ (\ref{etascalfact}) and Hubble parameter $H(\tau ,x^{i},t)$
(\ref{offdcosmp}) depend on spatial coordinates and a temperature-like
parameter $\tau .$ The model describes a locally anisotropic vacuum in GR
with a $\tau $-running effective cosmological constant $\underline{\Lambda }%
(\tau )$ and nonlinear symmetries (\ref{nonlinsymcosm}). Observational
constraints include SNe Ia, BAO, and CC data, with Hubble parameters $%
\mathring{H}(\mathring{a})\sim \mathring{H}(\mathring{z}),$ for prime f(R)
solutions (\ref{mgrmod}) and $H(z)\simeq H(\tau ,x^{i},z)$ for target
configurations (\ref{targcosmpar}). While $H(z)$ exhibits local anisotropy
and $\tau $-dependence, the model remains within GR, whereas $\mathring{H}(%
\mathring{z})$ pertains to the modified gravity framework. %%%%%%

We begin with the Pantheon database \cite{pantheon21}. The distance moduli $%
\ ^{obs}\mu _{\lbrack i]}$ (at red-shifts $z_{[i]}$ for $[i]$ labeling 1550
spectroscopically SNe Ia) are used for computing the so-called $\chi
_{SN}^{2}$ function for $N_{SN}=1701$ data points:%
\begin{equation}
\chi _{SN}^{2}(\theta _{\lbrack
i]},...)=\min_{H_{0}}\sum\nolimits_{[i],[j]=1}^{N_{SN}}\bigtriangleup \mu
_{\lbrack i]}(C_{SN}^{-1})_{[i][j]}\bigtriangleup \mu _{\lbrack j]},%
\mbox{
where }\bigtriangleup \mu _{\lbrack i]}=\ ^{th}\mu (z_{[i]},\theta _{\lbrack
i]},...)-\ ^{obs}\mu _{\lbrack i]}.  \label{chiparam}
\end{equation}%
In these formulas, $\theta _{\lbrack i]}$ are free model parameters and $%
C_{SN}$ is the $N_{SN}\times N_{SN}$ covariance matrix. The theoretical
values $\ ^{th}\mu $ can be computed for $\mathring{H}(\mathring{z})$ or $%
H(z)$ as follows:%
\begin{eqnarray}
\ ^{th}\mathring{\mu}(\mathring{z}) &=&5\log _{10}\frac{(1+\mathring{z})%
\mathring{D}_{M}(\mathring{z})}{10pc},\mbox{ for }\mathring{D}_{M}(\mathring{%
z})=c\int\nolimits_{0}^{\mathring{z}}\frac{d\ ^{1}z}{\mathring{H}(^{1}z)},%
\mbox{ or }  \notag \\
\ ^{th}\mu (z) &=&5\log _{10}\frac{(1+z)D_{M}(z)}{10pc},\mbox{ for }%
D_{M}(z)=c\int\nolimits_{0}^{z}\frac{d\ ^{1}z}{H(^{1}z)}.  \label{auxaaa}
\end{eqnarray}%
Using $\ ^{th}\mathring{\mu}(\mathring{z})$ or $\ ^{th}\mu (z)$ in (\ref%
{chiparam}), we can define the corresponding prime and target values of $%
\chi _{SN}^{2}$. These values are evaluated with $\ _{0}H$ (or equivalently $%
\ _{0}^{\ast }H$) treated as a nuisance parameter. In this framework, one
can analyze locally anisotropic data behavior, potentially dependent on a
temperature parameter $\tau$. Prime configurations allow the distinction of
modified logarithmic $f(R)$ models. %%%%%%

\vskip5pt At the next step, we consider the BAO new data from DESI \cite%
{desi24}. We can compare the results for diagonal or off-diagonal
configurations by calculating in both cases two distances:%
\begin{eqnarray}
\mathring{d}_{z}(\mathring{z}) &=&\frac{\mathring{r}_{s}(\mathring{z}_{d})}{%
\mathring{D}_{V}(\mathring{z})}\mbox{ and }\mathring{A}(\mathring{z})=\frac{%
\ _{0}\mathring{H}\sqrt{\ _{m}^{0}\Omega }}{c\mathring{z}}\mathring{D}_{V}(%
\mathring{z}),\mbox{ where }\mathring{D}_{V}(\mathring{z})=\left[ \frac{c%
\mathring{z}\mathring{D}_{M}^{2}(\mathring{z})}{\mathring{H}(\mathring{z})}%
\right] ^{1/3};\mbox{ and }  \notag \\
d_{z}(z) &=&\frac{r_{s}(z_{d})}{D_{V}(z)}\mbox{ and }A(z)=\frac{\ _{0}H\sqrt{%
\ _{m}^{0}\Omega }}{cz}D_{V}(z),\mbox{ where }D_{V}(z)=\left[ \frac{%
czD_{M}^{2}(z)}{H(z)}\right] ^{1/3}.  \label{distanis}
\end{eqnarray}%
In these formulas, $\mathring{z}{d}=z{d}$ -- corresponding to the ratio of
baryons to photons, ${b}^{0}\Omega /{\gamma }\Omega$, fixed by the Planck
2018 data \cite{agh18} -- is defined as the redshift at the end of the
baryon drag era. The co-moving sound horizons in the prime and target models
are computed following \cite{od24a} (see also Table I in \cite{od24} and
references therein): 
\begin{eqnarray}
\mathring{r}_{s}(\mathring{z}) &=&\int\nolimits_{\mathring{z}}^{\infty }%
\frac{\mathring{c}_{s}(\ ^{1}z)d\ ^{1}z}{\mathring{H}(^{1}z)}=\frac{1}{\sqrt{%
3}}\int\nolimits_{\mathring{z}}^{1/(1+\mathring{z})}\frac{d\mathring{a}}{%
\mathring{a}^{2}\mathring{H}(\mathring{a})\sqrt{1+[3_{b}^{0}\mathring{\Omega}%
/4_{\gamma }\mathring{\Omega}]\mathring{a}}}\mbox{ and }  \notag \\
r_{s}(z) &=&\int\nolimits_{z}^{\infty }\frac{c_{s}(\ ^{1}z)d\ ^{1}z}{H(^{1}z)%
}=\frac{1}{\sqrt{3}}\int\nolimits_{z}^{1/(1+z)}\frac{da}{a^{2}H(a)\sqrt{%
1+[3_{b}^{0}\Omega /4_{\gamma }\Omega ]a}}.  \label{comovhor}
\end{eqnarray}%
%
%
%
%
%%%%%%%

For off-diagonal configurations with the corresponding effective redshift $\
^{eff}z$, TABLE I lists the observational ranges for $z\left( x^{i},t\right)
=a^{-1}-1$, computed for any class of off-diagonal cosmological solutions
with effective scale factor $a=\ ^{\eta }a\mathring{a}$ (\ref{etascalfact})
and for $d_{z}(z)$ (\ref{distanis}): 
\begin{equation*}
\begin{array}{cc}
\mbox{TABLE I.} & \mbox{DESI DR1 BAO ranges of observational data} \\ 
\begin{array}{c}
\ ^{eff}z \\ 
- \\ 
z\left( x^{i},t\right) \subset \\ 
- \\ 
d_{z}(z\left( x^{i},t\right) )\subset \\ 
\end{array}
& 
\begin{array}{cccccc}
\begin{array}{c}
0.295 \\ 
-%
\end{array}
& 
\begin{array}{c}
0.51 \\ 
-%
\end{array}
& 
\begin{array}{c}
0.706 \\ 
-%
\end{array}
& 
\begin{array}{c}
0.93 \\ 
-%
\end{array}
& 
\begin{array}{c}
1.317 \\ 
-%
\end{array}
& 
\begin{array}{c}
2.33 \\ 
-%
\end{array}
\\ 
\begin{array}{c}
\lbrack 0.1-0.4] \\ 
-%
\end{array}
& 
\begin{array}{c}
\lbrack 0.4-0.6] \\ 
-%
\end{array}
& 
\begin{array}{c}
\lbrack 0.5-0.8] \\ 
-%
\end{array}
& 
\begin{array}{c}
\lbrack 0.8-1.1] \\ 
-%
\end{array}
& 
\begin{array}{c}
\lbrack 1.1-1.6] \\ 
-%
\end{array}
& 
\begin{array}{c}
\lbrack 1.77-4.16] \\ 
-%
\end{array}
\\ 
\begin{array}{c}
\lbrack 0.1237- \\ 
0.1285]%
\end{array}
& 
\begin{array}{c}
\lbrack 0.0778- \\ 
0.1303]%
\end{array}
& 
\begin{array}{c}
\lbrack 0.0615- \\ 
0.0643]%
\end{array}
& 
\begin{array}{c}
\lbrack 0.05026- \\ 
0.0542]%
\end{array}
& 
\begin{array}{c}
\lbrack 0.04133- \\ 
0.04155]%
\end{array}
& 
\begin{array}{c}
\lbrack 0.03100- \\ 
0.03246]%
\end{array}%
\end{array}%
\end{array}%
\end{equation*}
In this work, we label the corresponding dataset as TABLE I so as to clearly
differentiate it from Table I presented in \cite{od24}.

In formulas (\ref{comovhor}), the observational data \cite{desi24} include
BAO for red-shift range $0.1<z<4.16.$ Then, the $\chi ^{2}$ function fits
the BAO data as%
\begin{eqnarray*}
\mathring{\chi}_{BAO}^{2}(\ _{m}^{0}\mathring{\Omega},\theta _{\lbrack
i]},...) &=&\bigtriangleup \mathring{d}\cdot C_{d}^{-1}(\bigtriangleup 
\mathring{d})^{T}+\bigtriangleup \mathring{A}_{[i]}\cdot
C_{A}^{-1}(\bigtriangleup \mathring{A}_{[i]})^{T}, \\
\chi _{BAO}^{2}(\ _{m}^{0}\Omega ,\theta _{\lbrack i]},...)
&=&\bigtriangleup d\cdot C_{d}^{-1}(\bigtriangleup d)^{T}+\bigtriangleup
A_{[i]}\cdot C_{A}^{-1}(\bigtriangleup A_{[i]})^{T},
\end{eqnarray*}%
where the one-line target matrix $\bigtriangleup d=[\bigtriangleup d_{[i]}=
\ ^{obs}d_{z}(z_{[i]})-\ ^{th}d_{z}(z_{[i]},...)]$ can be written also as a
transposition $(\bigtriangleup d)^{T}=[\bigtriangleup
d_{[i]}]^{T};\bigtriangleup A=[\bigtriangleup A_{[i]}];$ and $\bigtriangleup
A=[\bigtriangleup A_{[i]}=\ ^{obs}A_{z}(z_{[i]})-\ ^{th}A_{z}(z_{[i]},...)].$
The prime values are similarly defined and computed, for instance, using $%
\bigtriangleup \mathring{d}=[\bigtriangleup \mathring{d}_{[i]}=\ ^{obs}d_{z}(%
\mathring{z}_{[i]})-\ ^{th}d_{z}(\mathring{z}_{[i]},...)].$ The covariance
matrices, $C_{d}^{-1}$ and $C_{A}^{-1},$ for the correlated BAO data are
defined as in \cite{perc09,blake11}. %%%%%%%

Now, we can compute the $\chi ^{2}$ functions of the Cosmic Chronometers
(CC) of the respective prime and target Hubble parameter data:%
\begin{equation*}
\mathring{\chi}_{H}^{2}(\theta _{1},...)=\sum\nolimits_{[i]=1}^{N_{H}}\left[%
\frac{\mathring{H}(\mathring{z}_{[i]},\theta _{1},...) - \ ^{obs}H(z_{[i]}}{%
\sigma _{\lbrack i]}}\right] ^{2}\mbox{ and }\chi _{H}^{2}(\theta
_{1},...)=\sum\nolimits_{[i]=1}^{N_{H}}\left[ \frac{H(z_{[i]},\theta
_{1},...)-\ ^{obs}H(z_{[i]}}{\sigma _{\lbrack i]}}\right] ^{2},
\end{equation*}%
where the data points $N_{H}=32$ can be taken from \cite{borghi21} (we
provide the most recent data). For such measurements and computations, the
formulas of type $H(z)=a^{\diamond }/a\simeq -(1+z)^{-1}\bigtriangleup
z/\bigtriangleup t$ are used. %%%%%

Then, the CMB observational parameters from Planck 2018 data \cite%
{agh18,chen18} are considered as 
\begin{eqnarray}
\mathring{R} &=&\sqrt{\ _{m}^{0}\mathring{\Omega}}\frac{\mathring{H}_{0}%
\mathring{D}_{M}(\mathring{z}_{\ast })}{c},\mathring{\ell}_{A}=\frac{\pi 
\mathring{D}_{M}(\mathring{z}_{\ast })}{\mathring{r}_{s}(\mathring{z}_{\ast
})},\ _{b}\mathring{\omega}=\ _{b}^{0}\mathring{\Omega}\mathring{h}^{2}; 
\notag \\
R &=&\sqrt{\ _{m}^{0}\Omega }\frac{H_{0}D_{M}(z_{\ast })}{c},\ell _{A}=\frac{%
\pi D_{M}(z_{\ast })}{r_{s}(z_{\ast })},\ _{b}\omega =\ _{b}^{0}\Omega h^{2},
\label{auxaab}
\end{eqnarray}%
In these formulas, the co-moving distances $\mathring{D}_{M}(\mathring{z}%
_{\ast })$ and $D_{M}(z_{\ast })$ are as in (\ref{auxaaa}); $\mathring{h}=%
\mathring{H}_{0}/[100$kms$^{01}M$pc$^{01}]$ and $h=H_{0}/[100$kms$^{01}M$pc$%
^{01}]$, and the co-moving sound horizons are computed as (\ref{comovhor}).
Such values allow us to compute the corresponding $\chi ^{2}$ functions for
the CMB data: 
\begin{equation*}
\mathring{\chi}_{CMB}^{2}=\min_{\ _{b}\omega }\bigtriangleup \mathbf{%
\mathring{x}\cdot }C_{MBX}^{-1}(\bigtriangleup \mathbf{\mathring{x}})^{T}%
\mbox{ and }\chi _{CMB}^{2}=\min_{\ _{b}\omega }\bigtriangleup \mathbf{%
x\cdot }C_{MBX}^{-1}(\bigtriangleup \mathbf{x})^{T},
\end{equation*}%
where (for instance, using respective prime and target values) $%
\bigtriangleup \mathbf{x=x-x}^{Pl},$ for $\mathbf{x}= (R,\ell _{A},\
_{b}\omega )$ from (\ref{auxaab}) and observational values considered in the
above cited works on Planck 2018 data, 
\begin{equation*}
\mathbf{\mathring{x}}^{Pl}=\mathbf{x}^{Pl}=(1.7428\pm 0.0053,301.406\pm
0.090,0.02259\pm 0.00017).
\end{equation*}%
We note that the covariance matrix $C_{MBX}$ is described in \cite{chen18}.

So, we conclude that the observational data may allow us to distinguish
between prime exponential f(R) cosmological models and the locally
anisotropic $\tau $-running families of off-diagonal cosmological NES
configurations.

Thus, observational data may permit distinguishing between prime exponential
f(R) cosmological models and the locally anisotropic, $\tau $-dependent
families of off-diagonal cosmological NES configurations. %%%%%%%%%

\subsubsection{$\protect\chi ^{2}$ functions for prime f(R) and target
off-diagonal cosmological models}

The off-diagonal cosmological solutions (\ref{lacosm3}) and the associated
prime (\ref{primecosmpar}) and target (\ref{targcosmpar}) configurations can
be subjected to experimental tests. For simplicity, we fix $\alpha =1$ in (%
\ref{mgrmod}); more generally, it can be treated as a free parameter for
MGTs or as physical constant related to an integration constant of the prime
d-metric. By summarizing the $\chi ^{2}$ functions from the previous
subsection, we can compute and analyze the contributions from SN Ia, BAO,
CC, and CMB data: 
\begin{equation}
\mathring{\chi}^{2}=\mathring{\chi}_{SN}^{2}+\mathring{\chi}_{BAO}^{2}+%
\mathring{\chi}_{H}^{2}+\mathring{\chi}_{CMB}^{2}\mbox{ and }\chi ^{2}(\tau
,x^{i})=\chi _{SN}^{2}+\chi _{BAO}^{2}+\chi _{H}^{2}+\chi _{CMB}^{2}.
\label{totchi}
\end{equation}%
The dependencies on $(\tau ,x^{i})$ for $\chi ^{2}$, as defined in locally
anisotropic and $\tau$-running flows, can be treated as parametric with
respect to certain N-adapted frames. Such behavior is relevant in DM
physics, particularly when considering filamentary structures and the
nontrivial vacuum structure of the accelerating Universe. Through the
nonlinear symmetries (\ref{nonlinsymcosm}), these locally anisotropic and
temperature-like dependencies generate DE configurations encoded in $%
\underline{\Lambda }(\tau ).$

The prime model for $\mathring{\chi}^{2}$, constructed with a fixed
radiation-matter ratio $\ _{r}\mathring{X}=\ _{r}^{0}\mathring{\rho}/\
_{m}^{0}\mathring{\rho}$, contains five free parameters: $\alpha ,\beta ,\
_{m}^{0}\mathring{\Omega}, \ _{\Lambda }\mathring{\Omega}$ and $\mathring{H}%
_{0}$ (reduced to four for $\alpha =1$). The fittings are performed using
the relations (\ref{3modcomp}). The corresponding contour plots and
observational constraints are presented and analyzed in Fig. 1 and Table II
of \cite{od24}. The same observational data can be employed for target
off-diagonal solutions, at least for small $\kappa $-deformations (\ref%
{nonlinsymcosm}) of $\chi ^{2}(\tau ,x^{i}),$ where the free parameters
acquire gravitational polarizations and $\tau $-running. Related bounds on
the generating and integration data for off-diagonal cosmologies are
summarized in TABLE II below. %%%%%

The prime exponential f(R) cosmological models are characterized by the
Hubble parameter: 
\begin{equation}
\mathring{H}^{2}=(\mathring{H}_{0})^{2}[\ _{m}^{0}\mathring{\Omega}(%
\mathring{a}^{-3}+\ _{r}\mathring{X}\mathring{a}^{-4})+1-\ _{m}^{0}\mathring{%
\Omega}-\ _{r}^{0}\mathring{\Omega}],  \label{scen1}
\end{equation}%
which contains two free parameters $\ _{m}^{0}\mathring{\Omega}$ and $%
\mathring{H}_{0}.$ Introducing $\eta $-polarizations with nontrivial $\
^{\eta }a$ and $\ ^{\eta }H$ as in (\ref{offdcosmp}), we can express the
corresponding target off-diagonal configurations in the analogous form 
\begin{equation}
H^{2}(\tau ,x^{i}.t)=(\ ^{\eta }H\mathring{H}_{0})^{2}[\ _{m}^{0}\mathring{%
\Omega}((\ ^{\eta }a\mathring{a})^{-3}+\ _{r}\mathring{X}(\ ^{\eta }a%
\mathring{a})^{-4})+1-\ _{m}^{0}\mathring{\Omega}-\ _{r}^{0}\mathring{\Omega}%
]  \label{scen1a}
\end{equation}%
also involving the same two free parameters. The $\Lambda $CDM scenario is
recovered in the limit $\beta \rightarrow +\infty $ for the prime model (\ref%
{scen1}) independently of any $\alpha >0.$ For the off-diagonal
configurations, however -- even for small $\kappa $-deformations in (\ref%
{scen1a}) -- the corresponding limits emerges only when $\zeta _{3}(\tau
)\simeq 1$ in $\chi $-polarized solutions of type (\ref{paramsoliton}). 
%%%%%%%%

Section IV of \cite{od24} provides a rigorous analysis of the conditions
under which exponential f(R) theories can exhibit large or closed deviations
from the standard $\Lambda $CDM; the same analysis can be extended for the
prime configurations considered here. Our new conceptual and theoretical
results demonstrate that target locally anisotropic cosmological d-metrics
and the corresponding models (\ref{scen1a}) are fully determined by two
generating functions $\psi (\tau )\simeq \psi (x^{k})$ and $\underline{\eta }%
\ \simeq \underline{\eta }_{3}(x^{k},t)$ as in (\ref{lacosm2}), together
with the generating and integration data $(\psi ,\underline{\eta }_{3};\ 
\underline{\Lambda }, \ ^{v}\underline{\Upsilon },_{1}n_{k},\ _{2}n_{k}).$
These quantities can be specified explicitly to fit modern cosmological
observations while remaining within the framework of GR and the associated $%
\tau $-running cosmological systems. %%%%%%

\subsubsection{Off-diagonal parametrization of the DE and EoS}

We obtain the best-fit values for both prime and target cosmological models
-- whether based on exponential f(R) models, GR, or relativistic geometric
flows -- by requiring $_{\Lambda }{\Lambda }\Omega (\tau )\subset
0.571_{-0.057}^{+0.058}$ for the target metrics (\ref{lacosm3}) and $%
_{\Lambda }\mathring{\Omega}\subset 0.570_{-0.007}^{+0.010}$ for the prime
metrics. Deviations from the standard $\Lambda $CDM model are quantified by
comparing the absolute minimum $\ ^{abs}m=\min \chi ^{2}$ and the number of
free parameters $N_{p}.$ These quantities enter the model-selection analysis
via the Akaike Information Criterion (AIC) \cite{akaike74},%
\begin{equation}
AIC=\min \chi ^{2}+2N_{p},  \label{aic}
\end{equation}%
where $\chi ^{2}$ is defined in formulas (\ref{totchi}).

For prime diagonalizable cosmological f(R) theories and the standard $%
\Lambda $CDM model, one may speculate on realistic cosmological scenarios
depending on $N_{p}.$ Although the latter involves fewer parameters (see
Table II in \cite{od24}), the main conclusion drawn from the Akaike
information criterion, 
\begin{equation}
\bigtriangleup AIC=\ ^{model}AIC-\ ^{\Lambda CDM}AIC,  \label{daic}
\end{equation}%
is that the standard model of cosmology is not statistically favored,
whereas the exponential modified gravity model (\ref{mgrmod}) exhibits
certain advantages. This conclusion, however, is not generally valid when
off-diagonal $\eta $-deformations (\ref{scen1a}) and possible $\tau $%
-running NESs configurations are included. Only for special nonholonomic $%
\kappa$-parametric data -- when such configurations can be effectively
diagonalized -- does the AIC remain applicable. In the general cases, one
must instead employ the geometric and quantum flow information criteria
developed in \cite{bsvv24}, adapted to associative and commutative solutions
and GR as in \cite{vv25a,vv25b}. %%%%%%

\vskip4pt With respect to AIC (\ref{aic}) in modern cosmology, we discuss
three important questions:

\begin{itemize}
\item[a/] \textbf{Nature of large differences:} How do the values of $%
\bigtriangleup AIC$ (\ref{aic}) reflect deviations from the standard $%
\Lambda $CDM model, and under what conditions might these deviations favor
MGTs or classes of off-diagonal cosmological solutions in NESs? We argue
that AIC is primarily relevant for diagonalizable cosmological solutions in
gravity theories characterized by a finite number of parameters. In the
context of modern accelerating cosmology, as well as DE and DM physics,
off-diagonal solutions become significant \cite%
{vv25a,vv25b,v25,bsssvv25,bsvv24}, since for such configurations, $N_{p}$
can effectively be "absorbed" into generating and integration functions. A
more refined analysis requires the use of EoS and geometric as well as
quantum information inspired thermodynamic variables.

\item[b/] \textbf{EoS parameterizations:} In this subsection, we outline how
parameterizations of the EoS can be employed to study MGTs and off-diagonal
cosmological solutions.

\item[c/] \textbf{Geometric flow thermodynamic approach:} The subsequent
subsection will extend this analysis, providing a brief discussion of how
Perelman's thermodynamic variables can be computed for off-diagonal
cosmological configurations.
\end{itemize}

%%%%%%

Let us test two widely used parameterizations of the DE EoS: the so-called $%
w $CDM and the Chevallier-Polarski-Linder, CPL, or $w_{0}w_{a}$CDM, models 
\cite{chev00,lind02}). In addition to the prime standard matter and DM
densities (\ref{primconserv}) and (\ref{dmd}), we introduce the DE pressure
and energy densities ($\ ^{de}\mathring{p},\ ^{de}\mathring{\rho})$, which
are related through the respective EoS,%
\begin{equation}
\ ^{de}\mathring{p}=w\ ^{de}\mathring{\rho},\mbox{ where }w=const,%
\mbox{ for
wCDM};w=w_{0}+w_{1}(1-\mathring{a}),\mbox{ for CPL}.  \label{deparm}
\end{equation}%
Respective generalizations of the prime $\Lambda $CDM (\ref{scen1}) and
off-diagonal (\ref{scen1a}) models are stated in agnostic ways:%
\begin{eqnarray}
\mathring{H}^{2} &=&(\mathring{H}_{0})^{2}[\ _{m}^{0}\mathring{\Omega}(%
\mathring{a}^{-3}+\ _{r}\mathring{X}\mathring{a}^{-4})+(1-\ _{m}^{0}%
\mathring{\Omega}-\ _{r}^{0}\mathring{\Omega})\breve{f}(\mathring{a})]%
\mbox{
and }  \label{scen2} \\
H^{2}(\tau ,x^{i},t) &=&(\ ^{\eta }H\mathring{H}_{0})^{2}[\ _{m}^{0}%
\mathring{\Omega}((\ ^{\eta }a\mathring{a})^{-3}+\ _{r}\mathring{X}(\ ^{\eta
}a\mathring{a})^{-4})+(1-\ _{m}^{0}\mathring{\Omega}-\ _{r}^{0}\mathring{%
\Omega})\breve{f}(^{\eta }a\mathring{a})],  \label{scen2a}
\end{eqnarray}%
\begin{equation*}
\mbox{ where, correspondingly, }\breve{f}(\mathring{a})=\left\{ 
\begin{array}{c}
\mathring{a}^{-3(1+w)},\mbox{ for wCDM}, \\ 
\mathring{a}^{-3(1+w_{0}+w_{1})}\mathring{a}^{3w_{1}(a-1)},\mbox{ for CPL}.%
\end{array}%
\right.
\end{equation*}

%%%%%
The conditions for generating and integrating data (\ref{intgendata}) for a
class of off-diagonal cosmological solutions (\ref{lacosm3}) are determined
from the comparison with observational data for f(R) and related
cosmological models, as summarized in TABLE II (for prime configurations, a
similar table is provided in \cite{od24}). 
\begin{equation*}
\begin{array}{cc}
\mbox{TABLE II.} & 
\mbox{limits on generating/integration data and the  best
fit values for free parameters } \\ 
\begin{array}{c}
\mbox{Model} \\ 
\\ 
e^{-\beta \mathcal{\mathring{R}}^{\alpha }} \\ 
Exp\ e^{-\beta \mathcal{\mathring{R}}} \\ 
\Lambda CDM \\ 
wCDM \\ 
\\ 
\mbox{CPL} \\ 
\\ 
\begin{array}{c}
\mbox{off-diag.} \\ 
\mbox{cosm.}%
\end{array}%
\end{array}
& 
\begin{array}{cccccc}
\mathring{\chi}^{2} & \mbox{AIC} & \bigtriangleup \mbox{AIC} & \ _{m}^{0}%
\mathring{\Omega} & \mathring{H} & 
\begin{array}{c}
\mbox{other} \\ 
\mbox{parameters}%
\end{array}
\\ 
2017.80 & 2027.80 & -22.99 & 0.3086-0.3190 & 64.47-67.67 & 
\begin{array}{c}
\beta \subset 0.46-1.06 \\ 
\alpha \subset 0.839-1.189%
\end{array}
\\ 
2017.80 & 2025.80 & -24.99 & 0.3089-0.3197 & 64.49-67.59 & \beta \subset
0.675-1.06 \\ 
2046.79 & 2025.79 & 0 & 0.2903-0.2923 & 66.98-70.07 & - \\ 
2029.93 & 2035.93 & -14.86 & 0.3058-0.3158 & 66.44-70.50 & w\subset
0.908-0.942 \\ 
2015.75 & 2023.72 & -27,07 & 0.3100-0.3206 & 64.45-67.62 & 
\begin{array}{c}
w_{0}\subset 0.688-0.795 \\ 
w_{1}\subset 0.460-0.800%
\end{array}
\\ 
\begin{array}{c}
\chi ^{2}(\tau ,x^{i}) \\ 
2015.75- \\ 
2046.79%
\end{array}
& 
\begin{array}{c}
\mbox{geom.} \\ 
\mbox{inform.} \\ 
\mbox{entropy}%
\end{array}
& 
\begin{array}{c}
\mbox{inform.} \\ 
\mbox{therm.}%
\end{array}
& 
\begin{array}{c}
\ _{\Lambda }\Omega ,\mbox{see}(\ref{3modcomp}), \\ 
0.2903-0.3206%
\end{array}
& 
\begin{array}{c}
H(\tau )\subset \\ 
64.45-70.50%
\end{array}
& 
\begin{array}{c}
\mbox{gener./integr.} \\ 
\mbox{data and} \\ 
\tau \mbox{-running}%
\end{array}%
\end{array}%
\end{array}%
\end{equation*}%
These data indicate that the absolute minimum of $\chi ^{2},$ the
corresponding AIC value, and the best-fit parameters $\ _{m}^{0}\mathring{%
\Omega}$ and $\mathring{H}_{0}$ lie between the $\Lambda $CDM and f(R)
predictions. Other data favor the CPL scenario, which attains the lowest $%
\chi ^{2}$ and AIC values. Overall, the best agreement with observations is
found for the CPL model and exponential f(R) gravities, though these fits
differ from the $\Lambda $CDM predictions.

It remains unclear which MGT best describes current data or whether
alternative gravity theories are needed. The last line in TABLE II shows
that generating and integration data (\ref{intgendata}) for (\ref{lacosm3})
can be chosen so that $\chi ^{2}(\tau ,x^{i})$ (\ref{totchi}), $_{\Lambda
}\Omega (\tau ,x^{i})$ (\ref{3modcomp}), and $H^{2}(\tau ,x^{i},t)$ (\ref%
{scen2a}) define off-diagonal and geometric flow deformations, unifying
cosmological scenarios across different MGTs within GR.

Off-diagonal models offer flexibility in (\ref{intgendata}), which can be
adapted or reparameterized for future observations. They can model
inhomogeneous and locally anisotropic DM distributions, structure formation
(e.g., quasi-periodic patterns, filaments), $\tau $-running of constants,
anisotropic polarizations, horizon deformations, and new nonlinear
symmetries. %%%%%%%
%%%%%%%%

The prime model (\ref{scen2}) compared to the standard $\Lambda $CDM model
contains an additional parameter $w$ for $w$CDM, but there are two extra
parameters $w_{0}$ and $w_{1}$ for CPL. The cosmological properties are
summarized in Tables II and III and Fig. 3 of \cite{od24}. For instance, it
was concluded that the value for AIC (\ref{aic}) and best fits for $\
_{m}^{0}\mathring{\Omega}$ and $\mathring{H}_{0}$ are between the $\Lambda $%
CDM and f(R) results. And the data strongly favours the CPL scenarios, when
the $\min \chi ^{2}$ and the AIC parameter achieve the smallest values. We
do not re-formulate incrementally those results using off-diagonal
deformations because the approach must be completely revised by using
geometric and quantum information flows of cosmological solutions as in \cite%
{bub23}.

The prime model (\ref{scen2}) introduces one additional parameter $w$
relative to the standard $\Lambda $CDM scenario for $w$CDM, while the CPL
parameterization involves two extra parameters, and $w_{0}$ and $w_{1}$. The
corresponding cosmological properties are summarized in Tables II and III
and Fig. 3 of \cite{od24}. For example, it was found that the AIC values (%
\ref{aic}) and the best-fit estimates for $\ _{m}^{0}\mathring{\Omega}$ and $%
\mathring{H}_{0}$ lie between those obtained for $\Lambda $CDM and for f(R)
models. Moreover, the data strongly favor the CPL scenario, for which both
the minimum $\min \chi ^{2}$ and the AIC reach their lowest values. In the
present work, we do not reproduce those results via incremental off-diagonal
deformations, because such an analysis must be reformulated from the outset
using geometric and quantum information flows of cosmological solutions,
following the framework developed in \cite{bub23}. %%%%%

The data in TABLE III (see below) on the large deviations $\bigtriangleup %
\mbox{AIC}$ favor the standard exponential gravity model and the CPL
parametrization. To clarify their roles, we also present and compare an
additional set of observational results in TABLE IV, as shown below. In
agreement with Table III and Fig. 3 of \cite{od24}, we conclude that the
generalized exponential model is disfavored, as its AIC value is
significantly larger than that of the standard exponential gravity model. 
\begin{equation*}
\begin{array}{cc}
\mbox{TABLE III.} & \mbox{Data: SN Ia + CC + CMP + 6 DESI BAO} \\ 
\begin{array}{c}
\mbox{Model} \\ 
- \\ 
Exp\ e^{-\beta \mathcal{\mathring{R}}} \\ 
\Lambda CDM \\ 
wCDM \\ 
\mbox{CPL} \\ 
\\ 
\begin{array}{c}
\mbox{off-diag.} \\ 
\mbox{cosm.}%
\end{array}%
\end{array}
& 
\begin{array}{ccccc}
\begin{array}{c}
\mbox{min }\mathring{\chi}^{2} \\ 
-%
\end{array}
& 
\begin{array}{c}
\mbox{AIC} \\ 
-%
\end{array}
& 
\begin{array}{c}
\bigtriangleup \mbox{AIC} \\ 
-%
\end{array}
& 
\begin{array}{c}
\ _{m}^{0}\mathring{\Omega} \\ 
-%
\end{array}
& 
\begin{array}{c}
\mathring{H} \\ 
-%
\end{array}
\\ 
2000.32 & 2008.38 & -27.83 & 0.3100-0.3219 & 64.42-67.63 \\ 
2032.15 & 2036.15 & 0 & 0.2911-0.2925 & 67.01-70.22 \\ 
2005.44 & 2011.44 & -24.71 & 0.3186-0.3330 & 64.61-67.83 \\ 
1998.82 & 2006.82 & -29.33 & 0.3078-0.3235 & 64.38-68.60 \\ 
\begin{array}{c}
\chi ^{2}(\tau ,x^{i}) \\ 
1998.92- \\ 
2032.15%
\end{array}
& 
\begin{array}{c}
\mbox{geom.} \\ 
\mbox{inform.} \\ 
\mbox{entropy}%
\end{array}
& 
\begin{array}{c}
\mbox{inform.} \\ 
\mbox{therm.}%
\end{array}
& 
\begin{array}{c}
\ _{\Lambda }\Omega ,\mbox{see}(\ref{3modcomp}), \\ 
0.2911-0.3330%
\end{array}
& 
\begin{array}{c}
H(\tau )\subset \\ 
64.38-70.22%
\end{array}%
\end{array}%
\end{array}%
\end{equation*}

Then, using the DESI BAO data set presented in TABLE IV, we find that a
slight shift in the estimated model parameters is possible; however, this
does not alter the large values of $\bigtriangleup\mathrm{AIC}$ that
distinguish the $\Lambda$CDM model from the exponential, $w$CDM, and CPL
scenarios. 
\begin{equation*}
\begin{array}{cc}
\mbox{TABLE IV} & \mbox{Data without BAO:\ SN Ia + CC + CMP} \\ 
\begin{array}{c}
\mbox{Model} \\ 
- \\ 
Exp\ e^{-\beta \mathcal{\mathring{R}}} \\ 
\Lambda CDM \\ 
wCDM \\ 
\mbox{CPL} \\ 
\\ 
\begin{array}{c}
\mbox{off-diag.} \\ 
\mbox{cosm.}%
\end{array}%
\end{array}
& 
\begin{array}{ccccc}
\begin{array}{c}
\mbox{min }\mathring{\chi}^{2} \\ 
-%
\end{array}
& 
\begin{array}{c}
\mbox{AIC} \\ 
-%
\end{array}
& 
\begin{array}{c}
\bigtriangleup \mbox{AIC} \\ 
-%
\end{array}
& 
\begin{array}{c}
\ _{m}^{0}\mathring{\Omega} \\ 
-%
\end{array}
& 
\begin{array}{c}
\mathring{H} \\ 
-%
\end{array}
\\ 
1997.98 & 2005.98 & -28.24 & 0.3094-0.3217 & 64.58-67.84 \\ 
2030.22 & 2034.22 & 0 & 0.2900-0.2924 & 67.12-70.36 \\ 
2001.59 & 2007.59 & -26.63 & 0.3221-0.3383 & 64.61-67.83 \\ 
1995.68 & 2003.68 & -30.54 & 0.2968-0.317 & 64.14-68.14 \\ 
\begin{array}{c}
\chi ^{2}(\tau ,x^{i}) \\ 
1995.68- \\ 
2030.22%
\end{array}
& 
\begin{array}{c}
\mbox{geom.} \\ 
\mbox{inform.} \\ 
\mbox{entropy}%
\end{array}
& 
\begin{array}{c}
\mbox{inform.} \\ 
\mbox{therm.}%
\end{array}
& 
\begin{array}{c}
\ _{\Lambda }\Omega ,\mbox{see}(\ref{3modcomp}), \\ 
0.2900-0.3383%
\end{array}
& 
\begin{array}{c}
H(\tau )\subset \\ 
64.14-70.36%
\end{array}%
\end{array}%
\end{array}%
\end{equation*}%
The last line of TABLE IV emphasizes that the recent observational data can
be explained by an appropriate choice of generating and integration data (%
\ref{intgendata}) for generic off-diagonal cosmological solutions in GR,
together with an analysis of possible $\tau$-running effects and nonlinear
polarizations of the cosmological constants. %%%%%

In TABLES I-IV of this section, we present explicit computed intervals for
observational quantities such as $\mathring{H}\subset 64.58-67.84$ and $%
H(\tau ,x^{i})\subset 64.14-70.36$ , which differ from the values reported
in other works -- for example, in Tables I-III of \cite{od24}, where one
finds $H_{0}=66.43_{-1.63}^{+1.71}$. For generic off-diagonal solutions, our
notation is more suitable for describing possible variations of generating
and integration data that remain compatible with the observational sets
under consideration. This flexibility allows us to remain within the
standard GR cosmological paradigm (at least for many cosmological
configurations and related geometric evolution scenarios), without invoking
MGTs for updated or alternative observational data.

Within our off-diagonal and conservative approach, it becomes necessary to
re-define certain boundary and initial conditions, introduce appropriate
generating functions and effective sources, with certain parametric
decompositions, and clarify the nonlinear symmetries of the corresponding
systems of nonlinear PDEs. This ensures a consistent formulation of
realistic models of cosmological dynamics when working with generic
off-diagonal configurations. %%%%%%

The above TABLES summarize the fit results of certain cosmological models in
the framework of f(R) theories and update, for instance, the results,
discussion and tables from \cite{npsa16}. For nonholonomic Einstein
cosmological systems studied in this work (and \cite{vv25a,vv25b,v25}), the
nonholonomic torsion can be nontrivial. The optimization and observational
constraints of cosmological models with nontrivial torsion were studied in
the case of f(T) gravity (as a teleparalled equivalent of GR) in recent
works \cite{pgg25,pgmz25}. In this paper, we do not extend that analysis
because our main goals is to prove that off-diagonal and geometric flow
deformations can be performed for cosmological models in GR, if we extract
LC configurations. Nevertheless, we emphasize that the AFCDM allows us to
elaborate on off-diagonal cosmological models in the framework of various
types of MGTs, for instance, in metric-affine gravity theories (in
particular, in f(Q) gravity and various nonassociative and noncommutative
string generalizations etc. \cite{gheorghiuap16,bsssvv25,bsvv24,v25}). 
%%%%%%%

We also note that local anisotropies of off-diagonal cosmological solutions (%
\ref{lacosm1}), (\ref{lacosm2}), or (\ref{lacosm3}) depending on local
coordinates $(x^{i},t)$ describe cosmological configurations which are
different from the local anisotropies in (generalized) Finsler-like theories
which are with generic dependencies on velocity/momentum like coordinates,
see details in \cite{bsssvv25}. In principle, we can state certain
generating data on our locally anisotropic cosmological models to reproduce
anistotropic/inhomogeneous cosmological theories with high symmetries,
involving algebraic constraints of Bianchi type, see examples in \cite%
{hawking73,misner73,wald82,kramer03}. Such theories are not compatible with
observational data in modern cosmology but can be nonholonomically deformed
into off-diagonal cosmological configurations, which encode in target
cosmological d-metrics a "memory" on prescribed primary anisotropic metrics
(certain results are reivewed in \cite{bsssvv25,bsvv24}).

Finally, in this subsection, we conclude that it is possible to remain
within the framework of GR by employing the effective model (\ref{scen2a})
together with the off-diagonal cosmological solutions (\ref{lacosm3}). The
generating data $^{\eta }a(\tau ,x^{i},t)$ and the corresponding $\ ^{\eta
}H(\tau ,x^{i},t)$, defined by appropriate integration data, effectively
absorb all prime parameters and can account for current observational
constraints as models of locally anisotropic cosmological evolution. Such
off-diagonal configurations, however, cannot be characterized
thermodynamically or informationally within the standard Bekenstein\^{a}%
\euro ``Hawking paradigm \cite{bek2,haw2}. Instead, they require a more
advanced geometric framework based on relativistic generalizations of the G.
Perelman W-entropy \cite{perelman1,vv25a,vv25b,bsvv24}. %%%%%%%

\subsection{Generalized G. Perelman thermodynamics for off-diagonal DE
configurations}

The theory of geometric flows of NESs and MGTs has been developed in detail
in \cite{gheorghiuap16,vv25a,vv25b,v25,bsssvv25,bsvv24}. The main
applications to GR were formulated for quasi-stationary off-diagonal
solutions, where cosmological configurations were generated by employing
certain geometric abstract duality transforms. In this subsection, we show
how those results can be reconsidered in order to formulate a generalized G.
Perelman thermodynamics for off-diagonal cosmological solutions (\ref%
{lacosm2}). For this purpose, we introduce an additional (3+1)-splitting: a
nonholonomic 2+2 decomposition remains essential for generating off-diagonal
solutions, while the equivalent representations (\ref{lacosm3}) play a
crucial role for analyzing the compatibility of such models with
experimental data. %%%%%%

The generalized relativistic R. Hamilton and D. Friedan geometric flow
equations \cite{hamilton82,friedan80,perelman1} can be derived in an
N-adapted variational form by employing the nonholonomic frame formalism and
the associated distortion relations for linear connections, as developed in 
\cite{vv25a,vv25b}. We note that the topological and geometric aspects of
geometric flows of Riemannian metrics are exhaustively reviewed in the
mathematical monographs \cite{kleiner06,morgan06,cao06}. In our approach, we
have constructed certain generalizations with applications to GR and MGTs by
using the AFCDM \cite{sv11,bsssvv25,vv25a,vv25b}, which provides a unified
scheme for the decoupling and integrability of physically relevant classes
of nonlinear PDEs. The thermodynamic properties of the resulting
off-diagonal geometric flow and gravitational configurations can also be
analyzed in terms of (modified) G. Perelman -- type thermodynamic variables.

%%%%%%%

For any $\tau $-family of d-metrics $\mathbf{g}(\tau )=\mathbf{g}(\tau,
x^{i},y^{a})$, we can introduce a statistical partition function of the form 
\begin{equation}
\widehat{Z}(\tau )=\exp [\int_{\widehat{\Xi }}[-\widehat{\zeta }+2]\ \left(
4\pi \tau \right) ^{-2}e^{-\widehat{\zeta }}\ \delta \widehat{V}(\tau )],
\label{spf}
\end{equation}%
where the volume element is defined and computed as 
\begin{equation}
\delta \widehat{V}(\tau )=\sqrt{|\mathbf{g}(\tau )|}\ dx^{1}dx^{2}\delta
y^{3}\delta y^{4}\ .  \label{volume}
\end{equation}%
In our approach, we used the canonical nonholonomic data $(\mathbf{g}(\tau )=%
\widehat{\mathbf{g}}(\tau ),\widehat{\mathbf{D}}(\tau ))$ (\ref{twocon})
instead of LC-data $(\mathbf{g}(\tau ),\mathbf{\nabla }(\tau ))$. This
choice allows us to define $\widehat{Z}$ (\ref{spf}) and the corresponding
functional $\widehat{\mathcal{W}}(\tau )$ (\ref{wf1}) for $\tau $-families
of exact or parametric solutions of (\ref{cdeq1}) or (\ref{cdeq1b}). %%%%%%

By applying the relativistic 4-dimensional canonical distortion (\ref%
{canondist}) to the geometric constructions presented in Section 5 of \cite%
{perelman1}, we can define and compute the corresponding (statistical)
thermodynamic variables: 
\begin{align}
\ \widehat{\mathcal{E}}\ (\tau )& =-\tau ^{2}\int_{\widehat{\Xi }}\ \left(
4\pi \tau \right) ^{-2}\left( \widehat{f}(\widehat{\mathbf{R}}sc)+|\ 
\widehat{\mathbf{D}}\ \widehat{\zeta }|^{2}-\frac{2}{\tau }\right) e^{-%
\widehat{\zeta }}\ \delta \widehat{V}(\tau ),  \label{qthermvar} \\
\ \ \widehat{S}(\tau )& =-\widehat{\mathcal{W}}(\tau )=-\int_{\widehat{\Xi }%
}\left( 4\pi \tau \right) ^{-2}\left( \tau ((\widehat{\mathbf{R}}sc)+|%
\widehat{\mathbf{D}}\widehat{\zeta }|^{2})+\widehat{\zeta }-4\right) e^{-%
\widehat{\zeta }}\ \delta \widehat{V}(\tau ),  \notag \\
\ \ \widehat{\sigma }(\tau )& =2\ \tau ^{4}\int_{\widehat{\Xi }}\left( 4\pi
\tau \right) ^{-2}|\ \widehat{\mathbf{R}}_{\alpha \beta }+\widehat{\mathbf{D}%
}_{\alpha }\ \widehat{\mathbf{D}}_{\beta }\widehat{\zeta }-\frac{1}{2\tau }%
\mathbf{g}_{\alpha \beta }|^{2}e^{-\widehat{\zeta }}\ \delta \widehat{V}%
(\tau ).  \notag
\end{align}%
Such (in general, nonassociative and noncommutative) metric and nonmetric
geometric thermodynamic variables were introduced in \cite%
{gheorghiuap16,vv25a,vv25b,v25,bsssvv25,bsvv24} for various classes of
modified gravity theories (MGTs) and off-diagonal solutions in general
relativity (GR). The fluctuation variable $\widehat{\sigma}(\tau)$ can be
expressed as a functional of $\widehat{\mathbf{R}}_{\alpha \beta}$, while
the quantities $\widehat{\mathcal{E}}(\tau)$ and $\widehat{S}(\tau)$ are
functionals of $\widehat{f}(\widehat{\mathbf{R}}sc)$ if the normalizing
functions are appropriately redefined, $\widehat{\zeta} \rightarrow \widehat{%
\zeta}{[1]}$. For simplicity, we omit these technical details here, as we do
not compute $\widehat{\sigma}(\tau)$ for the specific classes of
off-diagonal cosmological solutions considered in this work. %%%%%%

Fixing the temperature in (\ref{qthermvar}) at $\tau =\tau _{0}$, we can
compute the thermodynamic variables $[\widehat{\mathcal{E}}(\tau _{0}),%
\widehat{\mathcal{S}}(\tau _{0}),\widehat{\sigma }(\tau _{0})]$ for
relativistic Ricci solitons with respective nonholonomic distributions,
Killing symmetries along $\partial _{3}$, and nonlinear symmetries. Certain
classes of (off-diagonal) solutions may not be well-defined as physical
thermodynamic systems, for instance, when $\widehat{\mathcal{S}}(\tau
_{0})<0 $. To ensure physically viable solutions, we must restrict some
nonholonomic distributions and the distortions of linear connections. In
specific spacetime regions, off-diagonal deformations can lead to unphysical
models. Nevertheless, these new classes of solutions may be physically
relevant under other nonholonomic conditions. A detailed investigation is
therefore necessary for explicit classes of exact or parametric solutions of
physically significant nonlinear PDE systems in GR and modified gravity
theories. %%%%%%%%.

Many physical and observational properties of general $\tau$-families of
off-diagonal cosmological solutions -- encoding nontrivial topological and
quasi-periodic structures for dark energy (DE) and dark matter (DM)
configurations \cite{v16plb,bsssvv25,bsvv24} -- cannot be adequately studied
within the $\Lambda$CDM paradigm. Indeed, recent experimental data \cite%
{desi24,pantheon21,roy24,batic24,dival21} challenge the standard
cosmological model. To explain observations in accelerating cosmology and
the physics of DE and DM, a variety of alternative cosmological models based
on modified gravity theories (MGTs) have been developed \cite%
{hu07,appl07,nojiri08,cog07,yang10,od20,od24,v25}. In this work, we advocate
that prime diagonal cosmological solutions in exponential $f(R)$ theories
can be off-diagonally deformed into certain classes of exact or parametric
solutions in general relativity (GR). Such models can always be
characterized by G. Perelman thermodynamic variables, which can be computed
explicitly for all classes of solutions in geometric flow approaches, GR,
and various MGTs. The computation and analysis of these variables are
significantly simplified by employing nonlinear symmetries that transform
generating functions and sources into equivalent forms involving effective $%
\tau$-running cosmological constants $\underline{\Lambda}(\tau)$. %%%%%%%

Any $\tau $-family of off-diagonal cosmological solutions $\underline{%
\mathbf{g}}_{\alpha }[\underline{\Phi }(\tau )]\simeq \underline{\ \mathbf{g}%
}_{\alpha }[\underline{\eta }_{3}(\tau )]$ (\ref{lacosm2}) is determined by
generating sources 
\begin{equation*}
\underline{\Upsilon }_{\alpha }(\tau )=\ ^{m}\underline{\Upsilon }%
_{\alpha}(\tau )+\ ^{DEM}\underline{\Upsilon }_{\alpha }(\tau )= [\ _{h}^{m}%
\underline{\Upsilon }(\tau )+\ _{h}^{DEM}\underline{\Upsilon }(\tau ),\
_{v}^{m}\underline{\Upsilon }(\tau )+ \ _{v}^{DEM}\underline{\Upsilon }%
(\tau)],
\end{equation*}
where $\ ^{m}\underline{\Upsilon }{\alpha }$ corresponds to a real matter
source (\ref{emdt}) and $\ ^{DEM}\underline{\Upsilon }{\alpha }(\tau )$ is
associated with distortion tensors and other effective sources of geometric
or DE/DM origin. Under geometric flows, the $\tau$-families of generating
sources exhibit behavior analogous to (\ref{effrfs}) with a respective
decomposition $\widehat{\underline{\mathbf{J}}}{\alpha }(\tau)=\ ^{m}%
\widehat{\underline{\mathbf{J}}}{\alpha }(\tau)+\ ^{DEM}\widehat{\underline{%
\mathbf{J}}}_{\alpha }(\tau)$. For simplicity, we assume the same behavior
for horizontal and vertical $\tau$-dependent cosmological constants, $\ ^{h}%
\underline{\Lambda }(\tau)=\ ^{v}\underline{\Lambda}(\tau)=\underline{\Lambda%
}(\tau)$. Nonlinear symmetries of off-diagonal solutions (\ref{nonlintrsmalp}%
) and (\ref{nonlinsymcosm}) imply a possible decomposition of the effective
cosmological constants: 
\begin{equation}
\ \widehat{\underline{\mathbf{J}}}_{\alpha }(\tau )=\ ^{m}\ \widehat{%
\underline{\mathbf{J}}}_{\alpha }(\tau )+\ ^{DEM}\ \widehat{\underline{%
\mathbf{J}}}_{\alpha }(\tau )\rightarrow \underline{\Lambda }(\tau )=\ ^{m}%
\underline{\Lambda }(\tau )+\ ^{DE}\underline{\Lambda }(\tau ).
\label{nonlincosmsym}
\end{equation}%
Generic off-diagonal interactions of gravitational and matter fields, as
well as their $\tau$-evolution, mix nonlinearly, with possible contributions
from both metric and source terms, thereby generally polarizing the
geometric constants. One can always consider $\tau$-families of canonical
nonholonomic Einstein equations (\ref{cdeq1b}) with effective running
cosmological constants $\ ^{m}\underline{\Lambda}(\tau)+\ ^{DE}\underline{%
\Lambda}(\tau)$. Accordingly, the canonical Ricci scalar is given by $%
\widehat{\mathbf{R}}sc = 4,[\ ^{m}\underline{\Lambda}(\tau) + \ ^{DE}%
\underline{\Lambda}(\tau)]$.

%%%%%%

Our main goal is to compute explicitly the thermodynamic variables $\widehat{%
\underline{Z}}$ (\ref{spf}) and $\widehat{\underline{\mathcal{E}}}\ (\tau),\ 
\widehat{\underline{S}}(\tau )$ from (\ref{qthermvar}) for off-diagonal
cosmological solutions of cosmological solutions $\underline{\mathbf{g}}%
_{\alpha }[\underline{\Phi }(\tau )]\simeq \underline{\ \mathbf{g}}_{\alpha}[%
\underline{\eta }_{3}(\tau )]$ (\ref{lacosm2}), or equivalently $a(\tau ))=\
^{\eta}a(\tau ,x^{i},t)\mathring{a}(t)$ in (\ref{lacosm3}) and (\ref{scen1a}%
). To simplify the computations, we can chose a nonholonomic frame (and
coordinates) such that the normalizing functions satisfy the conditions $%
\widehat{\mathbf{D}}_{\alpha }\ \widehat{\zeta }=0$ and $\widehat{\zeta }%
\approx 0$. If necessary, the constructions can be redefined for arbitrary
frames and normalizing functions). With this choice, the thermodynamic
quantities take the form:%
\begin{align}
\ \widehat{\underline{Z}}(\tau )& =\exp [\int_{\widehat{\Xi }}\frac{1}{%
8\left( \pi \tau \right) ^{2}}\ \delta \underline{\mathcal{V}}(\tau )],\ 
\widehat{\underline{\mathcal{E}}}\ (\tau )=-\tau ^{2}\int_{\widehat{\Xi }}\ 
\frac{1}{\left( 2\pi \tau \right) ^{2}}[\ ^{m}\underline{\Lambda }(\tau )+\
^{DE}\underline{\Lambda }(\tau )-\frac{1}{2\tau }]\delta \underline{\mathcal{%
V}}(\tau ),  \label{thermvar1} \\
\ \ \widehat{\underline{S}}(\tau )& =-\ \widehat{\underline{W}}(\tau
)=-\int_{\widehat{\Xi }}\frac{1}{\left( 2\pi \tau \right) ^{2}}[\tau (\ ^{m}%
\underline{\Lambda }(\tau )+\ ^{DE}\underline{\Lambda }(\tau ))-1]\delta 
\underline{\mathcal{V}}(\tau ),  \notag
\end{align}%
where the details on prime and target cosmological configuration are encoded
in $\delta \ ^{q}\mathcal{V}(\tau )$, which is determined by the determinant
of corresponding off-diagonal metric solutions. %%%%%%

The volume form $\delta \underline{\mathcal{V}}(\tau )$ (\ref{volume}) for (%
\ref{thermvar1}) can be explicitly computed for cosmological d-metrics (\ref%
{lacosm2}) characterized by $\eta $--polarization functions, or for (\ref%
{paramsoliton}) under $\kappa $-parametric decompositions with $\chi $%
--polarization functions. The corresponding generating sources are encoded
indirectly in $\underline{\mathbf{g}}_{\alpha }[\underline{\Phi }(\tau )]$.
By employing (\ref{nonlinsymcosm}), we then obtain 
\begin{align}
\ \underline{\Phi }(\tau )& =2\sqrt{|[\ ^{m}\underline{\Lambda }(\tau )+\
^{DE}\underline{\Lambda }(\tau )]\ \underline{g}_{3}(\tau )|}=\ 2\sqrt{|[\
^{m}\underline{\Lambda }(\tau )+\ ^{DE}\underline{\Lambda }(\tau )]\ 
\underline{\eta }_{3}(\tau )\ \underline{\mathring{g}}_{3}(\tau )|}  \notag
\\
& \simeq 2\sqrt{|[\ ^{m}\underline{\Lambda }(\tau )+\ ^{DE}\underline{%
\Lambda }(\tau )]\ \underline{\zeta }_{3}(\tau )\ \ \underline{\mathring{g}}%
_{3}|}[1-\frac{\varepsilon }{2}\ \underline{\chi }_{3}(\tau )].
\label{genf1}
\end{align}%
For simplicity, we can study nonholonomic evolution models with trivial
integration functions $\ _{1}n_{k}=0$ and $\ _{2}n_{k}=0$ in (\ref{lacosm2})
and (\ref{lacosm3}). By introducing the approximations and expressions (\ref%
{genf1}) in (\ref{volume}), we compute: 
\begin{align*}
\ \delta \underline{\mathcal{V}}(\tau )& =\delta \mathcal{V}[\tau ,\ ^{m}%
\underline{\Lambda }(\tau )+\ ^{DE}\underline{\Lambda }(\tau );\ ^{m}\ 
\widehat{\underline{\mathbf{J}}}_{\alpha }(\tau ),\ ^{DEM}\ \widehat{%
\underline{\mathbf{J}}}_{\alpha }(\tau )];\psi (\tau ),\underline{h}%
_{3}(\tau )=\underline{\eta }_{3}(\tau )\underline{\mathring{g}}_{3}] \\
& =\frac{1}{\ ^{m}\underline{\Lambda }(\tau )+\ ^{DE}\underline{\Lambda }%
(\tau )}\ \delta \ _{\eta }\underline{\mathcal{V}},\mbox{ where }\ \delta \
_{\eta }\underline{\mathcal{V}}=\ \delta \ _{\eta }^{1}\underline{\mathcal{V}%
}\times \delta \ _{\eta }^{2}\underline{\mathcal{V}}.
\end{align*}%
Such volume forms, which encode off-diagonal cosmological prime and target
configurations under nonholonomic geometric evolutions, can be parameterized
as products of two functionals: 
\begin{align}
&\delta \ _{\eta }^{1}\underline{\mathcal{V}} =\delta \ _{\eta }^{1}\mathcal{%
V}[\ \ ^{m}\underline{\Lambda }(\tau )+\ ^{DE}\underline{\Lambda }(\tau
),\eta _{1}(\tau )\ \mathring{g}_{1}]=e^{\widetilde{\psi }(\tau
)}dx^{1}dx^{2}=\sqrt{|\ \ ^{m}\underline{\Lambda }(\tau )+\ ^{DE}\underline{%
\Lambda }(\tau )|}e^{\psi (\tau )}dx^{1}dx^{2},  \label{volumfuncts} \\
& \mbox{ for }\psi (\tau )\mbox{ being a solution of  }\partial
_{11}^{2}\psi +\partial _{22}^{2}\psi =2(\ \ _{h}^{m}\widehat{\underline{%
\mathbf{J}}}(\tau )+\ \ _{h}^{DEM}\widehat{\underline{\mathbf{J}}}(\tau )); 
\notag \\
& \delta \ _{\eta }^{2}\underline{\mathcal{V}} =\delta \ _{\eta }^{2}%
\mathcal{V}[\ \ _{v}^{m}\widehat{\underline{\mathbf{J}}}_{a}(\tau ),\ \ \
_{v}^{DEM}\ \widehat{\underline{\mathbf{J}}}_{a}(\tau ),\ \underline{\eta }%
_{3}(\tau )\ \underline{\mathring{g}}_{3}]=  \notag \\
& \frac{\partial _{t}|\underline{\eta }_{3}(\tau )\ \underline{\mathring{g}}%
_{3}|^{3/2}dy^{3}}{\ \sqrt{|\int dt\ [\ _{v}^{m}\widehat{\underline{\mathbf{J%
}}}_{a}(\tau )+\ _{v}^{DEM}\ \widehat{\underline{\mathbf{J}}}_{a}(\tau
)]\{\partial _{t}|\ \ \underline{\eta }_{3}(\tau )\ \underline{\mathring{g}}%
_{3}|\}^{2}|}} [dt+\frac{\partial _{i}\left( \int dt\ [\ \ _{v}^{m}\widehat{%
\underline{\mathbf{J}}}_{a}(\tau )+\ _{v}^{DEM}\ \widehat{\underline{\mathbf{%
J}}}_{a}(\tau )]\partial _{t}|\ \ \underline{\eta }_{3}(\tau )\ \underline{%
\mathring{g}}_{3}|\right) dx^{i}}{\ [\ _{v}^{m}\widehat{\underline{\mathbf{J}%
}}_{a}(\tau )+\ _{v}^{DEM}\ \widehat{\underline{\mathbf{J}}}_{a}(\tau
)]\partial _{t}|\ \ \underline{\eta }_{3}(\tau )\ \underline{\mathring{g}}%
_{3}|}].  \notag
\end{align}%
In these formulas, we distinguish effective sources and cosmological
constants with the labels $m$ and $DE$ since such functionals may or not
induce various quasi-periodic, filamentary, or other types of cosmological
structures. The functions $\psi (\tau )$ can be defined as a $\tau $--family
of solutions $\widetilde{\psi }(\tau )$ of 2-d Poisson equations with
effective sources $\ ^{m}\underline{\Lambda }(\tau )+\ ^{DE}\underline{%
\Lambda }(\tau ).$ By integrating on a closed hypersurface $\widehat{\Xi }$
the products of $h$- and $v$-forms from (\ref{volumfuncts}), we obtain a $%
\tau$-running cosmological phase space volume functional 
\begin{equation}
\ _{\eta }\underline{\mathcal{\mathring{V}}}(\tau )=\int_{\ \widehat{\Xi }%
}\delta \ _{\eta }\underline{\mathcal{V}}(\ ^{m}\widehat{\underline{\mathbf{J%
}}}_{a}(\tau ),\ ^{DEM}\ \widehat{\underline{\mathbf{J}}}_{a}(\tau ),%
\underline{\ \mathring{g}}_{\alpha }).  \label{volumfpsp}
\end{equation}%
The explicit formulas for the volume forms $\mathcal{\mathring{V}}(\tau)$
depend on the choice of prime and target data prescribed for $\widehat{\Xi}$%
, as well as on the type of off-diagonal deformations, which are encoded
through $\eta$- or $\zeta$-polarizations, as discussed in section \ref{sec3}%
. We assume that it is always possible to compute, in a suitable parametric
form, $_{\eta}\mathcal{\mathring{V}}(\tau)$ for the corresponding generating
and integration data. In a general cosmological context, we can consider
that the thermodynamic variables depend explicitly on the $\tau$-dependent
effective cosmological constants, with different MGTs and classes of
solutions distinguished by their respective dependencies. Certain values can
be computed in explicit form, providing a framework to interpret
observational cosmological data and to describe the nonholonomic geometric
evolution of off-diagonal DE and DM configurations.

%%%%%%%%

Introducing functional (\ref{volumfpsp}) into the formulas (\ref{thermvar1}%
), we compute 
\begin{align}
\ \ \underline{\widehat{Z}}(\tau )& =\exp [\frac{\ _{\eta }\underline{%
\mathcal{\mathring{V}}}(\tau )}{8\left( \pi \tau \right) ^{2}}]\ ,\underline{%
\widehat{\mathcal{E}}}\ (\tau )=[\frac{1}{\tau }-2(\ ^{m}\underline{\Lambda }%
(\tau )+\ ^{DE}\underline{\Lambda }(\tau ))]\frac{\ _{\eta }\underline{%
\mathcal{\mathring{V}}}(\tau )}{8\pi ^{2}}\ ,  \label{thermvar2} \\
\ \ \underline{\widehat{S}}(\tau )& =-\ ^{q}\widehat{W}(\tau )=[1-\tau (\
^{m}\underline{\Lambda }(\tau )+\ ^{DE}\underline{\Lambda }(\tau ))]\frac{%
_{\eta }\underline{\mathcal{\mathring{V}}}(\tau )}{4\left( \pi \tau \right)
^{2}}.  \notag
\end{align}

We can define the effective volume functionals (\ref{volumfuncts}) and
geometric thermodynamic variables (\ref{thermvar2}) for $\kappa $-parametric
decompositions using approximations (\ref{genf1}) and small polarizations of
cosmological constants and find parametric formulas for $\tau $-flows and
off-diagonal deformations of prime metrics as for d-metrics (\ref%
{paramsoliton}). Fixing a $\tau _{0}$, the above formulas can be used for
explicit computing of G. Perelman's thermodynamic variables for large
classes of Ricci cosmological solitons, and respective off-diagonal
solutions in GR.

\section{Conclusions}

\label{sec4} During the last 25 years, researchers in modern gravity and
accelerating cosmology have attempted almost every year to address key
challenges in modifying General Relativity (GR), refining the standard $%
\Lambda$CDM paradigm, and improving the fit with experimental data. Numerous
modified gravity theories (MGTs) have been developed, including those with
non-minimal couplings, nontrivial torsion and nonmetricity fields, as well
as quantum and string-inspired corrections. Among these, exponential $f(R)$
gravities have attracted considerable attention in recent years \cite%
{od24,od20,od24a}, since such models can be directly confronted with the
latest observational datasets. This includes the Pantheon+ SN Ia
compilation, the new BAO measurements from DESI DR1, cosmic chronometer (CC)
determinations of the Hubble parameter, and the most recent CMB data \cite%
{desi24,roy24,batic24,dival21}.

\vskip4pt In this paper, we elaborated on a conservative approach based on
the main Hypothesis (formulated in the Introduction) that accelerating
cosmological models and various dark energy (DE) and dark matter (DM)
effects in modified gravity theories (MGTs) can be modelled by generic
off-diagonal solutions in GR. To this end, we applied the AFCDM formalism to
construct exact and parametric cosmological solutions. This framework
enables the decoupling and integration, in general form, of physically
relevant systems of nonlinear PDEs. The coefficients of such off-diagonal
metrics, (non)linear connections, nonholonomic frames, and the corresponding
connection distortions are determined by generating functions and effective
sources, which may, in principle, depend on all spacetime coordinates. The
geometric technique of generating off-diagonal solutions is applicable both
in GR and in various MGTs. Furthermore, by exploiting the underlying
nonlinear symmetries, we establish explicit criteria under which certain
classes of solutions defining one cosmological model can be equivalently
reformulated to reproduce other models, including those with inflationary
and late-time acceleration dynamics. %%%%%

\vskip4pt The majority of researchers working on gravity and cosmology
continue to focus on modifying GR and the $\Lambda$CDM model, often by
considering diagonalizable cosmological configurations with additional
parameters, new physical constants, and alternative forms of the Lagrange
densities for gravitational and matter fields. Within such approximations
and modifications of gravity theories, the underlying nonlinear PDE systems
of physical relevance are usually reduced to certain nonlinear ODE systems.
Almost every year, new classes of Lagrange densities, parameter
configurations, and even entire cosmological paradigms are introduced in
order to fit the latest experimental data. In our work, we have also
elaborated on various MGTs, since the AFCDM framework can be applied both to
GR and its modifications. Using this approach, we have constructed various
classes of exact and parametric solutions, including quasi-stationary black
hole, wormhole, and toroidal configurations, as well as cosmological models
with anisotropic polarizations of fundamental constants \cite{sv11,bsssvv25}%
. Nevertheless, our analysis shows that by making suitable choices of
nonholonomic frame structures, together with appropriate classes of
generating data and integration functions, it is possible to encode modern
observational cosmological data in such a way that certain MGTs and even the 
$\Lambda$CDM paradigm can be modeled as specific prime metrics. The
corresponding target off-diagonal solutions can then be constructed as exact
or parametric solutions in GR formulated with nonholonomic dyadic variables. 
%%%%%%%%

\vskip4pt The long-standing rivalry among GR, MGTs, and various accelerating
cosmology and DE/DM theories may be resolved in favor of GR if generic
off-diagonal cosmological solutions are considered within a corresponding
nonholonomic geometric framework. This approach focuses on four-dimensional
modifications of GR, excluding higher-dimensional string/brane models and
generalized Finsler-like MGTs. The new classes of off-diagonal solutions
cannot be described thermodynamically in the standard Bekenstein--Hawking
paradigm. For instance, such solutions generally lack horizons, holographic
structures, or related duality properties. Instead, these cosmological
models, and their quasi-stationary duals, are characterized by specific
nonlinear symmetries. For nonholonomic Einstein manifolds, these new classes
of solutions can be naturally distinguished and described in thermodynamic,
classical, and quantum information-theoretic forms \cite%
{gheorghiuap16,vv25a,vv25b,v25,bsssvv25,bsvv24}, by employing relativistic
and nonholonomic generalizations of G. Perelman's concept of W-entropy \cite%
{perelman1}. We have already demonstrated how geometric thermodynamic
variables can be defined and computed in modern cosmological contexts (see
previous section). Developing this approach further particularly in refining
such results and achieving closer consistency with observational data
remains an open task for future research on off-diagonal cosmological
models. %%%%%

\vskip5pt \textbf{Acknowledgement:} This work was conducted within the
framework of a visiting fellowship at Kocaeli University in T\"{u}rkiye and
builds upon previous volunteer research programs at California State
University, Fresno, USA, and Taras Shevchenko National University of Kyiv,
Ukraine. The author is grateful to the referees whose valuable critical
remarks and suggestions allowed him to extend the paper and make the
geometric methods and AFCDM more accessible to researchers in modern
cosmology. %%%%

\newpage

\appendix\setcounter{equation}{0} 
\renewcommand{\theequation}
{A.\arabic{equation}} \setcounter{subsection}{0} 
\renewcommand{\thesubsection}
{A.\arabic{subsection}}

\section{Tables and ansatz for generating off-diagonal cosmologies}

\label{appendixa}

In this appendix, we summarize the procedure for the general decoupling and
integration of (modified) Einstein equations with generic off-diagonal
quasi-stationary and locally anisotropic cosmological metrics in GR \cite%
{vv25a,vv25b}. Detailed geometric constructions and rigorous proofs,
including various generalizations for MGTs, are reviewed in \cite%
{sv11,bsssvv25}. The AFCDM for GR, formulated in canonical nonholonomic
variables and introduced in section \ref{sec2} for generating off-diagonal
solutions, is outlined below in Tables A1 and A2. We also illustrate the use
of 2+2 nonholonomic variables and the corresponding ansatz for $\tau$%
-families of cosmological d-metrics, with references to earlier works on
more general constructions for metric-affine MGTs in \cite%
{gheorghiuap16,bsvv24}.

\subsection{Tables A1 and A2 for constructing locally anisotropic
cosmological solutions}

We employ a system of notations that allows us to generate, in abstract
geometric and N-adapted coefficient forms, two classes of generic
off-diagonal solutions: quasi-stationary and locally anisotropic
cosmological configurations. We then demonstrate how to construct, in full
generality, off-diagonal cosmological metrics. Primary cosmological metrics
can be chosen in diagonal form to model solutions in GR or MGTs, while $\tau 
$-families of target metrics are generated to define exact or parametric
cosmological solutions in GR and certain generalizations for relativistic
geometric flows of NESs.

\subsubsection{Off-diagonal ansatz and nonlinear PDEs}

In Table A1, we present the two essential types of parameterizations for
frames and coordinates on 4-dimensional Lorentz manifolds equipped with an
N-connection structure, featuring h- and v-splitting.

%\newpage

%%%%%%%%%%%
% Table A1
%%%%%%%%%%%

%\vskip5pt
%\begin{table*}[h]
{\scriptsize 
\begin{eqnarray*}
&&%
\begin{tabular}{l}
\hline\hline
\begin{tabular}{lll}
& {\ \textsf{Table A1:\ Diagonal and off-diagonal ansatz for systems of
nonlinear ODEs and PDEs} } &  \\ 
& to apply the Anholonomic Frame and Connection Deformation Method, \textbf{%
AFCDM}, &  \\ 
& \textit{for constructing }$\tau $-families of\textit{\ generic
off-diagonal exact and parametric solutions} & 
\end{tabular}%
\end{tabular}
\\
&&{%
\begin{tabular}{lll}
\hline
diagonal ansatz: PDEs $\rightarrow $ \textbf{ODE}s &  & AFCDM: \textbf{PDE}s 
\textbf{with decoupling; \ generating functions} \\ 
radial coordinates $u^{\alpha }=(r,\theta ,\varphi ,t)$ & $u=(x,y):$ & 
\mbox{ nonholonomic 2+2
splitting, } $u^{\alpha }=(x^{1},x^{2},y^{3},y^{4}=t)$ \\ 
LC-connection $\mathring{\nabla}$ & [connections] & $%
\begin{array}{c}
\mathbf{N}:T\mathbf{V}=hT\mathbf{V}\oplus vT\mathbf{V,}\mbox{ locally }%
\mathbf{N}=\{N_{i}^{a}(x,y)\} \\ 
\mbox{ canonical connection distortion }\widehat{\mathbf{D}}=\nabla +%
\widehat{\mathbf{Z}};\widehat{\mathbf{D}}\mathbf{g=0,} \\ 
\widehat{\mathcal{T}}[\mathbf{g,N,}\widehat{\mathbf{D}}]%
\mbox{ canonical
d-torsion}%
\end{array}%
$ \\ 
$%
\begin{array}{c}
\mbox{ diagonal ansatz  }g_{\alpha \beta }(u) \\ 
=\left( 
\begin{array}{cccc}
\mathring{g}_{1} &  &  &  \\ 
& \mathring{g}_{2} &  &  \\ 
&  & \mathring{g}_{3} &  \\ 
&  &  & \mathring{g}_{4}%
\end{array}%
\right)%
\end{array}%
$ & $\mathbf{g}(\tau )\Leftrightarrow $ & $%
\begin{array}{c}
g_{\alpha \beta }(\tau )=%
\begin{array}{c}
g_{\alpha \beta }(\tau ,x^{i},y^{a})\mbox{ general frames / coordinates} \\ 
\left[ 
\begin{array}{cc}
g_{ij}(\tau )+N_{i}^{a}(\tau )N_{j}^{b}(\tau )h_{ab}(\tau ) & N_{i}^{b}(\tau
)h_{cb}(\tau ) \\ 
N_{j}^{a}(\tau )h_{ab}(\tau ) & h_{ac}(\tau )%
\end{array}%
\right] ,\mbox{ 2 x 2 blocks }%
\end{array}
\\ 
\mathbf{g}_{\alpha \beta }(\tau )=[g_{ij}(\tau ),h_{ab}(\tau )], \\ 
\mathbf{g}(\tau )=\mathbf{g}_{i}(\tau ,x^{k})dx^{i}\otimes dx^{i}+\mathbf{g}%
_{a}(\tau ,x^{k},y^{b})\mathbf{e}^{a}(\tau )\otimes \mathbf{e}^{b}(\tau )%
\end{array}%
$ \\ 
$\mathring{g}_{\alpha \beta }=\left\{ 
\begin{array}{cc}
\mathring{g}_{\alpha }(r) & \mbox{ for BHs} \\ 
\mathring{g}_{\alpha }(t) & \mbox{ for FLRW }%
\end{array}%
\right. $ & [coord.frames] & $g_{\alpha \beta }(\tau )=\left\{ 
\begin{array}{cc}
g_{\alpha \beta }(\tau ,x^{i},y^{3}) & 
\mbox{ quasi-stationary
configurations} \\ 
\underline{g}_{\alpha \beta }(\tau ,x^{i},y^{4}=t) & 
\mbox{ locally anisotr.
cosmology}%
\end{array}%
\right. $ \\ 
&  &  \\ 
$%
\begin{array}{c}
\mbox{coord. transf. }e_{\alpha }=e_{\ \alpha }^{\alpha ^{\prime }}\partial
_{\alpha ^{\prime }}, \\ 
e^{\beta }=e_{\beta ^{\prime }}^{\ \beta }du^{\beta ^{\prime }},\mathring{g}%
_{\alpha \beta }=\mathring{g}_{\alpha ^{\prime }\beta ^{\prime }}e_{\ \alpha
}^{\alpha ^{\prime }}e_{\ \beta }^{\beta ^{\prime }} \\ 
\begin{array}{c}
\mathbf{\mathring{g}}_{\alpha }(x^{k},y^{a})\rightarrow \mathring{g}_{\alpha
}(r),\mbox{ or }\mathring{g}_{\alpha }(t), \\ 
\mathring{N}_{i}^{a}(x^{k},y^{a})\rightarrow 0.%
\end{array}%
\end{array}%
$ & [N-adapt. fr.] & $\left\{ 
\begin{array}{cc}
\begin{array}{c}
\mathbf{g}_{i}(\tau ,x^{k}),\mathbf{g}_{a}(\tau ,x^{k},y^{3}), \\ 
\mbox{ or }\mathbf{g}_{i}(\tau ,x^{k}),\underline{\mathbf{g}}_{a}(\tau
,x^{k},t),%
\end{array}
& \mbox{ d-metrics } \\ 
\begin{array}{c}
N_{i}^{3}(\tau )=w_{i}(\tau ,x^{k},y^{3}),N_{i}^{4}=n_{i}(\tau ,x^{k},y^{3}),
\\ 
\mbox{ or }\underline{N}_{i}^{3}(\tau )=\underline{n}_{i}(\tau ,x^{k},t),%
\underline{N}_{i}^{4}=\underline{w}_{i}(\tau ,x^{k},t),%
\end{array}
& 
\end{array}%
\right. $ \\ 
$\mathring{\nabla},$ $Ric=\{\mathring{R}_{\ \beta \gamma }\}$ & Ricci tensors
& $\widehat{\mathbf{D}}(\tau ),\ \widehat{\mathcal{R}}ic(\tau )=\{\widehat{%
\mathbf{R}}_{\ \beta \gamma }(\tau )\}$ \\ 
$~^{m}\mathcal{L[\mathbf{\phi }]\rightarrow }\ ^{m}\mathbf{T}_{\alpha \beta }%
\mathcal{[\mathbf{\phi }]}$ & 
\begin{tabular}{l}
generating \\ 
sources%
\end{tabular}
& $%
\begin{array}{cc}
\widehat{\mathbf{J}}_{\ \nu }^{\mu }(\tau )=\mathbf{e}_{\ \mu ^{\prime
}}^{\mu }\mathbf{e}_{\nu }^{\ \nu ^{\prime }}\mathbf{J}_{\ \nu ^{\prime
}}^{\mu ^{\prime }}[\ ^{m}\mathcal{L}(\mathbf{\varphi ),}T_{\mu \nu }(\tau
),\Lambda (\tau )] &  \\ 
=diag[\ ^{h}J(\tau ,x^{i})\delta _{j}^{i},\ ^{v}J(\tau ,x^{i},y^{3})\delta
_{b}^{a}], & \mbox{ quasi-stat. conf.} \\ 
=diag[\ ^{h}J(\tau ,x^{i})\delta _{j}^{i},\ ^{v}\underline{J}(\tau
,x^{i},t)\delta _{b}^{a}], & \mbox{locally anisot. cosmology}%
\end{array}%
$ \\ 
trivial equations for $\mathring{\nabla}$-torsion & LC-conditions & $%
\widehat{\mathbf{D}}_{\mid \widehat{\mathcal{T}}\rightarrow 0}(\tau )=%
\mathbf{\nabla }(\tau )\mbox{ extracting new classes of solutions in GR}.$
\\ \hline\hline
\end{tabular}%
}
\end{eqnarray*}%
}This table can be extended for higher dimension Lorentz manifolds and (co)
tangent Lorentz bundles as considered in{\scriptsize \ }\cite{bsssvv25}.

\subsubsection{$\protect\tau $-families of off-diagonal locally anisotropic
cosmological solutions}

We summarize below the main steps for constructing off-diagonal, locally
anisotropic solutions of the (modified) Einstein equations using the AFCDM
approach: %%%%%%%%%%%
% Table A2
%%%%%%%%%%%
%\vskip5pt
%\begin{table*}[t]
{\scriptsize 
\begin{eqnarray*}
&&%
\begin{tabular}{l}
\hline\hline
\begin{tabular}{lll}
& {\large \textsf{Table A2:\ Off-diagonal locally anisotropic cosmological
models}} &  \\ 
& Exact solutions of $\widehat{\mathbf{R}}_{\mu \nu }=\underline{\mathbf{%
\Upsilon }}_{\mu \nu }$ (\ref{cdeq1}) or $\widehat{\mathbf{R}}_{\mu \nu
}(\tau )=\underline{\Lambda }(\tau )\underline{\mathbf{g}}_{\mu \nu }(\tau )$%
(\ref{cdeq1b}) transformed into a system of nonlinear PDEs (\ref{eq1}) - (%
\ref{e2c}); & 
\end{tabular}
\\ 
\mbox{ cosmological nonholonomic Ricci-solitons, }$\tau =\tau _{0},%
\mbox{
which reproduces the formulas from Table 3 in Appedix B to }$\cite%
{vv25a,vv25b}%
\end{tabular}
\\
&&%
\begin{tabular}{lll}
\hline\hline
$%
\begin{array}{c}
\mbox{d-metric ansatz with} \\ 
\mbox{Killing symmetry }\partial _{3}=\partial _{\varphi }%
\end{array}%
$ &  & $%
\begin{array}{c}
d\underline{s}^{2}=g_{i}(x^{k})(dx^{i})^{2}+\underline{g}%
_{a}(x^{k},y^{4})(dy^{a}+\underline{N}_{i}^{a}(x^{k},y^{4})dx^{i})^{2},%
\mbox{ for } \\ 
g_{i}=e^{\psi {(x}^{k}{)}},\,\,\,\,\underline{g}_{a}=\underline{h}_{a}({x}%
^{k},t),\ \underline{N}_{i}^{3}=\underline{n}_{i}({x}^{k},t),\,\,\,%
\underline{\,N}_{i}^{4}=\underline{w}_{i}({x}^{k},t),%
\end{array}%
$ \\ 
&  &  \\ 
Effective matter sources &  & $\underline{\mathbf{\Upsilon }}_{\ \nu }^{\mu
}=[~\ _{h}\Upsilon ({x}^{k})\delta _{j}^{i},~\ _{v}\underline{\Upsilon }({x}%
^{k},t)\delta _{b}^{a}];x^{1},x^{2},y^{3},y^{4}=t$ \\ \hline
Nonlinear PDEs &  & $%
\begin{array}{c}
\psi ^{\bullet \bullet }+\psi ^{\prime \prime }=2\ ^{h}\Upsilon ; \\ 
\underline{\varpi }^{\diamond }\ \underline{h}_{3}^{\diamond }=2\underline{h}%
_{3}\underline{h}_{4}\ ^{v}\underline{\Upsilon }; \\ 
\underline{n}_{k}^{\diamond \diamond }+\underline{\gamma }\underline{n}%
_{k}^{\diamond }=0; \\ 
\underline{\beta }\underline{w}_{i}-\underline{\alpha }_{i}=0;%
\end{array}%
$ for $%
\begin{array}{c}
\underline{\varpi }{=\ln |\partial _{t}\underline{{h}}_{3}/\sqrt{|\underline{%
h}_{3}\underline{h}_{4}|}|,} \\ 
\underline{\alpha }_{i}=(\partial _{t}\underline{h}_{3})\ (\partial _{i}%
\underline{\varpi }),\ \underline{\beta }=(\partial _{t}\underline{h}_{3})\
(\partial _{t}\underline{\varpi }), \\ 
\ \underline{\gamma }=\partial _{t}\left( \ln |\underline{h}_{3}|^{3/2}/|%
\underline{h}_{4}|\right) , \\ 
\partial _{1}q=q^{\bullet },\partial _{2}q=q^{\prime },\partial
_{4}q=\partial q/\partial t=q^{\diamond }%
\end{array}%
$ \\ \hline
$%
\begin{array}{c}
\mbox{ Generating functions:}\ \underline{h}_{4}({x}^{k},t), \\ 
\underline{\Psi }(x^{k},t)=e^{\underline{\varpi }},\underline{\Phi }({x}%
^{k},t); \\ 
\mbox{integr. functions:}\ h_{4}^{[0]}(x^{k}),\ _{1}n_{k}(x^{i}),\  \\ 
_{2}n_{k}(x^{i});\mbox{\& nonlinear symmetries}%
\end{array}%
$ &  & $%
\begin{array}{c}
\ (\underline{\Psi }^{2})^{\diamond }=-\int dt\ ^{v}\underline{\Upsilon }%
\underline{h}_{3}^{\diamond }, \\ 
\underline{\Phi }^{2}=-4\ \underline{\Lambda }\underline{h}_{3}; \\ 
\underline{h}_{3}=\underline{h}_{3}^{[0]}-\underline{\Phi }^{2}/4\ 
\underline{\Lambda },\underline{h}_{3}^{\diamond }\neq 0,\ \underline{%
\Lambda }\neq 0=const%
\end{array}%
$ \\ \hline
Off-diag. solutions, $%
\begin{array}{c}
\mbox{d--metric} \\ 
\mbox{N-connec.}%
\end{array}%
$ &  & $%
\begin{array}{c}
\ g_{i}=e^{\ \psi (x^{k})}\mbox{ as a solution of 2-d Poisson eqs. }\psi
^{\bullet \bullet }+\psi ^{\prime \prime }=2\ ^{h}\underline{\Upsilon }; \\ 
\overline{h}_{4}=-(\overline{\Psi }^{2})^{\diamond }/4\ ^{v}\underline{%
\Upsilon }^{2}\underline{h}_{3}; \\ 
\underline{h}_{3}=h_{3}^{[0]}-\int dt(\underline{\Psi }^{2})^{\diamond }/4\
^{v}\underline{\Upsilon }=h_{3}^{[0]}-\underline{\Phi }^{2}/4\ \underline{%
\Lambda }; \\ 
\underline{n}_{k}=\ _{1}n_{k}+\ _{2}n_{k}\int dt(\underline{\Psi }^{\diamond
})^{2}/\ ^{v}\underline{\Upsilon }^{2}\ |h_{3}^{[0]}-\int dt(\underline{\Psi 
}^{2})^{\diamond }/4\ ^{v}\underline{\Upsilon }|^{5/2}; \\ 
\underline{w}_{i}=\partial _{i}\ \underline{\Psi }/\ \partial _{t}\underline{%
\Psi }=\partial _{i}\underline{\Psi }^{2}/\ \partial _{t}\underline{\Psi }%
^{2}. \\ 
\end{array}%
$ \\ \hline
LC-configurations &  & $%
\begin{array}{c}
\partial _{t}\underline{w}_{i}=(\partial _{i}-\underline{w}_{i}\partial
_{t})\ln \sqrt{|\underline{h}_{4}|},(\partial _{i}-\underline{w}_{i}\partial
_{4})\ln \sqrt{|\underline{h}_{3}|}=0, \\ 
\partial _{k}\underline{w}_{i}=\partial _{i}\underline{w}_{k},\partial _{t}%
\underline{n}_{i}=0,\partial _{i}\underline{n}_{k}=\partial _{k}\underline{n}%
_{i}; \\ 
\underline{\Psi }=\underline{\check{\Psi}}(x^{i},t),(\partial _{i}\underline{%
\check{\Psi}})^{\diamond }=\partial _{i}(\underline{\check{\Psi}}^{\diamond
})\mbox{ and } \\ 
\ ^{v}\underline{\Upsilon }(x^{i},t)=\underline{\Upsilon }[\underline{\check{%
\Psi}}]=\underline{\check{\Upsilon}},\mbox{ or }\underline{\Upsilon }=const.
\\ 
\end{array}%
$ \\ \hline
N-connections, zero torsion &  & $%
\begin{array}{c}
\underline{n}_{k}=\underline{\check{n}}_{k}=\partial _{k}\underline{n}(x^{i})
\\ 
\mbox{ and }\underline{w}_{i}=\partial _{i}\underline{\check{A}}=\left\{ 
\begin{array}{c}
\partial _{i}(\int dt\ \underline{\check{\Upsilon}}\ \underline{\check{h}}%
_{3}^{\diamond }])/\underline{\check{\Upsilon}}\ \underline{\check{h}}%
_{3}^{\diamond }{}; \\ 
\partial _{i}\underline{\check{\Psi}}/\underline{\check{\Psi}}^{\diamond };
\\ 
\partial _{i}(\int dt\ \underline{\check{\Upsilon}}(\underline{\check{\Phi}}%
^{2})^{\diamond })/\underline{\check{\Phi}}^{\diamond }\underline{\check{%
\Upsilon}};%
\end{array}%
\right. .%
\end{array}%
$ \\ \hline
$%
\begin{array}{c}
\mbox{polarization functions} \\ 
\mathbf{\mathring{g}}\rightarrow \underline{\widehat{\mathbf{g}}}\mathbf{=}[%
\underline{g}_{\alpha }=\underline{\eta }_{\alpha }\underline{\mathring{g}}%
_{\alpha },\underline{\eta }_{i}^{a}\underline{\mathring{N}}_{i}^{a}]%
\end{array}%
$ &  & $%
\begin{array}{c}
ds^{2}=\underline{\eta }_{i}(x^{k},t)\underline{\mathring{g}}%
_{i}(x^{k},t)[dx^{i}]^{2}+\underline{\eta }_{3}(x^{k},t)\underline{\mathring{%
h}}_{3}(x^{k},t)[dy^{3}+\underline{\eta }_{i}^{3}(x^{k},t)\underline{%
\mathring{N}}_{i}^{3}(x^{k},t)dx^{i}]^{2} \\ 
+\underline{\eta }_{4}(x^{k},t)\underline{\mathring{h}}_{4}(x^{k},t)[dt+%
\underline{\eta }_{i}^{4}(x^{k},t)\underline{\mathring{N}}%
_{i}^{4}(x^{k},t)dx^{i}]^{2}, \\ 
\end{array}%
$ \\ \hline
$%
\begin{array}{c}
\mbox{ Prime metric defines } \\ 
\mbox{ a cosmological solution}%
\end{array}%
$ &  & $%
\begin{array}{c}
\lbrack \underline{\mathring{g}}_{i}(x^{k},t),\underline{\mathring{g}}_{a}=%
\underline{\mathring{h}}_{a}(x^{k},t);\underline{\mathring{N}}_{k}^{3}=%
\underline{\mathring{w}}_{k}(x^{k},t),\underline{\mathring{N}}_{k}^{4}=%
\underline{\mathring{n}}_{k}(x^{k},t)] \\ 
\mbox{diagonalizable by frame/ coordinate transforms.} \\ 
\end{array}%
$ \\ 
$%
\begin{array}{c}
\mbox{Example of a prime } \\ 
\mbox{ cosmological metric }%
\end{array}%
$ &  & $%
\begin{array}{c}
\mathring{g}_{1}=a^{2}(t)/(1-kr^{2}),\mathring{g}_{2}=a^{2}(t)r^{2}, \\ 
\underline{\mathring{h}}_{3}=a^{2}(t)r^{2}\sin ^{2}\theta ,\underline{%
\mathring{h}}_{4}=c^{2}=const,k=\pm 1,0; \\ 
\mbox{ any frame transform of a FLRW or a Bianchi metrics} \\ 
\end{array}%
$ \\ \hline
Solutions for polarization funct. &  & $%
\begin{array}{c}
\eta _{i}=e^{\ \psi (x^{k})}/\mathring{g}_{i};\underline{\eta }_{4}%
\underline{\mathring{h}}_{4}=-\frac{4[(|\underline{\eta }_{3}\underline{%
\mathring{h}}_{3}|^{1/2})^{\diamond }]^{2}}{|\int dt\ ^{v}\underline{%
\Upsilon }[(\underline{\eta }_{3}\underline{\mathring{h}}_{3})]^{\diamond
}|\ };\mbox{ gener. funct. }\underline{\eta }_{3}=\underline{\eta }%
_{3}(x^{i},t); \\ 
\underline{\eta }_{k}^{3}\ \underline{\mathring{N}}_{k}^{3}=\ _{1}n_{k}+16\
\ _{2}n_{k}\int dt\frac{\left( [(\underline{\eta }_{3}\underline{\mathring{h}%
}_{3})^{-1/4}]^{\diamond }\right) ^{2}}{|\int dt\ ^{v}\underline{\Upsilon }[(%
\underline{\eta }_{3}\underline{\mathring{h}}_{3})]^{\diamond }|\ };\ 
\underline{\eta }_{i}^{4}\ \underline{\mathring{N}}_{i}^{4}=\frac{\partial
_{i}\ \int dt\ ^{v}\underline{\Upsilon }(\underline{\eta }_{3}\underline{%
\mathring{h}}_{3})^{\diamond }}{\ ^{v}\underline{\Upsilon }(\underline{\eta }%
_{3}\underline{\mathring{h}}_{3})^{\diamond }},%
\end{array}%
$ \\ \hline
Polariz. funct. with zero torsion &  & $%
\begin{array}{c}
\eta _{i}=e^{\ \psi }/\mathring{g}_{i};\underline{\eta }_{4}=-\frac{4[(|%
\underline{\eta }_{3}\underline{\mathring{h}}_{3}|^{1/2})^{\diamond }]^{2}}{%
\underline{\mathring{g}}_{4}|\int dt\ ^{v}\underline{\Upsilon }[(\underline{%
\eta }_{3}\underline{\mathring{h}}_{3})]^{\diamond }|\ };%
\mbox{ gener.
funct. }\underline{\eta }_{3}=\underline{\check{\eta}}_{3}({x}^{i},t); \\ 
\underline{\eta }_{k}^{4}=\partial _{k}\underline{\check{A}}/\mathring{w}%
_{k};\underline{\eta }_{k}^{3}=(\partial _{k}\underline{n})/\mathring{n}_{k}.
\\ 
\end{array}%
$ \\ \hline\hline
\end{tabular}%
\end{eqnarray*}%
}

Tables A1 and A2 can be used to generate $\tau$-families of off-diagonal
exact and parametric cosmological solutions in GR for various prescribed
generating functions and (effective) sources. Typically, such solutions
involve six independent components of the Lorentzian metric (out of ten),
depending on at least three spacetime coordinates. They describe generic
nonlinear off-diagonal geometric flow evolution and gravitational field
dynamics under nonholonomic constraints, distortion relations, and effective
sources.

The presence of nonlinear symmetries allows the introduction of effective,
generally $\tau $-dependent, cosmological constants. The physical properties
of these off-diagonal solutions differ significantly from those obtained
using diagonalizable ansatze, revealing a new class of nonlinear phenomena
with applications in GR, modified gravity theories, accelerating cosmology,
and DE and DM physics (see \cite{vv25a,vv25b,bsssvv25,bsvv24}). Some
physically relevant cosmological examples are discussed in section \ref{sec3}%
.

\subsection{An ansatz for generating parametric off-diagonal cosmological
solutions}

In this subsection, we summarize the results of Sections 3.1 and 3.2 in \cite%
{bsssvv25} (including the respective tables); see also \cite{vv25a,vv25b}
and references therein. Using the methods presented, we can generate $\tau $%
-families of off-diagonal cosmological solutions of (\ref{cdeq1b}) with
small $\kappa $-parametric deformations in (\ref{lacosm2}) by employing
nonlinear symmetries and transformations (cf. (\ref{htransf}), (\ref%
{nonlintrsmalp}), and (\ref{nonlinsymcosm})). These d-metrics are expressed
in terms of $\chi $-polarization functions, 
\begin{equation*}
d\ \widehat{s}^{2}(\tau )=\widehat{g}_{\alpha \beta }(\tau ,r,\theta ,t;\psi
,\underline{\Lambda }(\tau ),\ ^{v}\underline{J}(\tau ))du^{\alpha
}du^{\beta }=e^{\psi _{0}(\tau ,r,\theta )}[1+\kappa \ ^{\psi }\chi (\tau
,r,\theta )][(dx^{1}(r,\theta ))^{2}+(dx^{2}(r,\theta ))^{2}]
\end{equation*}%
\begin{eqnarray}
&&+\zeta _{3}(\tau )(1+\kappa \ \underline{\chi }(\tau ))\underline{%
\mathring{g}}_{3}\{d\phi +[(\underline{\mathring{N}}_{k}^{3})^{-1}[\
_{1}n_{k}(\tau )+16\ _{2}n_{k}(\tau )[\int dt\frac{\left( \partial _{t}[(%
\underline{\zeta }_{3}(\tau )\underline{\mathring{g}}_{3})^{-1/4}]\right)
^{2}}{|\int dt\partial _{t}[\ \ ^{v}\underline{J}(\tau )(\underline{\zeta }%
_{3}(\tau )\ \underline{\mathring{g}}_{3})]|}]  \label{paramsoliton} \\
&&+\kappa \frac{16\ _{2}n_{k}(\tau )\int dt\frac{\left( \partial _{t}[(%
\underline{\zeta }_{3}(\tau )\ \underline{\mathring{g}}_{3})^{-1/4}]\right)
^{2}}{|\int dt\partial _{t}[\ ^{v}\underline{J}(\tau )(\underline{\zeta }%
_{3}(\tau )\ \underline{\mathring{g}}_{3})]|}(\frac{\partial _{t}[(%
\underline{\zeta }_{3}(\tau )\ \underline{\mathring{g}}_{3})^{-1/4}%
\underline{\chi }(\tau )\ )]}{2\partial _{t}[(\underline{\zeta }_{3}(\tau )\ 
\underline{\mathring{g}}_{3})^{-1/4}]}+\frac{\int dt\partial _{t}[\ ^{v}%
\underline{J}(\tau )(\underline{\zeta }_{3}(\tau )\underline{\chi }(\tau )\ 
\underline{\mathring{g}}_{3})]}{\int dt\partial _{t}[\ ^{v}\underline{J}%
(\tau )(\underline{\zeta }_{3}(\tau )\ \underline{\mathring{g}}_{3})]})}{\
_{1}n_{k}(\tau )+\ 16\ _{2}n_{k}(\tau )[\int dt\frac{\left( \partial _{t}[(%
\underline{\zeta }_{3}(\tau )\ \underline{\mathring{g}}_{3})^{-1/4}]\right)
^{2}}{|\int dt\partial _{t}[\ \ ^{v}\underline{J}(\tau )(\underline{\zeta }%
_{3}(\tau )\ \underline{\mathring{g}}_{3})]|}]}]\underline{\mathring{N}}%
_{k}^{3}dx^{k}\}^{2}.  \notag
\end{eqnarray}%
\begin{eqnarray*}
&&-\{\frac{4[\partial _{t}(|\underline{\zeta }_{3}(\tau )\ \underline{%
\mathring{g}}_{3}|^{1/2})]^{2}}{\ \underline{\mathring{g}}_{4}|\int dt\{\
^{v}\underline{J}(\tau )\partial _{t}(\underline{\zeta }_{3}(\tau )\ 
\underline{\mathring{g}}_{3})\}|}-\kappa \lbrack \frac{\partial _{t}(%
\underline{\chi }(\tau )|\underline{\zeta }_{3}(\tau )\ \underline{\mathring{%
g}}_{3}|^{1/2})}{4\partial _{t}(|\underline{\zeta }_{3}(\tau )\ \underline{%
\mathring{g}}_{3}|^{1/2})}-\frac{\int dt\{\ ^{v}\underline{J}(\tau )\partial
_{t}[(\underline{\zeta }_{3}(\tau )\ \underline{\mathring{g}}_{3})\underline{%
\chi }(\tau )]\}}{\int dt\{\ ^{v}\underline{J}(\tau )\partial _{t}(%
\underline{\zeta }_{3}(\tau )\ \underline{\mathring{g}}_{3})\}}]\}\ \ 
\underline{\mathring{g}}_{4} \\
&&\{dt+[\frac{\partial _{i}\ \int dt\ ^{v}\underline{J}(\tau )\ \partial _{t}%
\underline{\zeta }_{3}(\tau )}{(\underline{\mathring{N}}_{i}^{3})\ ^{v}%
\underline{J}(\tau )\partial _{t}\underline{\zeta }_{3}(\tau )}+\kappa (%
\frac{\partial _{i}[\int dt\ ^{v}\underline{J}(\tau )\ \partial _{t}(%
\underline{\zeta }_{3}(\tau )\ \underline{\mathring{g}}_{3})]}{\partial
_{i}\ [\int dt\ \ ^{v}\underline{J}(\tau )\partial _{t}\underline{\zeta }%
_{3}(\tau )]}-\frac{\partial _{t}(\underline{\zeta }_{3}(\tau )\ \underline{%
\mathring{g}}_{3})}{\partial _{t}\underline{\zeta }_{3}(\tau )})]\underline{%
\mathring{N}}_{i}^{4}dx^{i}\}^{2}
\end{eqnarray*}%
In these formulas, $\psi _{0}(\tau ,r,\theta )$ and $\ ^{\psi }\chi (\tau
,r,\theta )$ are solutions of 2-d Poisson equations. If we fix $\tau =\tau
_{0}$ in (\ref{paramsoliton}), one can generate parametric cosmological
solutions of (\ref{cdeq1}) that can be interpreted either as nonholonomic
Einstein equations, or as some cases, as examples of relativistic Ricci
solitons. The prime metric $(\ \underline{\mathring{g}}_{\alpha },\underline{%
\mathring{N}}_{k}^{a})$ can be chosen in the FLRW form (\ref{pflrw}). For
other types of cosmological models, one may consider the evolution of prime
metrics of the form (\ref{lacosm2}) or (\ref{lacosm3}).

\end{document}